\documentclass[aps,prb,twocolumn,shortbibliography,superscriptaddress]{revtex4-2}
\usepackage{epsfig}
\usepackage{epstopdf}
\usepackage{amsmath}
\usepackage{amsfonts}
\usepackage{amssymb}
\usepackage{hyperref}
\usepackage{bm}
\usepackage{makecell}
\usepackage{rotating}
\usepackage{hyperref}

\usepackage{graphicx}
\usepackage{dcolumn}
\usepackage{bm}
\usepackage{color}

\usepackage{tikz,xcolor,hyperref}

\definecolor{lime}{HTML}{A6CE39}
\DeclareRobustCommand{\orcidicon}{%
	\begin{tikzpicture}
	\draw[lime, fill=lime] (0,0)
	circle [radius=0.16]
	node[white] {{\fontfamily{qag}\selectfont \tiny ID}};
	\draw[white, fill=white] (-0.0625,0.095)
	circle [radius=0.007];
	\end{tikzpicture}
	\hspace{-2mm}
}

\foreach \x in {A, ..., Z}{%
	\expandafter\xdef\csname orcid\x\endcsname{\noexpand\href{https://orcid.org/\csname orcidauthor\x\endcsname}{\noexpand\orcidicon}}
}

\begin{document}

\title{{\em Ab-initio} overestimation of the topological region in Eu-based compounds}

\author{Giuseppe Cuono\orcidD}
\email{gcuono@magtop.ifpan.edu.pl}
\affiliation{International Research Centre MagTop, Institute of Physics, Polish Academy of Sciences,
Aleja Lotnik\'ow 32/46, PL-02668 Warsaw, Poland}

\author{Raghottam M. Sattigeri\orcidC}
\email{rsattigeri@magtop.ifpan.edu.pl}
\affiliation{International Research Centre MagTop, Institute of Physics, Polish Academy of Sciences,
Aleja Lotnik\'ow 32/46, PL-02668 Warsaw, Poland}

\author{Carmine Autieri\orcidA}
\email{autieri@magtop.ifpan.edu.pl}
\affiliation{International Research Centre MagTop, Institute of Physics, Polish Academy of Sciences,
Aleja Lotnik\'ow 32/46, PL-02668 Warsaw, Poland}

\author{Tomasz Dietl\orcidB}
\affiliation{International Research Centre MagTop, Institute of Physics, Polish Academy of Sciences,
Aleja Lotnik\'ow 32/46, PL-02668 Warsaw, Poland}


\date{\today}
\begin{abstract}
An underestimation of the fundamental band gap values by the density functional theory within the local density approximation and associated approaches is a well-known challenge of {\em ab-initio} electronic structure computations. Motivated by recent optical experiments [D. Santos-Cottin {\em et al.}, arXiv:2301.08014], we have revisited first-principle results obtained earlier for EuCd$_2$As$_2$ and extended the computational studies to the whole class of systems EuCd$_2$X$_2$ (X = P, As, Sb, Bi), to EuIn$_2$X$_2$ (X = P, As, Sb), and to nonmagnetic AEIn$_2$As$_2$ (AE= Ca, Sr, Ba) employing a hybrid functional method. We find that our approach provides the magnitude of the energy gap for EuCd$_2$As$_2$ in agreement with the experimental value. Actually, our results indicate that EuSn$_2$As$_2$, BaIn$_2$As$_2$, EuCd$_2$Bi$_2$ and EuCd$_2$SbBi are robust topological insulators, while all other compounds are topologically trivial semiconductors. The trivial band gaps of EuCd$_2$P$_2$, EuCd$_2$As$_2$ and EuCd$_2$Sb$_2$ are in the range of 1.38-1.48 eV, 0.72-0.79 eV and 0.46-0.49 eV, respectively. The topologically trivial Eu-based systems are antiferromagnetic semiconductors with a strong red shift of the energy gap in a magnetic field caused by the exchange coupling of the band states to spins localized on the 4$f$-shell of Eu ions. Additionally, the EuIn$_2$X$_2$ (X = P, As) compounds show altermagnetic exchange-induced band spin-splitting, particularly noticeable in the case of states derived from 5$d$-Eu orbitals.
\end{abstract}


\maketitle

\section{Introduction}

Surprising physics of magnetic topological materials and possible applications in sensors, metrology, computing, and catalysis \cite{Bernevig:2022_N} have triggered experimental and computational search for compounds with robust topological functionalities coexisting with or brought about by a magnetic order. In particular, high-throughput first-principles calculations, implementing the density functional theory (DFT) within the generalized gradient approximation (GGA)+Hubbard $U$, indicated that 130 compounds out of 430 magnetic materials studied have topological phases when scanning $U$ \cite{Xu:2020_N}. This outcome is encouraging and makes timely the question to what extent more computationally expensive but more reliable approaches will modify the theory predictions. In particular, it has been established that  Hedin's $GW$ approach \cite{Bechstedt:2009_pssb} or the hybrid functional method \cite{Bechstedt:2009_pssb,Schlipf13,Garza:2016_JPChL} describe correctly band gaps of semiconductors and insulators, alleviating the deficiency of the conventional DFT implementations predicting too small or even negative fundamental energy gaps for narrow-gap semiconductors.

One of the relevant material families is  EuCd$_2$X$_2$ (X = P, As, Sb) pnictides that show antiferromagnetic (AFM) ordering below 11~K \cite{Schellenberg:2011_ZAAC} and crystallize in the CaAl$_2$Si$_2$-type structure (space group P$\bar{3}$m1), in which Cd$_2$As$_2$ layers are separated by the trigonal Eu layers.  According to experimental and theoretical results, the ground state of the considered systems shows an in-plane A-type AFM ordering \cite{Rahn18,Krishna:2018_PRB}. Within GGA + $U$, if the N\'eel vector is out-of-plane, we have a Dirac band \cite{Rahn18,Hua:2018_PRB,Ma:2020_AM}, which becomes gapped for the in-plane spin directions \cite{Rahn18,Krishna:2018_PRB,Hua:2018_PRB,Ma:2020_AM,Cao_2022_PRR}. The resulting gap is in a single meV range, and the system can then be classified as an antiferromagnetic crystalline topological insulator characterized by the topological index  $Z_4 = 2$ \cite{Ma:2020_AM}. Interestingly, if non-zero magnetization appears, the GGA+$U$ computations indicate that the Dirac states evolve into a pair of Weyl bands \cite{Wang:2019_PRB,Ma:2019_SA,Ma:2020_AM,Taddei22,Wang:2022_PRB,Cao_2022_PRR}.


Experimentally observed colossal negative magnetoresistance, showing a sharp maximum at N\'eel temperature, was usually linked to that topological phase transition and the chiral anomaly specific to the Weyl systems \cite{Taddei22}. The picture implied by the GGA+$U$ computations appeared also consistent with ARPES results, though only the valence band portion of the band structure was visualized in available samples \cite{Ma:2019_SA,Ma:2020_AM,Wang:2022_PRB,Cao_2022_PRR}.
Furthermore, EuCd$_2$As$_2$ in a form of ferromagnetic quintuple layers was theoretically studied, and the presence of a topological transition between the quantum anomalous Hall insulator and quantum spin Hall insulator under an electric field predicted \cite{Niu19}.

Motivated by the recent optical studies on high-purity EuCd$_2$As$_2$ \cite{santoscottin2023eucd2as2}, we have carried out {\em ab-initio} computations for EuCd$_2$X$_2$ (X = P, As, Sb, Bi) employing the Heyd-Scuseria-Ernzerhof (HSE) hybrid functional \cite{Paier06} and compared the results to those obtained within the standard GGA+$U$ approach or with the use of the strongly constrained and appropriately normed (SCAN) functional \cite{PhysRevLett.115.036402}. Our findings confirm that the system energy is lower for the AFM spin configuration compared to the ferromagnetic spin arrangement in the whole pnictide series. Furthermore, we find that EuCd$_2$Bi$_2$ and EuCd$_2$SbBi are antiferromagnetic topological semimetals, whereas the remaining three compounds are magnetic semiconductors with a direct gap $E_g$ at the Brillouin zone center, and with the 4$f$-band at least 1.5~eV below the top of the valence band. Importantly, the determined range of $E_g=0.72$-0.79\,eV for EuCd$_2$As$_2$ within HSE+$U$, is in good agreement with the experimental value $E_g = 0.77$\,eV \cite{santoscottin2023eucd2as2}. Finally, we show that a change of $E_g$ on going from AFM to FM spin arrangements is in accord with the observed
red-shift of the band gap in a magnetic field \cite{santoscottin2023eucd2as2}. We conclude that DFT results within GGA+$U$ approach tend to overestimate the abundance of topological materials as shown for other material classes \cite{Hussain2022electronic,D3CP00005B}. At the same time, however,  the material family of EuCd$_2$X$_2$ (X = P, As, Sb) pnictides constitutes a worthwhile playground for studying the interplay of localized magnetism and quantum localization in semiconductors, which results in colossal negative magnetoresistance that is not well understood \cite{Dietl:2008_JPSJ}, despite that several decades have elapsed since its observation in AFM EuTe \cite{Shapira:1972_PRB}. At the same time, encouraging results found here on the topological robustness of EuCd$_2$Bi$_2$ in either bulk or 2D form \cite{Wang:2021_MH}, call for experimental verifications. 

Furthermore, we have also investigated  other materials with similar compositions and magnetism but different space groups. In particular, EuIn$_2$As$_2$ (space group P6$_3$/mmc) is known to be one of the few bulk axion insulators not protected by inversion symmetry proposed until now \cite{Riberolles2021,Islam2023}.
Non-magnetic AEIn$_2$As$_2$ (AE = Ca, Sr, Ba) compounds (also space group P6$_3$/mmc) were found to be topological insulators within GGA \cite{Guo2022the,PhysRevMaterials.6.044204}. EuSn$_2$As$_2$ (space group R$\bar{3}$m) was shown experimentally to be an axion insulator \cite{Xu2019higher,PhysRevX.9.041039}.
Establishing topological classes of those compounds is certainly relevant for future investigations. If not mentioned otherwise, the calculations presented here include the spin-orbit coupling (SOC). Further details on the computational methodology are given in Appendix A.


\begin{figure}[t!]
\centering
\includegraphics[width=6.3cm,angle=270]{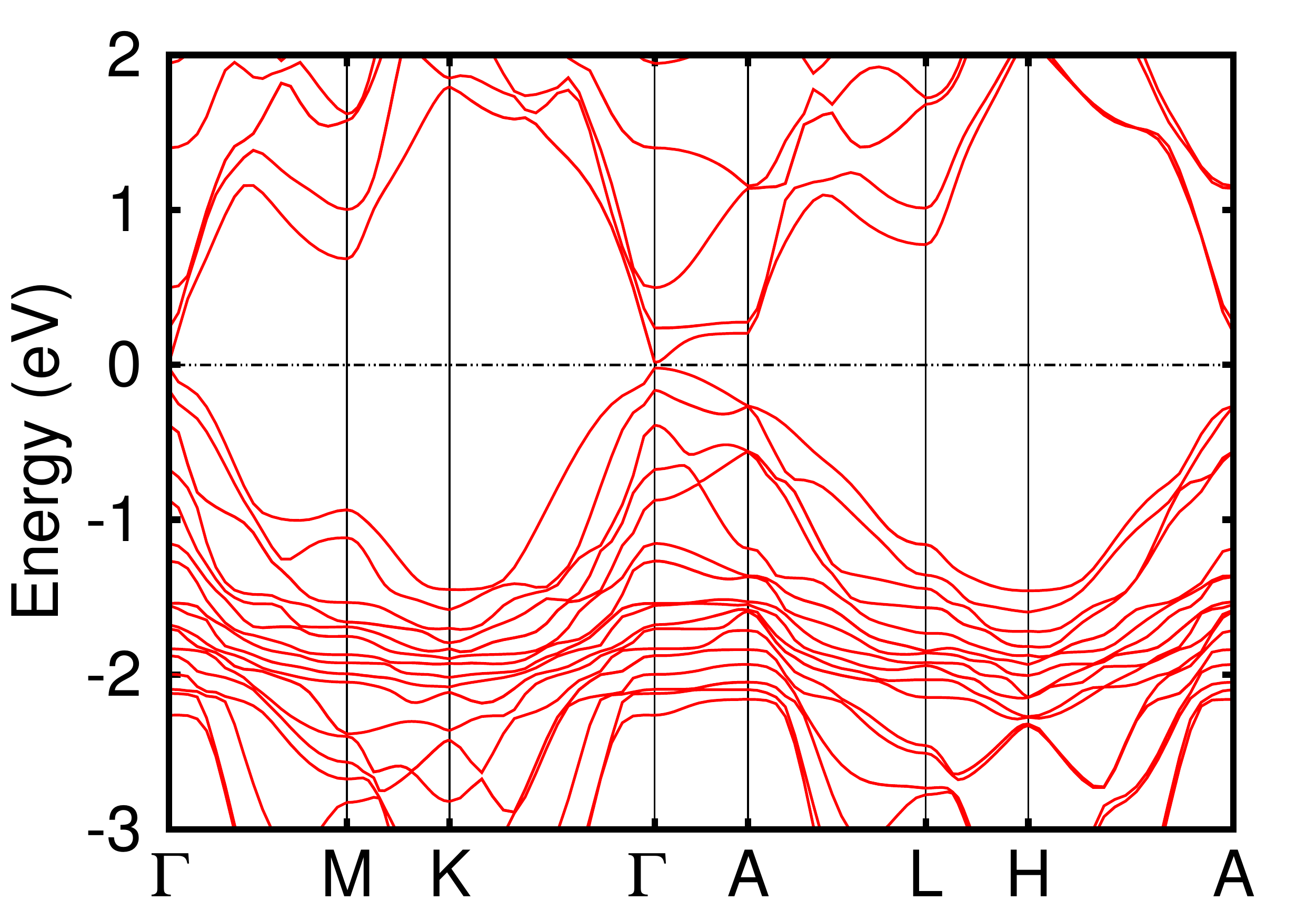}
\caption{Band structure in the AFM$a$ configuration of EuCd$_2$As$_2$ by using GGA+$U$ with $U=7$\,eV on the $f$-orbitals of Eu. The Fermi energy is set to zero. The $k$-path is described in the Appendix A.
}
\label{Band_structure}
\end{figure}

\section{Electronic properties of E\lowercase{u}C\lowercase{d}$_2$A\lowercase{s}$_2$ within GGA+$U$}

Regarding E\lowercase{u}C\lowercase{d}$_2$A\lowercase{s}$_2$, we have used the experimental lattice constants $a=b=4.44$\,{\AA} and $c=7.33$\,{\AA} and the atomic positions reported in Ref.~\onlinecite{Rahn18}.
We start by investigating the electronic properties of EuCd$_2$As$_2$ in the AFM ground state within the standard GGA+$U$ approach employing, however, a larger $U$ value  than in the previous studies, $U= 7$\,eV, which appears more appropriate for open $4f$ shells,  as discussed in Appendix B. As shown in Fig.~\ref{Band_structure},
for this value of $U=7$\,eV, the $f$-states of Eu form flat bands located at around 1.5-2.0\,eV below the Fermi level.  Furthermore, the system is in a trivial phase, though the band gap is rather small, $E_g = 40$\,meV.

\begin{figure}[t!]
\centering
\includegraphics[width=6.3cm,angle=270]{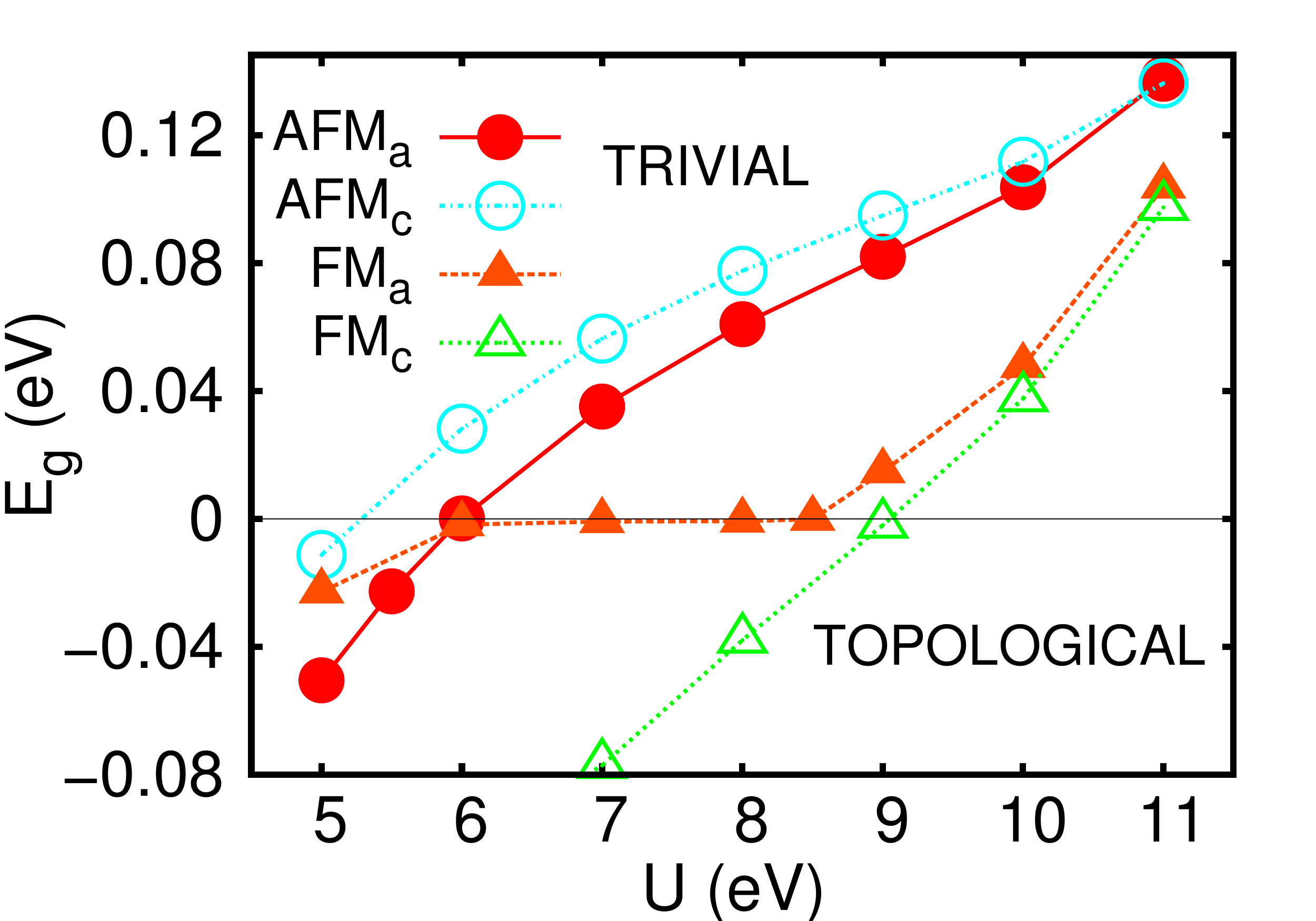}
\caption{Energy gap $E_g$ for EuCd$_2$As$_2$ as a function of the Coulomb repulsion energy $U$. The solid red line, the dot-dashed cyan line, the dashed orange line and the dotted green line are obtained in GGA+$U$ approximation and they are in the AFM phase with the spins oriented in the $a$-$b$ plane and along the $c$-axis and in the FM phases with spins oriented in the $a$-$b$ plane and along the $c$-axis, respectively. For the negative values of $E_g$ we are in the topological phase, while for positive values we are in the trivial phase.}
\label{GAP_SOC}
\end{figure}

Because of the exchange and spin-orbit interactions, the band gap strongly depends on the magnetic configuration of Eu spins. Therefore, we consider four different collinear magnetic configurations:  AFM and FM orderings with the Eu spins oriented in the $a$-$b$ plane (denoted as AFM$a$ and FM$a$)  and along the $c$-axis (AFM$c$ and FM$c$).
Figure \ref{GAP_SOC} shows the calculated band gap $E_g(\Gamma)=E_s^{\text{Cd}}(\Gamma)-E_p^{\text{As}}(\Gamma)$ as a function of the Coulomb repulsion energy $U$ for the four magnetic configurations.
In the trivial phase, the $E_g$ value is positive and indicates the trivial band arrangement such that $E_g=E_e-E_{hh}$, where $E_e$ is the energy of the conduction band bottom and $E_{hh}$ is the heavy-hole energy at the valence band top. If $E_g$ is negative, we are in the topological phase and $E_g$ indicates the magnitude of the band inversion.
For all the magnetic configurations, the gap increases as a function of $U$.
For the AFM$a$ ground state, there is a transition from the topological to the trivial phase for $U=6.0$\,eV, while it occurs around 5.5\,eV for the AFM$c$ case. For the FM phases, the transition is around $U=9.0$\,eV.
For all values of $U$, the FM$c$ phase has a lower gap than the AFM phases, therefore, an applied magnetic field along the $c$-axis will always decrease the band gap.
Within GGA+$U$, the low energy bands are of $s$-Cd, $s$-As, and $p$-As orbitals that are treated within a simple GGA. Therefore, additional effort is required to improve the band gap problem for these wide bands.


\section{Electronic properties of E\lowercase{u}C\lowercase{d}$_2$A\lowercase{s}$_2$ beyond GGA+$U$}

\begin{figure}[t!]
\centering
\includegraphics[width=1.0\linewidth,height=6.2cm]{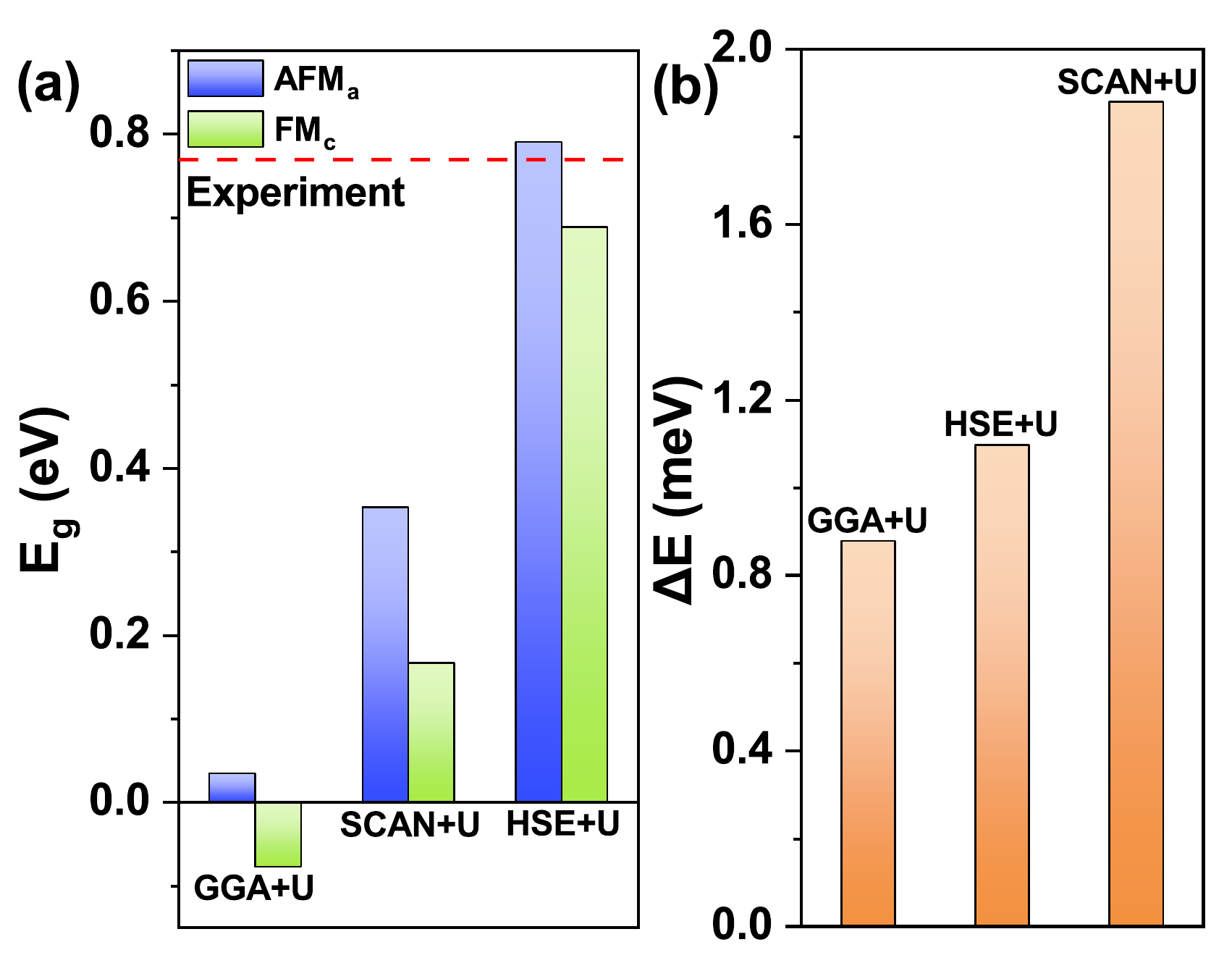}
\caption{Electronic properties of EuCd$_2$As$_2$ within different exchange-correlation functionals with the same $U= 7$\,eV. (a) $E_g$ obtained within GGA+$U$, SCAN+$U$, HSE+$U$ for the AFM$a$ and FM$c$ phases. The dashed red horizontal line indicates the experimental band gap \cite{santoscottin2023eucd2as2}. (b) Total energy difference per formula unit $\Delta E$ between the FM$c$ and AFM$a$ phases for different approximations.}
\label{GAPS_U_7}
\end{figure}

To handle the energy gap problem for wide bands originating from $s$, $p$, and $d$ cation and anion orbitals, we use the HSE and SCAN functionals. 
Within our approach, strong correlations within the $f$ shell are additionally described by  the Coulomb repulsion energy $U$.
In Fig.~\ref{GAPS_U_7}(a), we report $E_g$ magnitudes obtained within different computational methods at the same value of  $U=7$\,eV. We show the results in both AFM$a$ and FM$c$ spin configurations. Our value of $E_g$ within the GGA+$U$ approach matches previous computational results. However, the $E_g$ value within the HSE method is larger  by 0.7\,eV compared to the GGA+$U$ outcome. Interestingly, in the case of topologically nontrivial quintuple layers of  XMnY AFM materials series, the magnitude of the {\em inverted} bandgap appears also larger within HSE than that obtained employing  the GGA+$U$ method \cite{Niu20,Mao20}.

We also report the energy difference per formula unit ${\Delta}E$ between the FM$c$ and AFM$a$ spin arrangements in Fig.~\ref{GAPS_U_7}(b), where we show that the AFM$a$ is always the ground state, as observed experimentally. However, the value of ${\Delta}E$ is relatively small allowing to manipulate the magnetic phase from AFM to FM under the applied magnetic field along the $c$-axis. 
We note that the size of the gap is related to the energetic stability of the magnetic phases. When the difference between the gaps $E_g$ of the AFM and FM phases becomes larger for a given method, the energy difference $\Delta E$ configurations becomes larger too, attending 2\,meV for the SCAN+$U$ approach.

Assuming $U=7$\,eV, the band gap of 0.79\,eV, obtained within HSE+$U$, matches the experimental value from optical measurements, $E_g = 0.77$\,eV \cite{santoscottin2023eucd2as2}. The band gap decreases by $125$\,meV in the magnetic fields of 2\,T at which Eu spin magnetization saturates  \cite{santoscottin2023eucd2as2}. In our calculations with HSE+$U$ we have a $\Delta E_g=102$\,meV between the AFM$a$ and FM$c$ phases. In the SCAN+$U$ approach, we have $\Delta E_g=186$\,meV while in GGA+$U$ we find $\Delta E_g=112$\,meV.
This gap reduction can be entirely attributed to the exchange spin-splittings at the $\Gamma$ point of the highest valence band and lowest conduction band due to the presence of 4$f$-Eu in the highest valence band and 6$s$-Eu in the lowest conduction band (see Appendix C for more information).
Regarding the directions of Eu spins, in agreement with experimental observations \cite{Rahn18,Taddei22}, we find that the N\'eel vector is rather in-plane than parallel to the $c$ axis, $E(\mbox{AFM}a) < E(\mbox{AFM}c)$, with a tiny magnetocrystalline anisotropy of 0.2\,meV per the formula unit.

\begin{figure}[t!]
\centering
\includegraphics[width=2.99cm,angle=270]{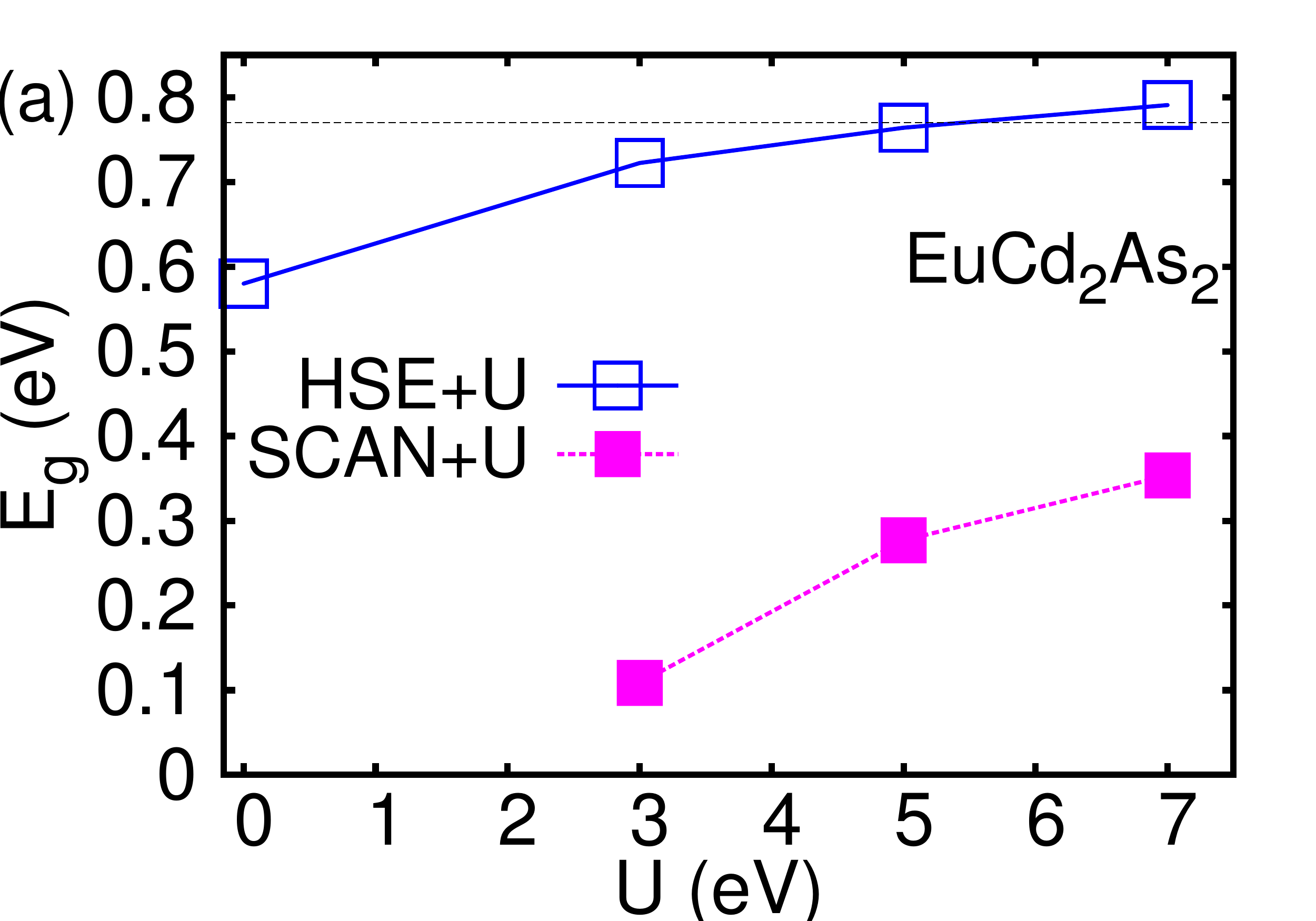}
\includegraphics[width=2.99cm,angle=270]{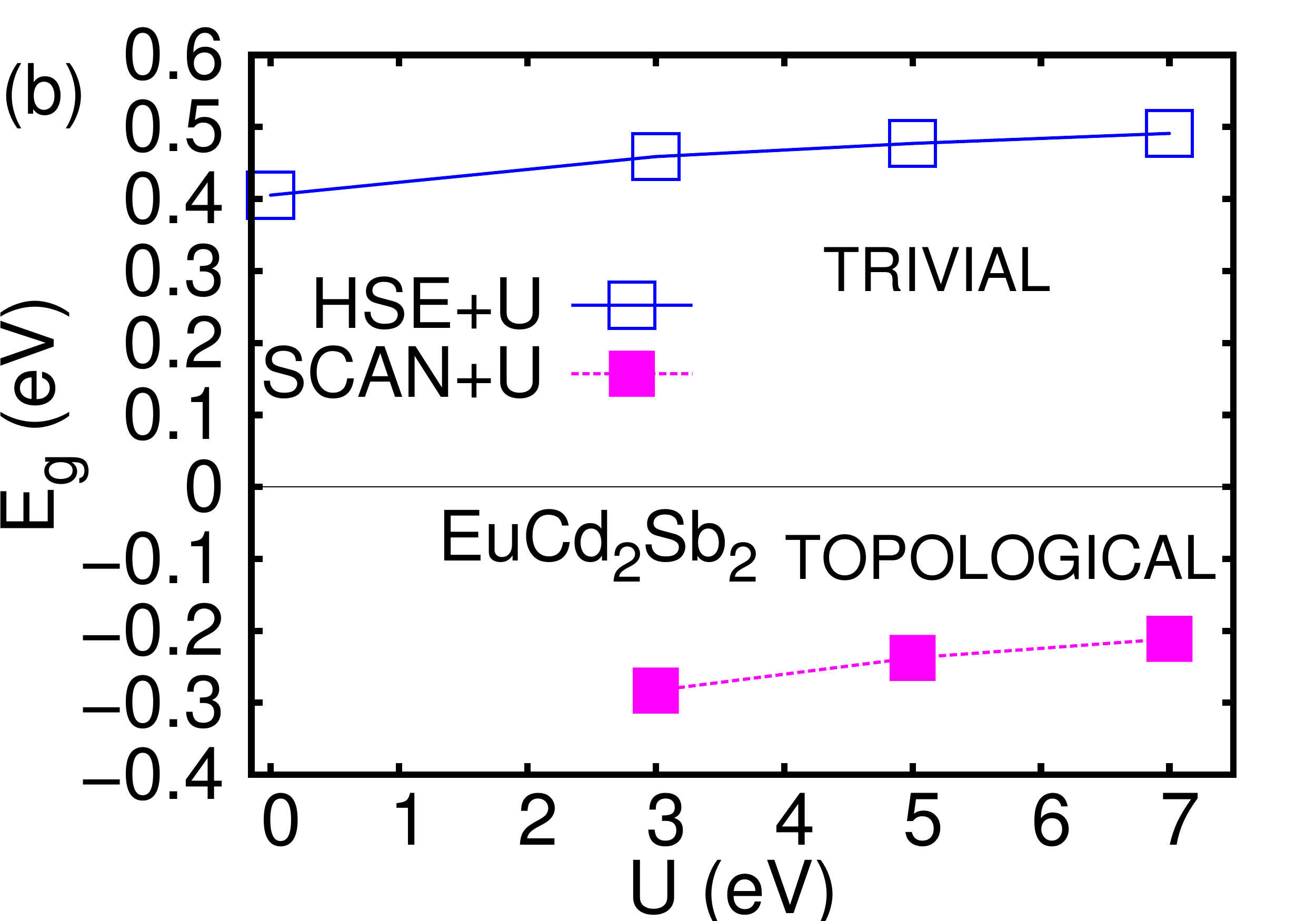}
\includegraphics[width=2.99cm,angle=270]{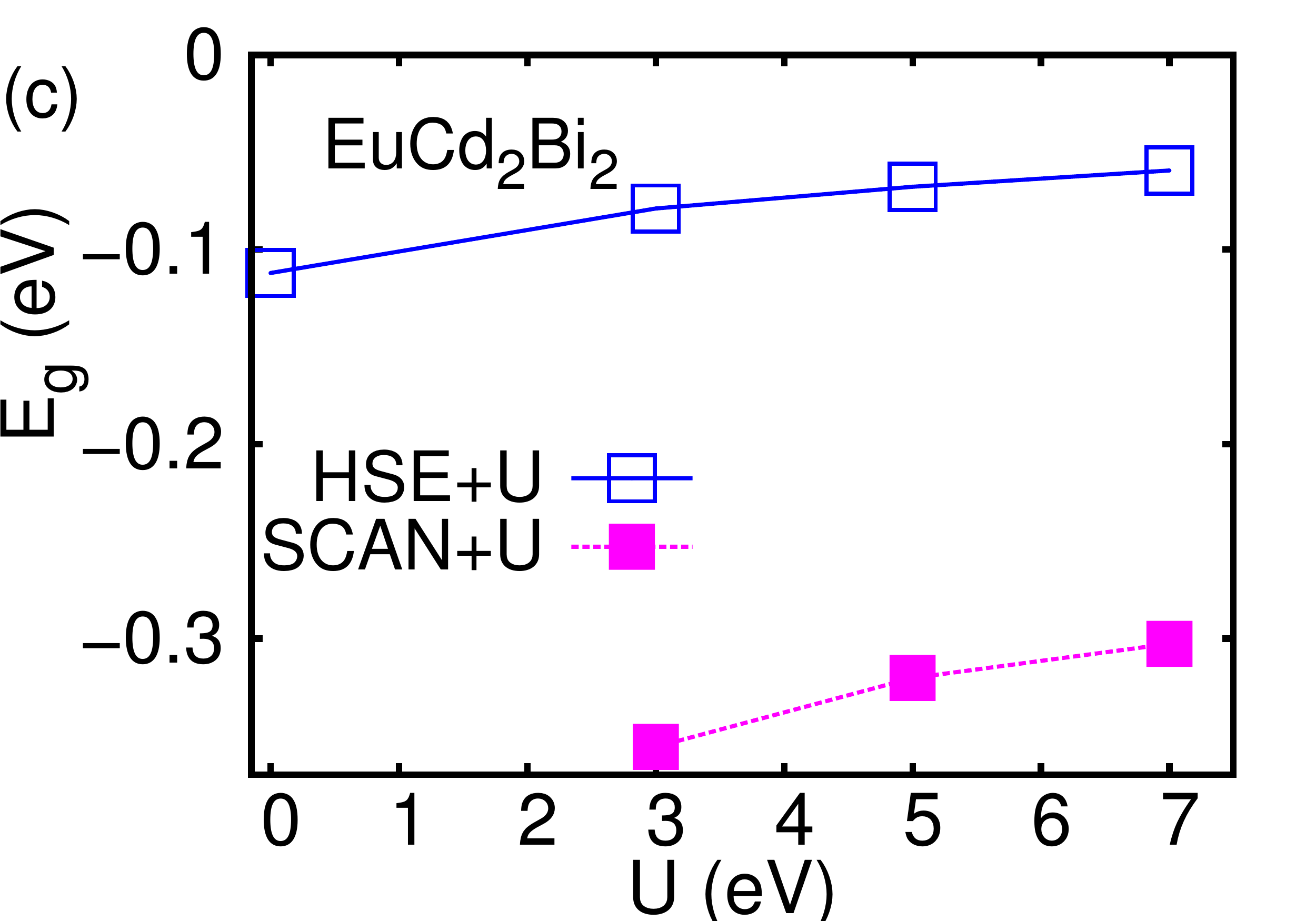}
\includegraphics[width=2.99cm,angle=270]{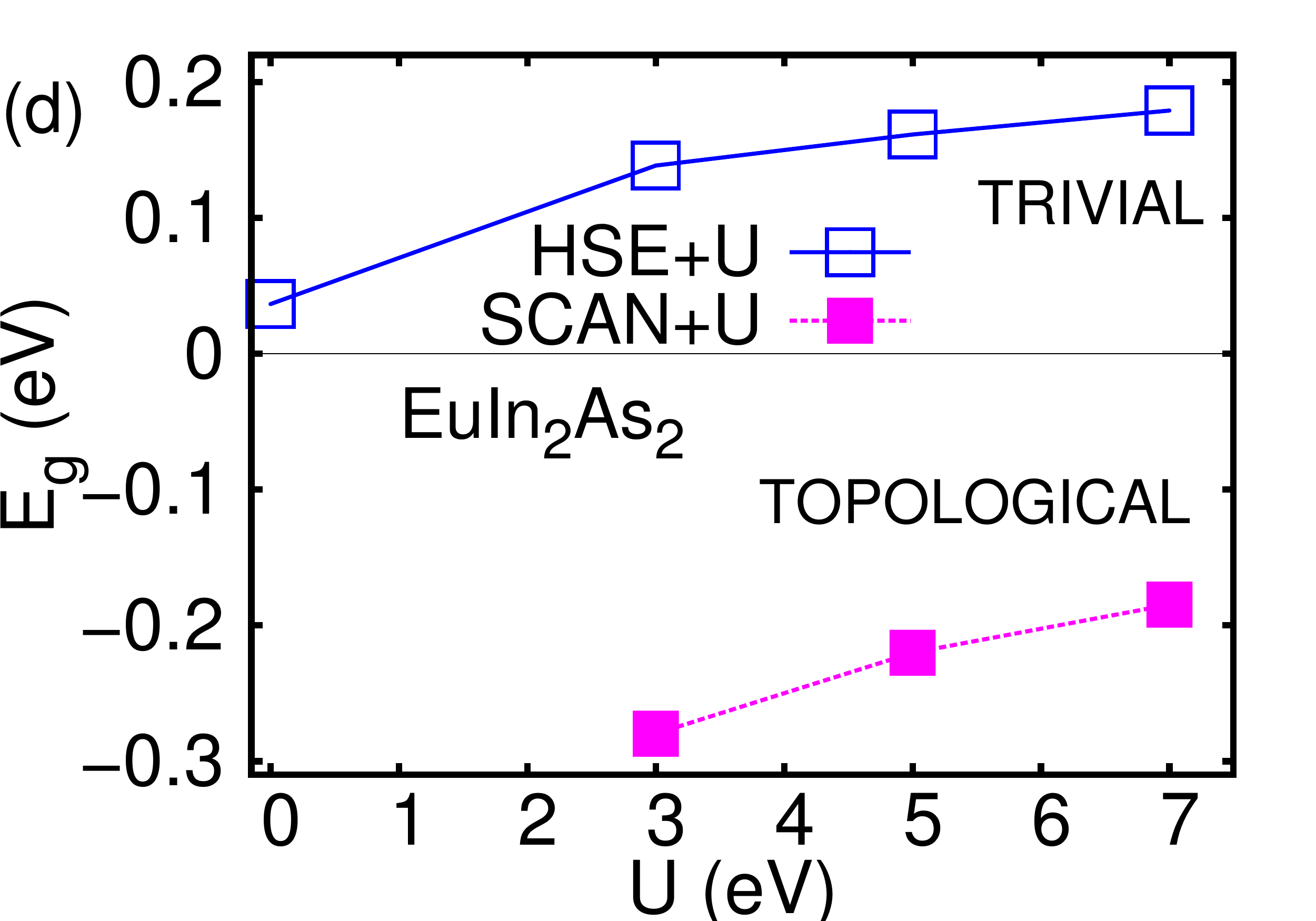}
\caption{$E_g$ for (a) EuCd$_2$As$_2$, (b) EuCd$_2$Sb$_2$, (c) EuCd$_2$Bi$_2$ and (d) EuIn$_2$As$_2$ as a function of the Coulomb repulsion in the AFM$a$ configuration by considering HSE+$U$ (blue solid line) and SCAN+$U$ (purple dashed line). The horizontal dashed line for EuCd$_2$As$_2$ represents the experimental value\cite{santoscottin2023eucd2as2}.}
\label{GAP_SOC_HYBRID_Eu}
\end{figure}

In Fig.~\ref{GAP_SOC_HYBRID_Eu}(a), we report $E_g$ for the AFM$a$ configuration as a function of $U$ obtained combining the HSE exchange-correlation functional with the addition of the Coulomb repulsion energy $U$ for $f$ electrons. For comparison, we also report the results within the SCAN+$U$ approach. We can see that for all the investigated $U$ values we are in the trivial phase.
Within SCAN+$U$ approach even a small value of $U=3.0$\,eV leads to a sizeable trivial band gap. The use of the SCAN without $U$ produces $f$-electrons very close to the Fermi level which is an unphysical result, therefore, we do not report the band gap for SCAN only.
With the use of the HSE functional, we have a trivial band gap of 0.59\,eV. This band gap increases up to 0.72-0.79\,eV within the HSE+$U$ approach and the $4f$ band moves down in energy (see Appendix D). Therefore, we have a sizeable trivial band gap for all values of $U$ which means that EuCd$_2$As$_2$ is a robust trivial insulator. The HSE method opens the trivial band gap between $s$-Cd, $s$-As, and $p$-As wide bands, however, it misses the position of the 4$f$-levels and the band gap becomes slightly underestimated. Thus, we need to use the HSE functional to correctly describe the wide bands and add $U$ for positioning properly the $f$-levels.  However, $U$ as small as 1.5\,eV is sufficient to reproduce the experimental positions of $f$ levels' within the HSE+$U$ approach (see Appendix D).

\section{Other members of the E\lowercase{u}C\lowercase{d}$_2$A\lowercase{s}$_2$ family}

We performed the calculations also for EuCd$_2$P$_2$, EuCd$_2$Sb$_2$, and EuCd$_2$Bi$_2$ that have the same space group as EuCd$_2$As$_2$, namely P$\bar{3}$m1.

Regarding EuCd$_2$P$_2$, we have used the lattice constants $a=b=4.32$\,{\AA} and $c=7.18$\,{\AA} and the atomic positions reported in Ref.~\onlinecite{Wang21Advanced}.
For EuCd$_2$Sb$_2$ we have used the lattice constants $a=b=4.69$\,{\AA} and $c=7.72$\,{\AA} and the atomic positions reported in Ref.~\onlinecite{Artmann96}, while for EuCd$_2$Bi$_2$ we have relaxed the volume and the atomic positions since the experimental values are not available. We have used $a=b=4.92$\,{\AA} and $c=7.95$\,{\AA} with a reasonable ratio of $c/a=1.62$ while for the experimental volume of the other compounds we get $c/a=1.64$-1.66.
In this material class, the theoretical volumes are larger by 1-2\% than the experimental volumes and larger volumes produce more topology.
Therefore, heavier atoms increase volume, SOC and bandwidth, all properties that increase the band inversion.

In Fig.~\ref{GAP_SOC_HYBRID_Eu}(b), we show $E_g$ for EuCd$_2$Sb$_2$ as a function of $U$ by considering HSE+$U$ and SCAN+$U$ functionals, EuCd$_2$Sb$_2$ is trivial in the HSE+$U$ case while it is topological in the SCAN+$U$ implementation.
Moving to EuCd$_2$Bi$_2$, we find this material to be topological. The inverted band gap is 0.11\,eV within the HSE approach. The ground state is always AFM with $\Delta E = 2.05$\,meV within GGA+$U$; $\Delta E = 4.35$\,meV within SCAN+$U$;
$\Delta E= 1.71$\,meV within HSE+$U$, with $U=7$\,eV, the easy axis is in-plane as for EuCd$_2$As$_2$ with the magnetocrystalline anisotropy of 0.5\,meV per formula unit. In Fig.~\ref{GAP_SOC_HYBRID_Eu}(c) we show $E_g$ for EuCd$_2$Bi$_2$ as a function of $U$ by considering HSE+$U$ and SCAN+$U$ functionals.

\begin{figure}[t!]
\centering
\includegraphics[width=1\linewidth]{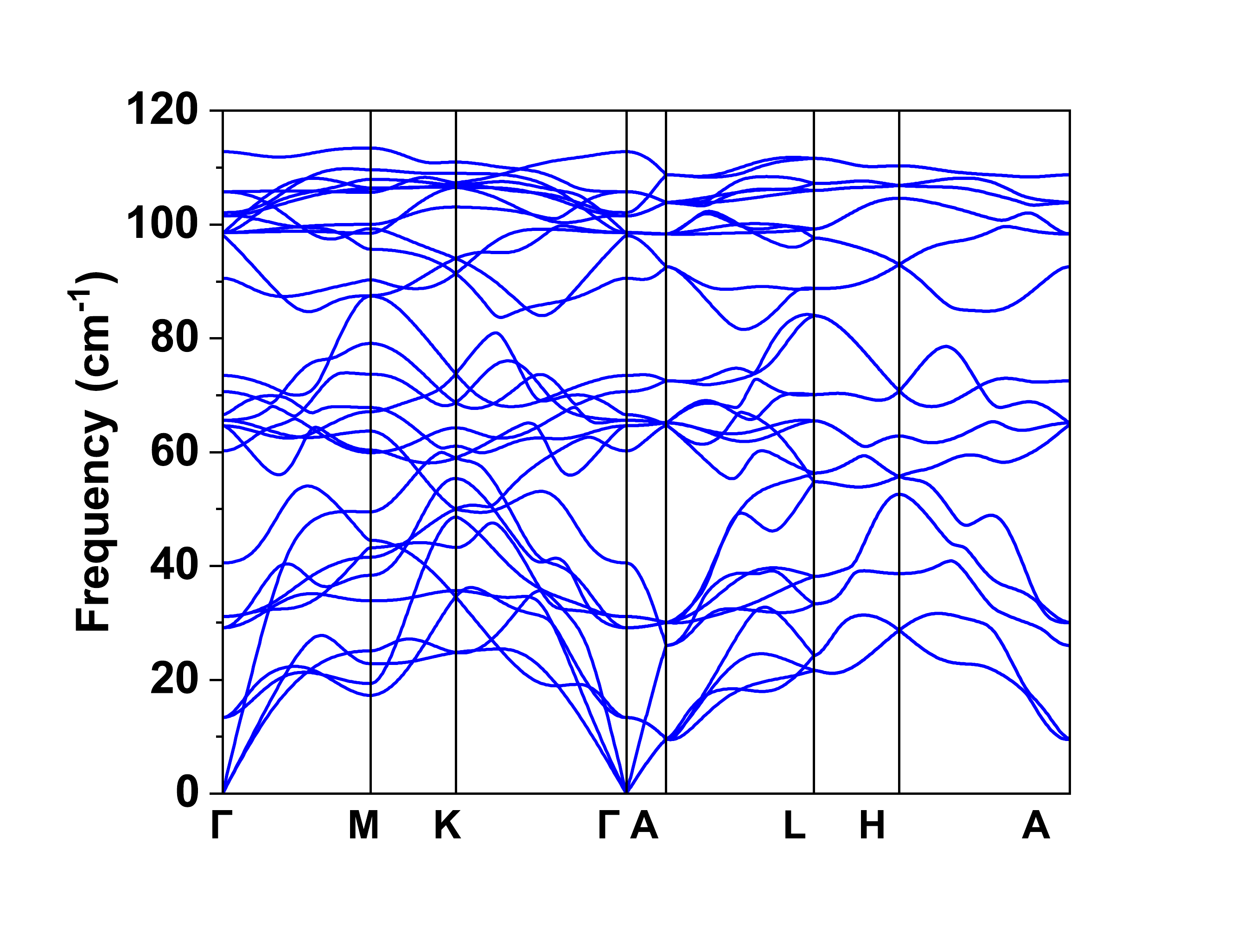}
\caption{Phonon dispersion curves of EuCd$_2$Bi$_2$ for the fully relaxed crystal structure within GGA+$U$ with $U = 7$\,eV.}
\label{Phonons}
\end{figure}

To check the structural stability of EuCd$_2$Bi$_2$, we performed phonon calculations within density functional perturbation theory approximation. The phonon dispersion curves are reported in Fig.~\ref{Phonons} which indicate the absence of imaginary frequencies in the entire Brillouin zone. This implies that the calculated crystal structure of EuCd$_2$Bi$_2$ is dynamically stable. In the literature, there is also a calculation of the phonons for non-magnetic phase with $f$-levels in the core for the EuCd$_2$As$_2$ \cite{krishna2019anisotropic,du2022comparative}.

\begin{figure}[t!]
\centering
\includegraphics[width=6.2cm,angle=270]{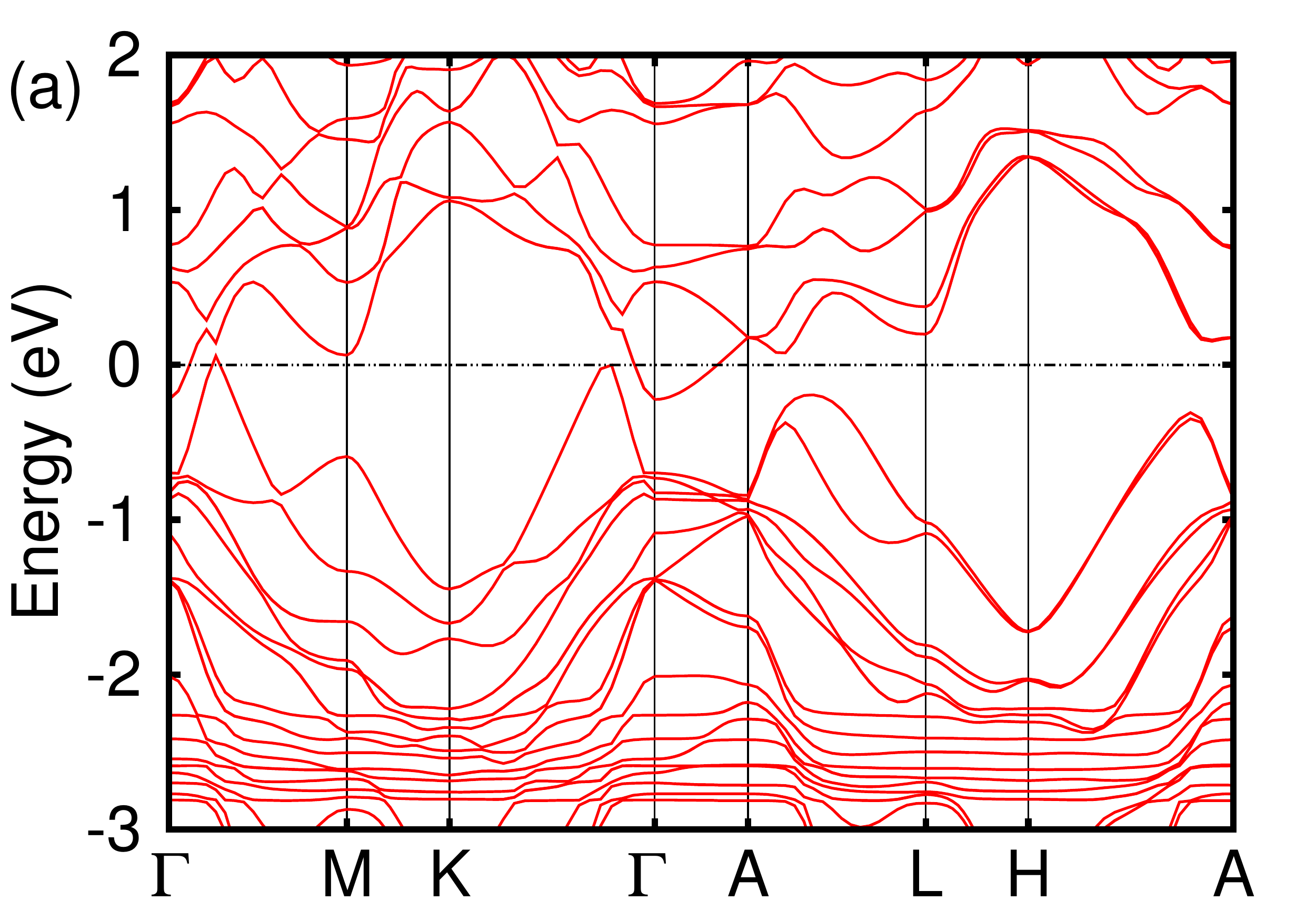}
\includegraphics[width=6.2cm,angle=270]{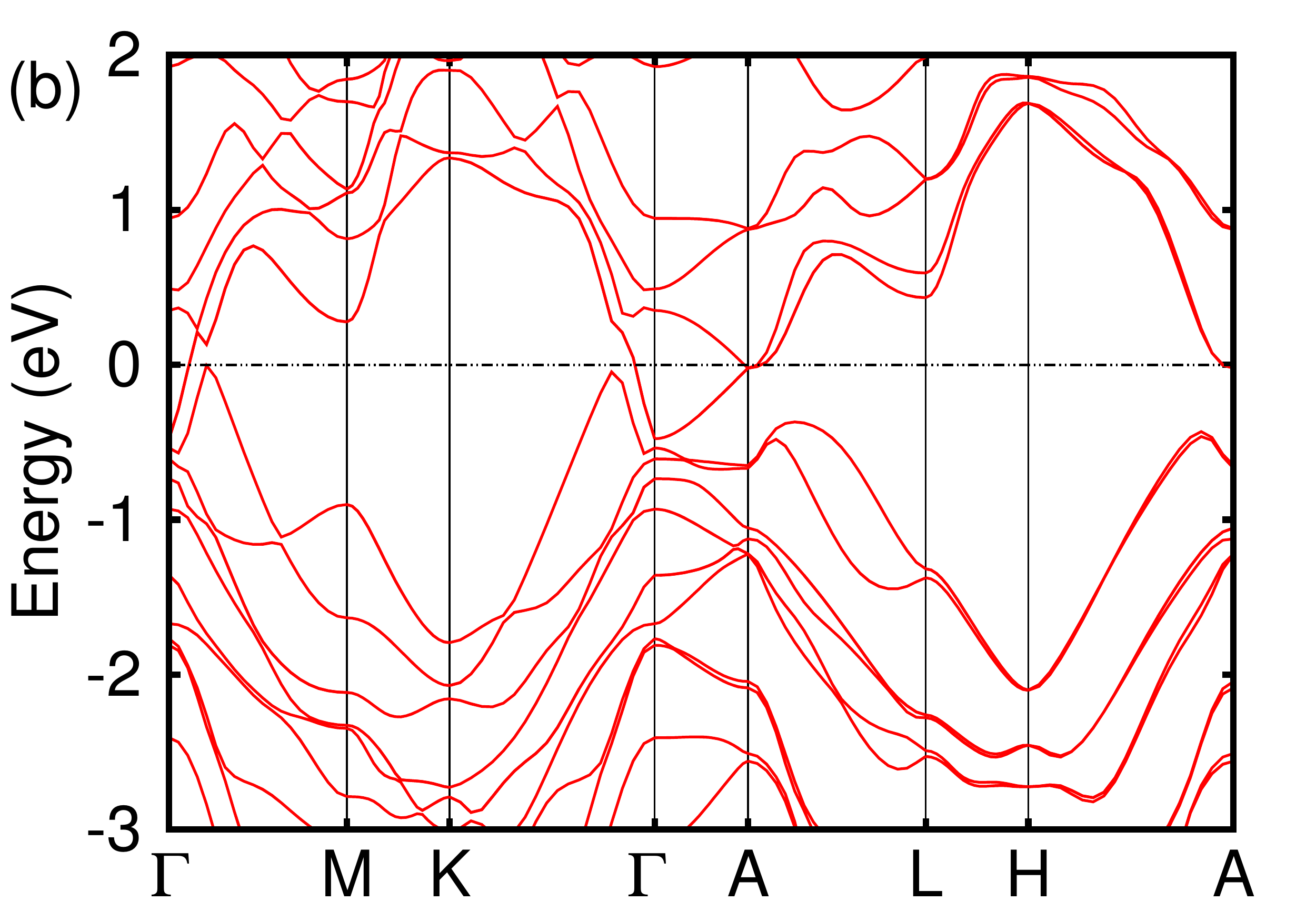}
\caption{Band structure of EuCd$_2$Bi$_2$ in the AFM$a$ configuration by considering (a) GGA+$U$ and (b) HSE+$U$ with $U=7$\,eV between -3 and 2\,eV. The Fermi level is set at zero energy.}
\label{BAND_SOC_HYBRID_EuCdBi}
\end{figure}

The band structure of EuCd$_2$Bi$_2$ for different exchange functionals is plotted in Fig.~\ref{BAND_SOC_HYBRID_EuCdBi}, we can see how there are very few differences between GGA+$U$ and HSE+$U$.

\begin{figure}[t!]
\centering
\includegraphics[width=2.99cm,angle=270]{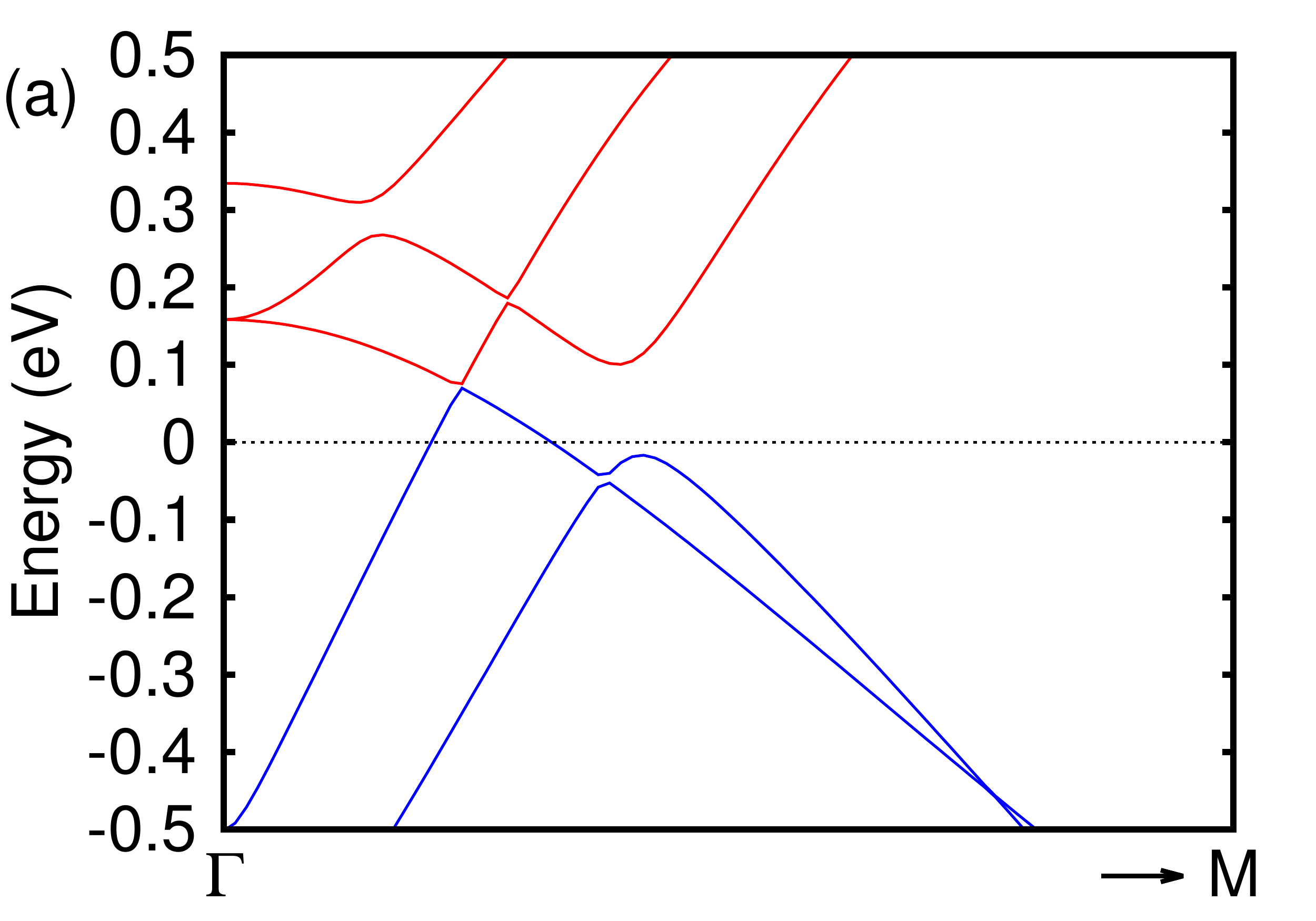}
\includegraphics[width=2.99cm,angle=270]{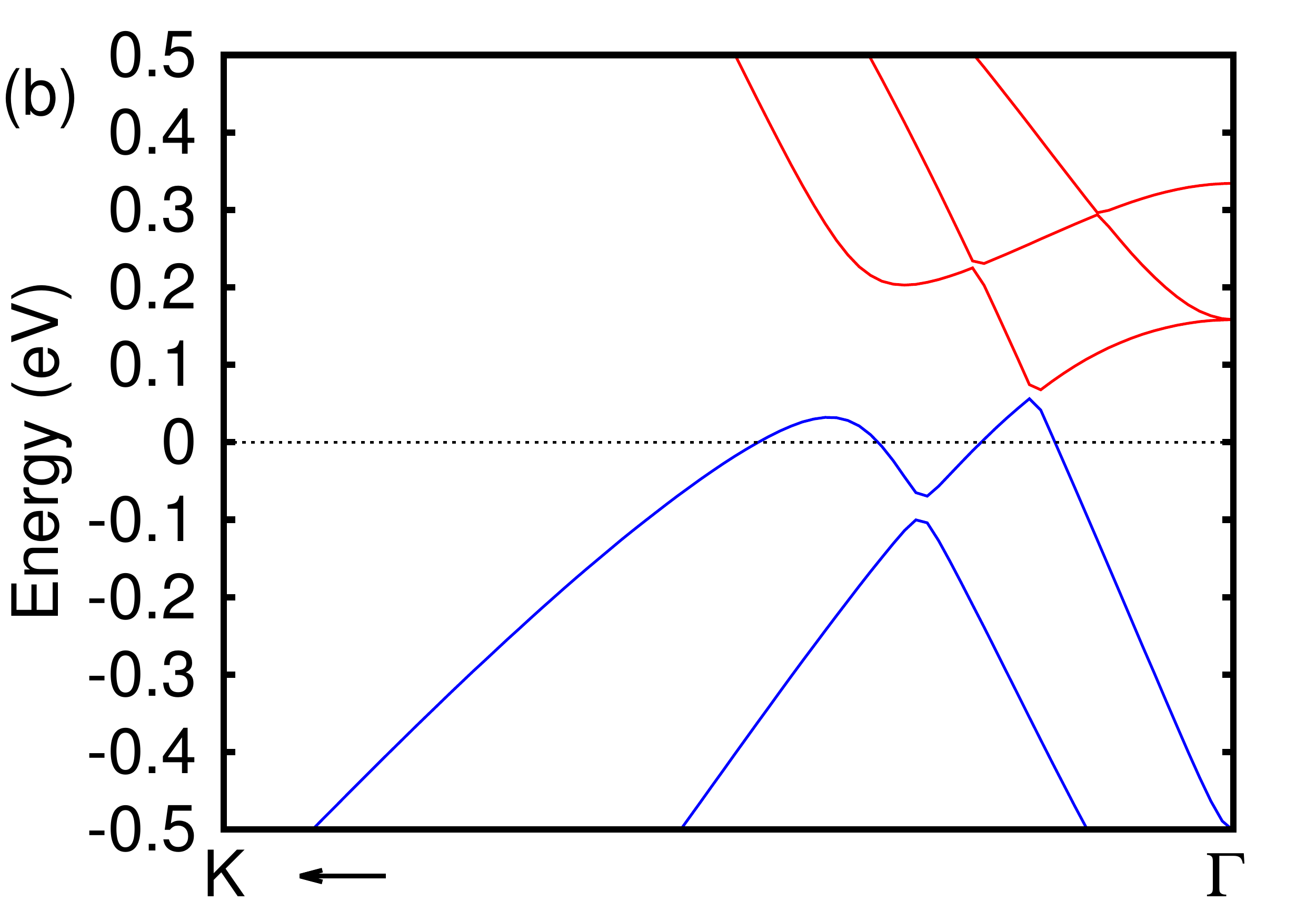}
\includegraphics[width=2.99cm,angle=270]{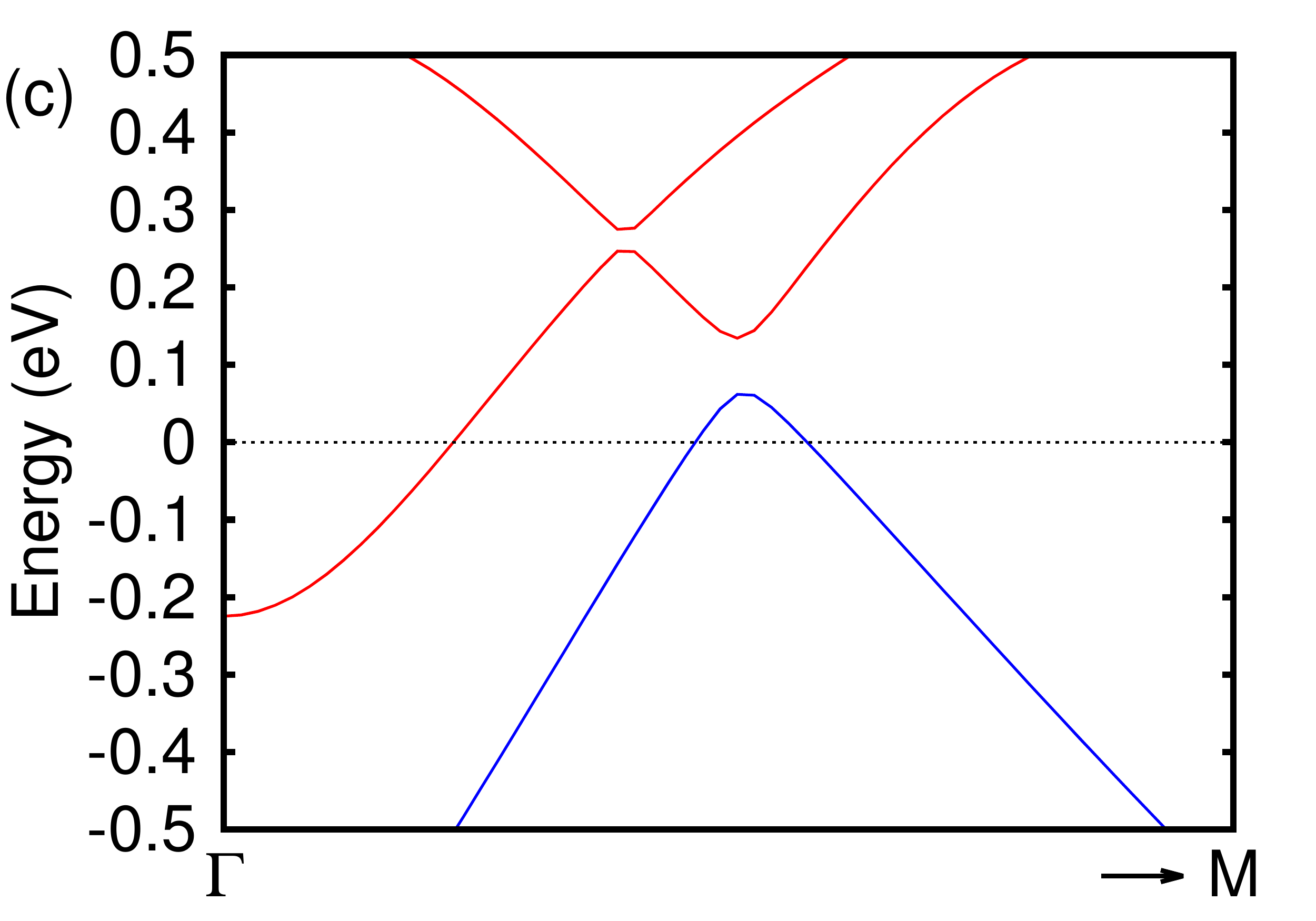}
\includegraphics[width=2.99cm,angle=270]{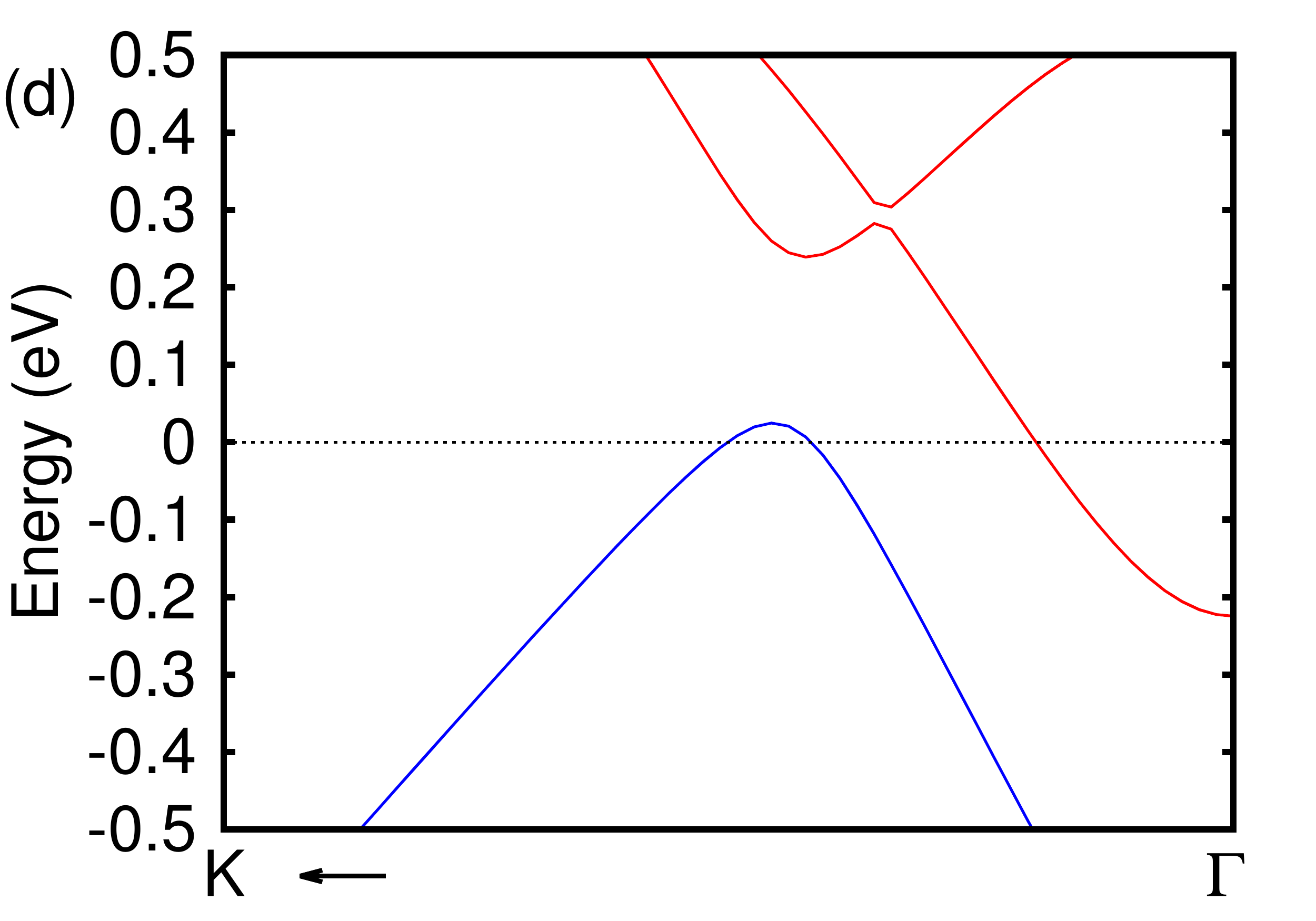}
\caption{Band structure of EuCd$_2$Bi$_2$ in the AFM$a$ configuration without [panels (a) and (b)] and with spin-orbit coupling [panels (c) and (d)] along the $\Gamma$-M  [panels (a) and (c)] and $\Gamma$-K directions [panels (b) and (d). The valence bands are plotted in blue, while the conduction bands are plotted in red. The Fermi level is set at zero energy.}
\label{NOSOC_SOC_EuCd2Bi2}
\end{figure}

There is a  change in the band order at the $\Gamma$ point in EuCd$_2$Bi$_2$ compared to EuCd$_2$As$_2$.  The band gap as a function of $U$  band gap increases as a function of $U$ and suggests a topological behavior.
We plot the band structure along $\Gamma$-M and $\Gamma$-K of the EuCd$_2$Bi$_2$ without and with SOC in Fig.~\ref{NOSOC_SOC_EuCd2Bi2}. We can observe that the SOC disentangles valence bands from the conduction bands as usual in topological systems creating in this case a topological semimetal. The trivial band gap decreases as the anion becomes heavier until the system becomes topological semimetal for the case of Bi.

Due to the large band inversion in the Bi-compound, the topological band arrangement persists substituting Bi with Sb. Indeed, properties of EuCd$_2$SbBi are similar to EuCd$_2$Bi$_2$, as shown in Appendix F.

\begin{figure}[t!]
\centering
\includegraphics[width=2.99cm,angle=270]{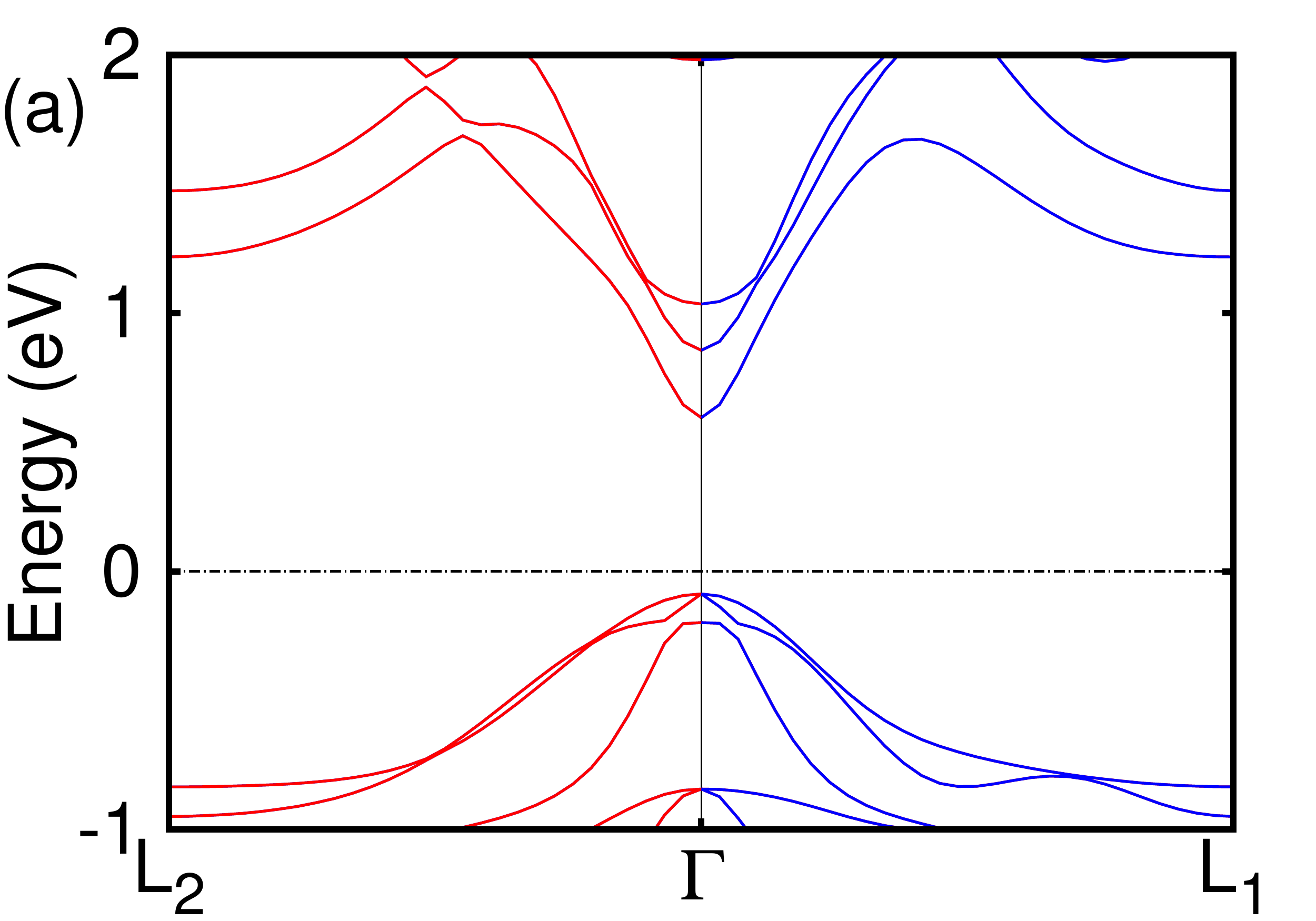}
\includegraphics[width=2.99cm,angle=270]{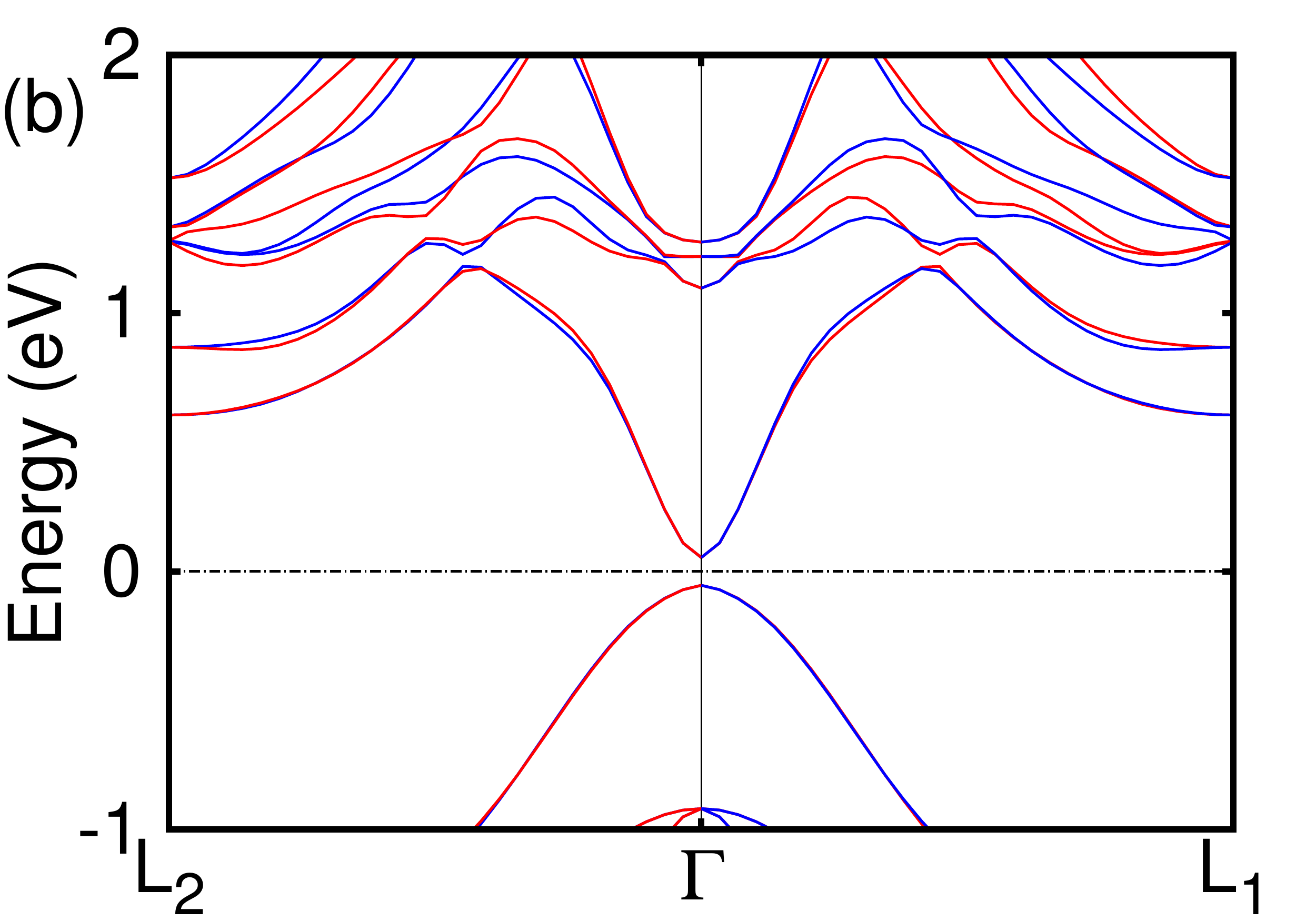}
\caption{Band structure of (a)
EuCd$_2$As$_2$ and (b)
EuIn$_2$As$_2$ along the path L$_2$-$\Gamma$-L$_1$ in the AFM phase without spin-orbit coupling and with the HSE functional. We plot in blue the spin-up channel and in red the spin-down channel. In panel (a), the red and blue lines perfectly overlap along both $k$-paths. The Fermi energy is set to zero.}
\label{ALTERMAGNETISM}
\end{figure}

Plotting the band structure along particular $k$-paths, including L$_1$ and L$_2$ directions (see the Appendix A), we can look for altermagnetic signatures, i.e., for the exchange-induced spin-splitting of bands in the AFM phase \cite{Smejkal:2022_PRX,Smejkal22beyond}. The compounds with space group P$\bar{3}$m1 are not altermagnetic as shown in Fig.~\ref{ALTERMAGNETISM}(a), while according to the data presented in Fig.~\ref{ALTERMAGNETISM}(b) the compounds with the space group P6$_3$/mmc show altermagnetism.

\begin{figure}[t!]
\centering
\includegraphics[width=1.0\linewidth]{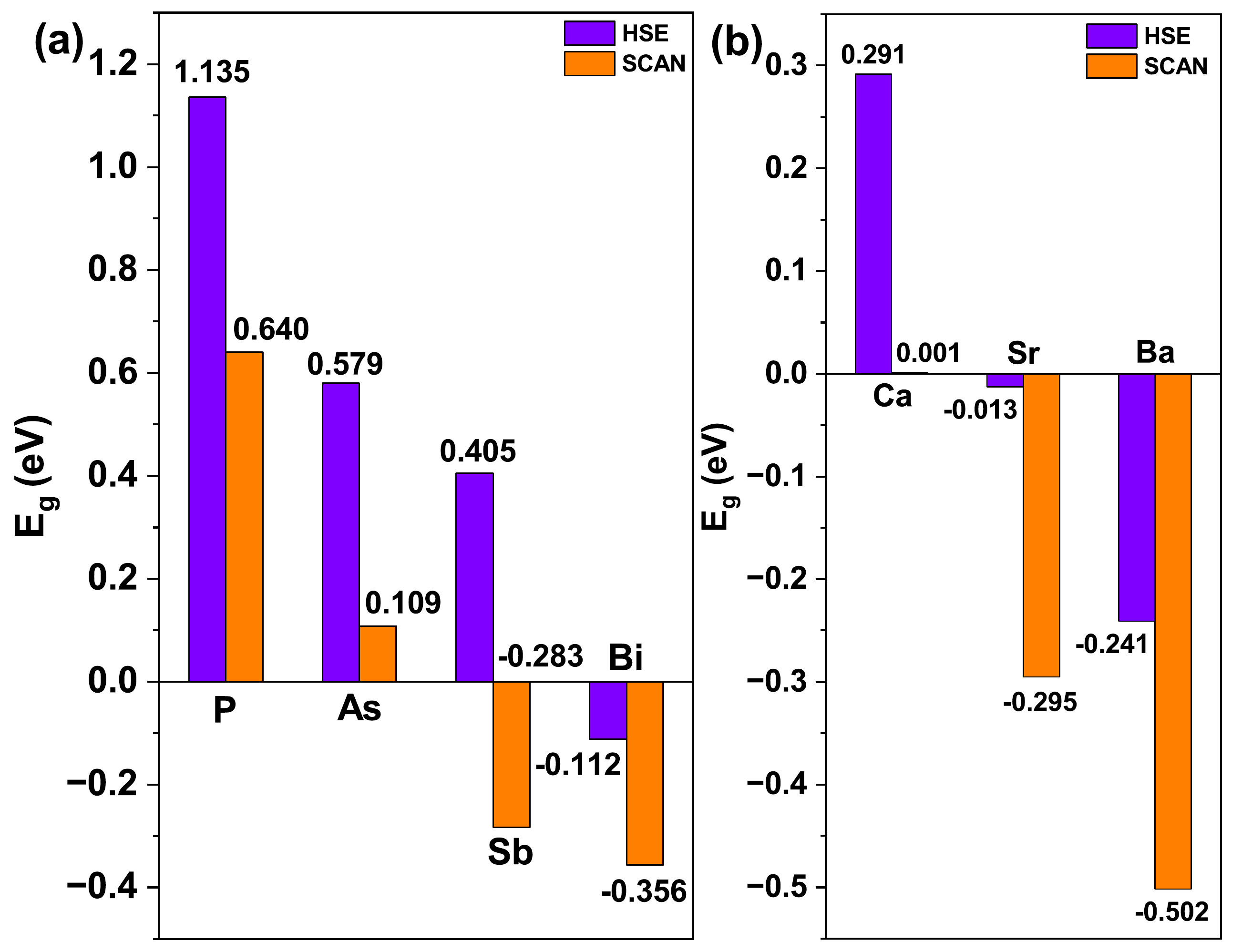}
\caption{$E_g$ for different compounds belonging to the same family by considering HSE and SCAN. (a) Band gap of EuCd$_2$X$_2$ (X = P, As, Sb, Bi) in the AFM$a$ phase, since SCAN does not return realistic results we plot SCAN+$U$ with
$U=3$\,eV. (b) Band gap of AEIn$_2$As$_2$ (AE=Ca, Sr, Ba).}
\label{GAP_families}
\end{figure}

As a summary, we report in Fig.~\ref{GAP_families}(a) the obtained band gap values for EuCd$_2$X$_2$ (X = P, As, Sb, Bi) compounds within the HSE approach.  As seen, the trivial band gap decreases as the X element becomes heavier and relativistic effects, such as spin-orbit splitting and the mass term, larger. The compound with the lightest atom P is the most trivial (see Appendix E).

\begin{table*}
\begin{center}
    \caption{Summary of topology classes for selected materials and computational methods. We divide the compounds into three columns: in the first column there are the compounds trivial within GGA+$U$, SCAN/SCAN+$U$ and HSE/HSE+$U$, in the second column there are the compounds that are trivial within HSE/HSE+$U$, in the last column there are compounds topological within any exchange-correlation functional. Going from left to right in the table, one of the atoms becomes heavier and the systems become more topological.\\}\label{topology}
    \begin{tabular}{|c|c|c|c|}
    \hline
     {Space group}  &  Robust Trivial Insulator & Trivial in HSE/HSE+$U$ & Topological in HSE/HSE+$U$ \\
      {} &  {} & Topological in SCAN/SCAN+$U$ & {}  \\
    \hline
    P$\bar{3}$m1 (No.\,164)   &  EuCd$_2$P$_2$, EuCd$_2$As$_2$ &   EuCd$_2$Sb$_2$  &  EuCd$_2$SbBi$^\dagger$, EuCd$_2$Bi$_2$$^\dagger$ \\
    \hline
    P6$_3$/mmc  (No.\,194)   &   EuIn$_2$P$_2$   &    EuIn$_2$As$_2$  & - \\
    \hline
    P6$_3$/mmc  (No.\,194)  nonmagnetic &   CaIn$_2$As$_2$  &  -  &  SrIn$_2$As$_2$*,  BaIn$_2$As$_2$  \\
    \hline
    R$\bar{3}$m  (No.\,166) FM phase & EuIn$_2$Sb$_2$   &  - &  EuSn$_2$As$_2$ \\
    \hline
    \multicolumn{4}{l}{*This compound has a tiny topological band gap of 13 meV within HSE, so it is not considered as a robust topological material.}\\
    \multicolumn{4}{l}{$^\dagger$ These compounds are topological semimetals within both GGA+$U$ and HSE.}
    \end{tabular}
\end{center}
\end{table*}


\section{Other similar families of compounds}

We move now to EuIn$_2$As$_2$ \cite{Gong22,Riberolles2021,Sato20} that crystallizes in a hexagonal Bravais lattice with
the space-group P6$_3$/mmc \cite{Regmi20,Goforth08,Rosa12}.
We have used the lattice parameters $a=b=4.21$\,{\AA}  and $c=17.89$\,{\AA} and the atomic positions reported in Ref.~\onlinecite{Regmi20}.
In Fig.~\ref{GAP_SOC_HYBRID_Eu}(d), we show $E_g$ for EuIn$_2$As$_2$ as a function of $U$.
It was proposed that EuIn$_2$As$_2$ could host higher-order topology \cite{Xu2019higher}, however, within the HSE approach this material is a trivial insulator.
For all the investigated $U$ values, the system is  trivial within the HSE+$U$ approach but topological employing the SCAN+$U$ method, so that
EuIn$_2$As$_2$ is closer to the topological regime compared to EuCd$_2$As$_2$. Regarding the band structure,
the P6$_3$/mmc symmetry shows altermagnetism \cite{Smejkal:2022_PRX} on the contrary to P$\bar{3}$m1, as depicted in Figs.~\ref{ALTERMAGNETISM}(a,b).
Interestingly, the largest altermagnetic spin-splitting, reaching 150\,meV in the conduction band, appears for band states derived from 5$d$-orbitals of Eu, which reflects a relatively strong intraatomic $d$-$f$ exchange. In general, we note that it is difficult to find magnetic topological insulators showing altermagnetism and this material class appears as a worthwhile candidate.

We have studied also the Eu-based systems in the R$\bar{3}$m phase. We have found that EuIn$_2$Sb$_2$ is always trivial, while EuSn$_2$As$_2$ is topological within the HSE approach. For EuIn$_2$Sb$_2$, we have used the lattice constants $a =b =4.58$\,{\AA} and $c=27.79$\,{\AA} \cite{Wang21npj}. For EuSn$_2$As$_2$, we have used the lattice constants $a= b=4.21$\,{\AA} and $c=26.46$\,{\AA} \cite{Arguilla17}. A more extended discussion is presented in Appendix E.


Regarding nonmagnetic compounds with space group P6$_3$/mmc,  the lattice constants that we have used are $a = b =4.22$\,{\AA} and $c =17.97$\,{\AA}  for CaIn$_2$As$_2$ ; $a = b=4.31$\,{\AA} and $c =18.36$\,{\AA}  for SrIn$_2$As$_2$; $a=b=4.39$\,{\AA} and $c =18.83$\,{\AA}  for BaIn$_2$As$_2$ obtained from Ref.~\onlinecite{Guo2022the}. As shown in Fig.~\ref{GAP_families}(b) both BaIn$_2$As$_2$ and SrIn$_2$As$_2$ are topological insulators within  the HSE and SCAN methods, while CaIn$_2$As$_2$ is topological only employing the GGA  \cite{Guo2022the}.
Also in this case, the inverted band gap increases with the atomic weight of the atoms.
Thus, GGA overestimates the abundance of topologically non-trivial systems also in nonmagnetic materials but less severely compared to the case of Eu-based compounds.

Topological phases implied by our results are summarized in Table I for the compounds analyzed in this paper. The compounds that future a topological band gap within both SCAN and HSE functionals are considered as robust topological insulators, while the compounds that are topological within SCAN+$U$ and trivial within HSE+$U$ are considered to be either trivial or on the verge of the topological regime. As SrIn$_2$As$_2$ has a very small topological gap, the only robust topological systems among the compounds considered in our work are EuCd$_2$Bi$_2$, EuSn$_2$As$_2$, BaIn$_2$As$_2$, and the new proposed compound EuCd$_2$SbBi. We could expect to find other topological compounds replacing light elements by heavier elements.


\section{Conclusions}

We have investigated the materials family EuCd$_2$X$_2$ (X = P, As, Sb, Bi) and related compounds with a focus on EuCd$_2$As$_2$.
We have shown that the DFT with local functionals, such as GGA+$U$, overestimates the topological region for the Eu-based compounds while the HSE+$U$ method constitutes a more realistic tool for predicting topological classes of particular materials and to determine energies of both wide and narrow bands. Considering EuCd$_2$As$_2$ as an example, we obtain a band gap in the range  0.72-0.79\,eV within HSE+U versus the experimental value of 0.77 eV \cite{santoscottin2023eucd2as2} and we obtain a gap reduction in the magnetic field ${\Delta}E_g$ = 0.102 eV to compare with the experimental value of ${\Delta}E_g$ = 0.125 eV \cite{santoscottin2023eucd2as2}. Within the same approach, the trivial band gaps of EuCd$_2$P$_2$ and EuCd$_2$Sb$_2$ are 1.38-1.48 eV and 0.46-0.49 eV, respectively. From the calculation of the magnetocrystalline anisotropy, the N\'eel vector is in-plane for both topologically trivial EuCd$_2$As$_2$ and nontrivial EuCd$_2$Bi$_2$ compounds belonging to the space group 164.
While the use of local functionals strongly overestimates the topological region in the case of Eu-based compounds, it only slightly overestimates the topological region in the corresponding nonmagnetic materials AEIn$_2$As$_2$ (AE= Ca, Sr, Ba). We conclude that the two robust topological systems among those investigated are EuSn$_2$As$_2$ and BaIn$_2$As$_2$. Additionally, we propose other two topological systems as EuCd$_2$Bi$_2$ and EuCd$_2$SbBi, which are antiferromagnetic topological semimetals.

\section*{Acknowledgments}
The authors thank A. Akrap, M. Orlita and J.-M. Zhang for useful discussions.
The work is supported by the Foundation for Polish Science through the International Research Agendas program co-financed by the European Union within the Smart Growth Operational Programme (Grant No.\,MAB/2017/1).
We acknowledge the access to the computing facilities of the Pozna\'n Supercomputing and Networking Center Grant No.\,609.
We acknowledge the access to the computing facilities of the Interdisciplinary Center of Modeling at the University of Warsaw, Grant G84-0, GB84-1 and GB84-7. We acknowledge the CINECA award under the ISCRA initiative  IsC85 "TOPMOST" and IsC93 "RATIO" grant, for the availability of high-performance computing resources and support.



\section*{Appendix A: Computational details}

We have performed density functional theory (DFT) first-principles calculations using the VASP package \cite{Kresse93,Kresse96,Kresse96b} based on plane-wave basis set and projector augmented wave method \cite{Kresse99}. Except for altermagnetism studies, our calculations take into account SOC. The relevant SOC in EuCd$_2$As$_2$ is one of the \emph{p}-As that is reported to be 164\,meV \cite{wadge2021electronic}.

We mainly focused on EuCd$_2$As$_2$ which grows in a trigonal crystal structure (space group No.\,164,  P$\bar{3}$m1). This space group has inversion symmetry, some of the inversion symmetry points are on the positions of the Eu atoms. To simulate the A-type AFM configuration with spins in the $a$-$b$ plane, we have doubled the primitive cell along the $c$-axis. 

The experimental lattice constants of EuCd$_2$As$_2$, which we used for our calculations, $a = b = 4.44$ {\AA} and $c = 7.33$ {\AA}, were quoted in studies of AFM spin ordering by resonant elastic x-ray scattering \cite{Rahn18}.
To see how the lattice parameters change for different functionals, we fixed $a$ to the experimental value and checked the trend of the energy as a function of $\Delta c/c$ employing GGA+$U$ and HSE+$U$ approaches in the experimental AFM$a$ spin configuration. In the GGA+$U$ case we fixed the value of $U = 10$\,eV to be sure that we are in the trivial phase. In HSE+$U$ we used $U=7$\,eV, which correctly reproduces the experimental energies. The energy minimum is found for $c = 7.49$ and $7.43$ {\AA} in the GGA+$U$ and HSE+$U$ approach, respectively, to be compared to the experimental value $c = 7.33$ {\AA}. We have also checked for the case of HSE+$U$ with $U = 7$\,eV that the band gap, for the experimental lattice constants, is 0.79 eV, while for the theoretical value of $c$, $E_g = 0.74$\,eV. The other compounds that we have studied belong to the same family and the results are similar. Therefore the results only slightly change if we use the theoretical lattice constants instead of the experimental ones.

\begin{figure}[t!]
\centering
\includegraphics[width=0.7\linewidth]{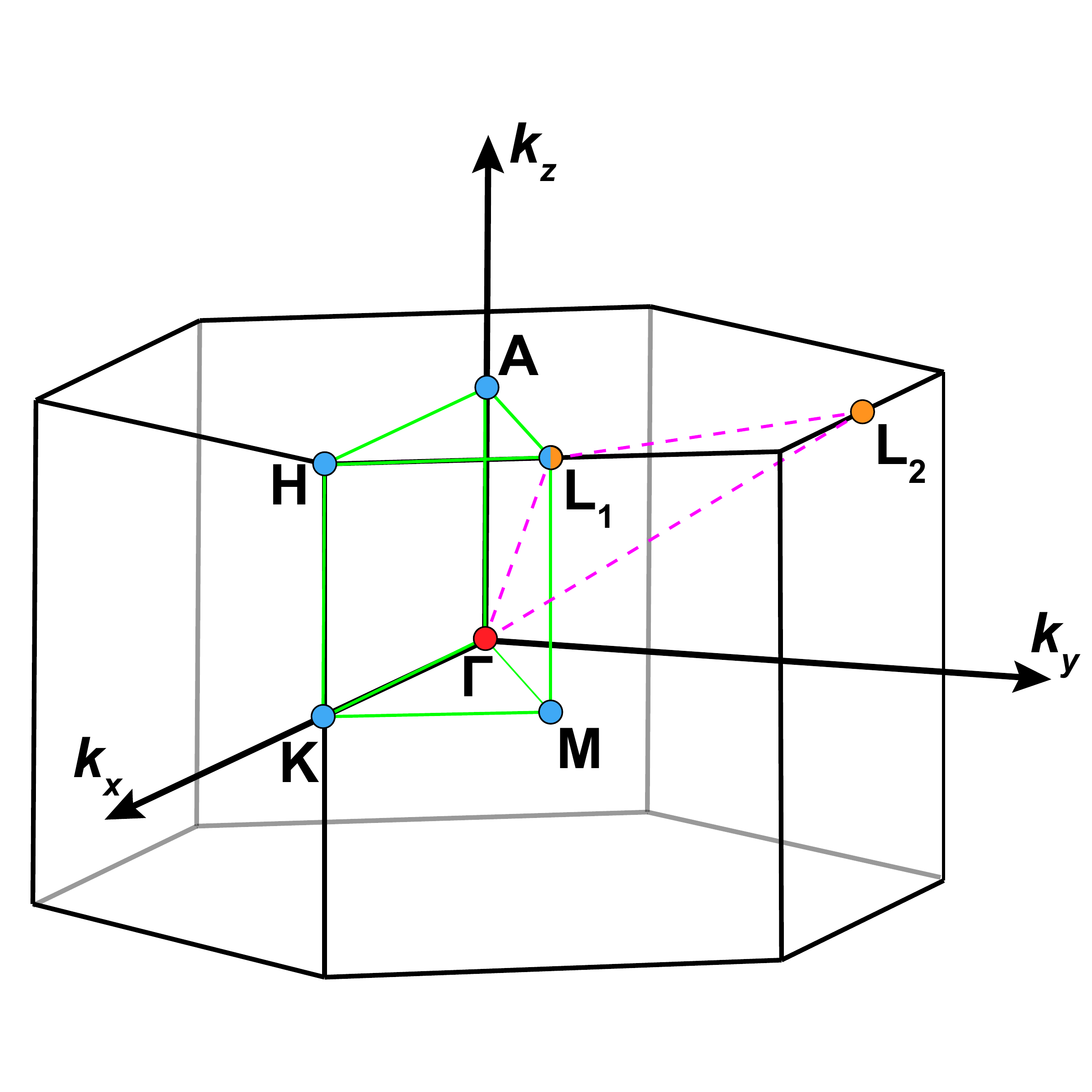}
\caption{Irreducible Brillouin zone of a typical hexagonal primitive cell with the high-symmetry momentum path for electronic structure highlighted in green. Dashed magenta line indicates the high-symmetry momentum path used to show the altermagnetism in space group P6$_3$/mmc in which the points L$_1$ and L$_2$ have the same eigenvalues but opposite spin.}
\label{bri_zone}
\end{figure}

We have used a plane-wave energy cutoff of 300\,eV and a 11$\times$11$\times$3 \textit{k}-points
grid centered at the $\Gamma$ point with 363 independent \textit{k}-points in the first Brillouin zone for the GGA+$U$ and SCAN+$U$ approaches. The \textit{k}-path of the Brillouin zone is depicted in Fig.~\ref{bri_zone}.
Increasing the energy cutoff up to 500\,eV just rises the trivial band gap by a very few meV but creates a problem in the convergence when employing the hybrid HSE functional without $U$. For the HSE method, we have used an 8$\times$8$\times$3 \textit{k}-grid. Increasing the \textit{k}-grid, the trivial band gap decreases by a few meV. To verify the structural stability of the newly proposed compound EuCd$_2$Bi$_2$, we have performed investigations of the lattice dynamics using phonon dispersion curves obtained within the density functional perturbation theory (DFPT) approximations using PHONOPY interface to VASP \cite{phonopy,phonopy-phono3py-JPSJ}. A 2$\times$2$\times$1 supercell of the AFM ground state was generated to compute the force constants within the GGA+$U$ approach with a \textit{k}-mesh of least 6$\times$6$\times$2.

\section*{Appendix B: Values of $U$ for $4f$-electrons}

We first consider the generalized gradient approximation
(GGA) of Perdrew, Burke, and Ernzerhof (PBE) \cite{Perdew96}  with  the Hubbard energy $U$  for the 4$f$-orbitals of Eu.
 A number of previous GGA+$U$ studies of EuCd$_2$As$_2$ in the in-plane AFM ground state employed $U$ up to 5\,eV \cite{Soh19,Wang19,Taddei22,Rahn18,Ma:2020_AM,Gati21,Ma:2019_SA} or 6\,eV \cite{Du22}, and found this compound to be topologically nontrivial.

However, according to several works, the value of $U$ for 4$f$ electrons in Eu$^{+2}$ is well above 5\,eV, e.g.,  $U_f-J_f =7.1$\,eV \cite{Yu2020} and $U_f= 7.397$\,eV, $J_f= 1.109$\,eV \cite{Larson_2006}. For the topological system EuB$_6$, the value of $U=7$\,eV has been used \cite{PhysRevLett.129.166402} while even larger values of $U$ were used in literature for the topological EuCd$_2$Sb$_2$ \cite{doi:10.1063/1.5129467}.
Thus, as reported in the main text, to evaluate the energy gap magnitude for EuCd$_2$As$_2$ with various magnetic orders, we have scanned $U$ from 5 to $11$\,eV assuming $J_H = 0.15U$.

The use of $U$ improves the description of correlations within $4f$-Eu narrow band but GGA is not effective in describing wide bands.
We could add $U$ to all orbitals close to the Fermi level including $s$-Cd, $s$-As, and $p$-As states, as suggested by the linear response \cite{Cococcioni2005} and the ABCN0 \cite{Agapito2015} methods. However, GGA+$U$ approach is not effective if the wide band is close to the Fermi level just in one $k$-point (the $\Gamma$ point in our case) \cite{Hussain2022electronic}.

\begin{figure}[t!]
\centering
\includegraphics[width=5.4cm,height=8.0cm,angle=270]{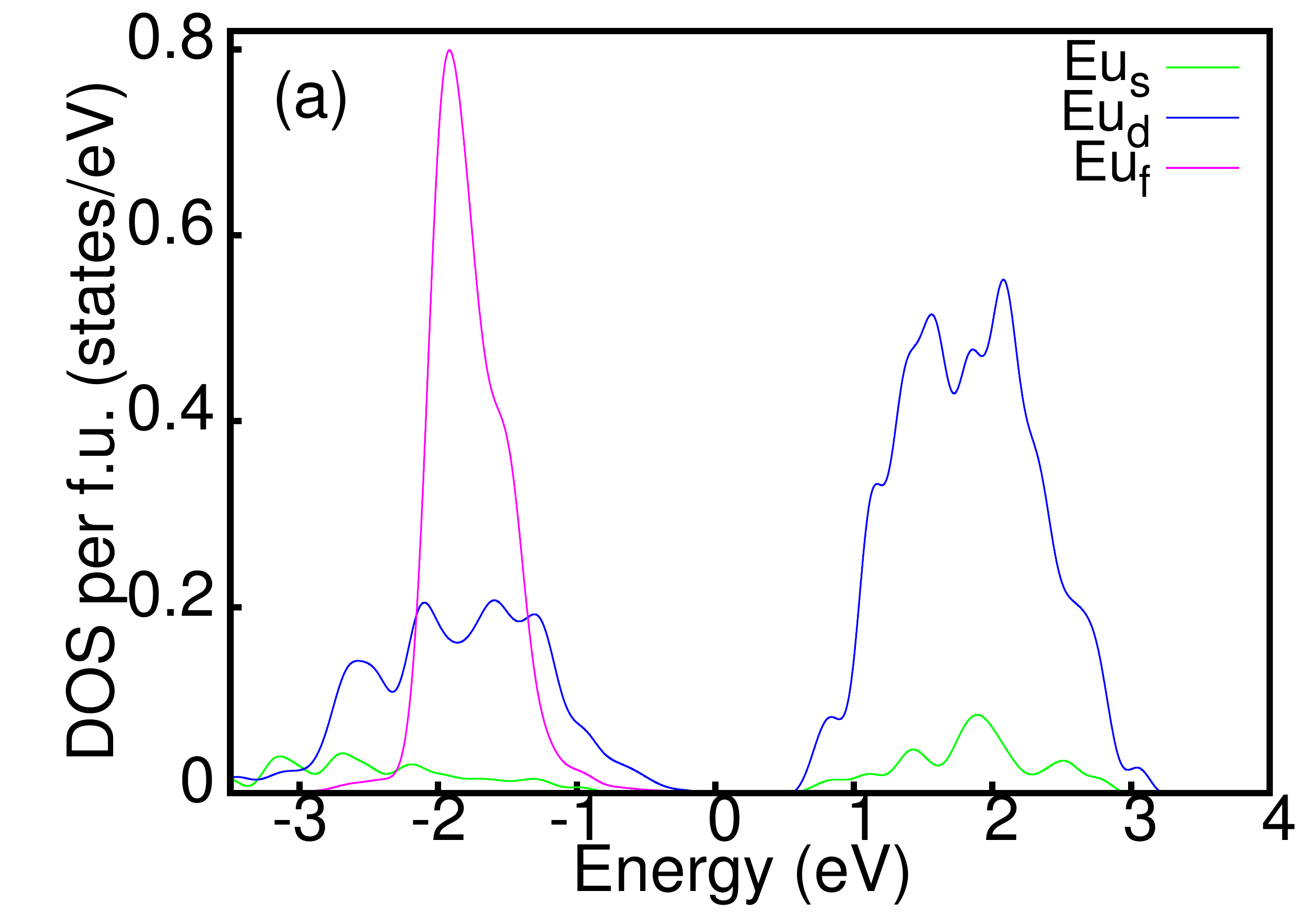}
\includegraphics[width=5.4cm,height=8.0cm,angle=270]{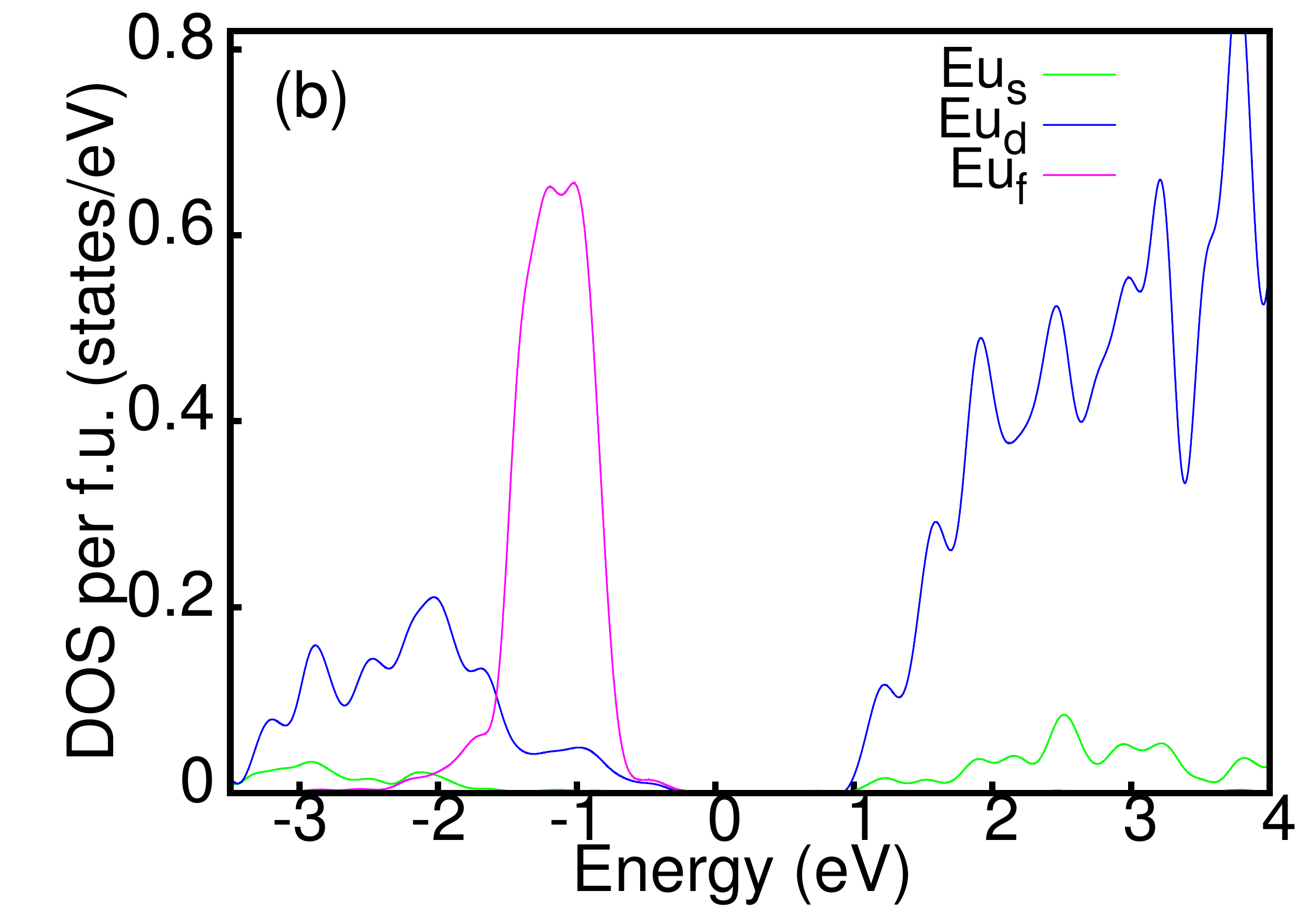}
\includegraphics[width=5.4cm,height=8.0cm,angle=270]{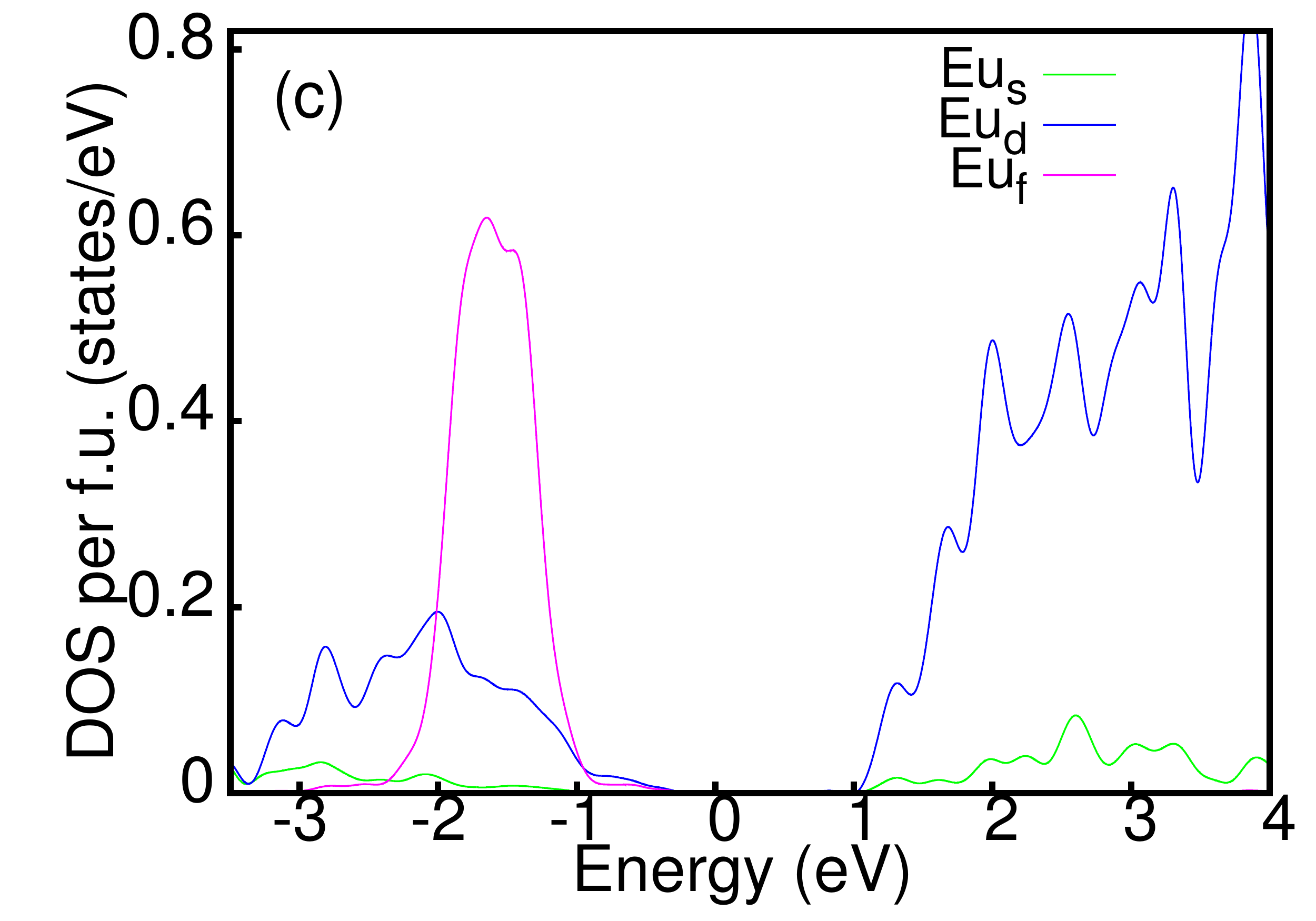}
\caption{Local density of states of the Eu atoms in the AFM$a$ configuration by using (a) GGA+$U$ with $U=7$\,eV on the $4f$-orbitals of Eu, (b) HSE hybrid functional and (c) HSE+$U$ with $U=1.5$\,eV. The $s$-, $d$- and $f$-states are plotted in green, blue and pink, respectively. The DOS of the $f$-states is divided by 20. The Fermi level is set at zero energy. The band gap seems larger because we are plotting just the Eu-states without Cd- and As-states.}\label{DOS}
\end{figure}

As underlined in the main text, the underestimation of the band gap value within DFT with local functionals is a well-known issue in {\em ab initio} studies of semiconductors.  Within the \textit{k}-space topology, the problem of the underestimation of the trivial band gap turns into the overestimation of the topological band gap \cite{Hussain2022electronic,D3CP00005B} meaning that we have an overestimation of the topological region in the phase diagram.

To go beyond GGA+$U$, we have used the strongly constrained and appropriately normed (SCAN) \cite{PhysRevLett.115.036402} and Heyd-Scuseria-Ernzerhof 2006 (HSE) hybrid functionals \cite{Paier06}.
 Using SCAN+$U$ and HSE+$U$, we take into account the Coulomb repulsion energy $U$ acting on 4$f$ electrons  and SCAN or HSE functionals acting on electrons in wide bands derived from $sp$ orbitals. Indeed, it was proposed to combine the HSE hybrid functional with the Hubbard $U$, and, by analyzing a set of II-VI semiconductors, it was shown that HSE+$U$ calculations reproduce the experimental band gap \cite{Aras14}. It was shown that also SCAN functional requires a Hubbard $U$ correction to reproduce the properties of some compounds \cite{Sai18,Long20}. Within SCAN+$U$ and HSE+$U$, we have usually used $U=7$\,eV with the Hund coupling $J_H= 0.15U$, if not mentioned otherwise. However, our conclusions concerning the topological phase diagram are weakly dependent on the $U$ value.


\section*{Appendix C: Wavefunction and spin-splittings in valence and conduction band for E\lowercase{u}C\lowercase{d}$_2$A\lowercase{s}$_2$}

We consider the AFMa configuration within the HSE approach with $U=7$\,eV and SOC. In this case, the fractional contributions of atomic orbitals at the $\Gamma$ point are 75\% 4$p$-As, 11\% 4$d$-Cd and 14\% 4$f$-Eu at the valence band top and 12\% 6$s$-Eu, 34\% 5$s$-Cd and 54\% 4$s$-As at the bottom of the conduction band. The Eu-4$f$ states are coupled to 6$s$-Eu via the intraatomic potential exchange interaction ${\cal{J}}_{6s-4f}$. Therefore, the conduction band is sensitive to the position of the 4$f$-levels. The 5$d$-electrons of Eu are absent at the $\Gamma$ point but their weight rises quickly with $k$ in both conduction and valence band band.

In the FMc configuration, the exchange spin-splitting of the valence band at the $\Gamma$ point with respect to Eu spin direction is antiferromagnetic, ${\Delta}E_v =- 107$\,meV, which points to the exchange energy $J_v = 2{\Delta}E_v/S = -61$\,meV for $S = 7/2$, whereas the interaction is ferromagnetic in the conduction band, where the spin-splitting is  $J_c$=${\Delta}E_c= +114$\,meV and $J_c = +65$\,meV.
Therefore, the sum of the two spin-splitting is 221\,meV, implying the a gap reduction of 110\,meV that is comparable with the value of ${\Delta}E_g =102$\,meV for HSE+$U$ discussed in the main text.

We have qualitatively similar results within GGA+$U$. At $U=10$\,eV with SOC, ${\Delta}E_v = -44$\,meV and ${\Delta}E_c=   104$\,meV. Therefore, the exchange spin-splittings have the same signs as in Mn-doped CdTe \cite{Autieri:2021_PRB}. Thus, it appears that $J_v$ is predominantly determined by hybridization of valence band states with $4f$ orbitals, i.e., by the kinetic exchange, whereas the non-zero $J_c$ value mainly results from the intra-atomic potential exchange described by ${\cal{J}}_{6s-4f} = 52$\,meV and, possibly, ${\cal{J}}_{5d-4f} = 215$\,meV \cite{Dietl:1994_PRB}.

\section*{Appendix D: Energy of the 4$\lowercase{f}$-electrons}
It was shown that the HSE hybrid functional does not reproduce well the position of the $f$-orbitals \cite{Schlipf13,Atta09}.
Even when we use HSE, an additional $U$ is required to reproduce the position of the $f$-bands. We report the local density of states (DOS) of the Eu-orbitals in the AFMa phase by using GGA+$U$, HSE and HSE+$U$ in Fig.~\ref{DOS}. Experimentally, the position of the majority 4$f$-Eu narrow bands is in the range 1.2-1.6\,eV \cite{Ma:2020_AM}.
To reproduce the position of the $f$-levels within HSE+$U$, we need a Coulomb repulsion around $1.5$\,eV as we can see from the local DOS in Fig.~\ref{DOS}. Therefore,  we conclude that HSE alone is not able to reproduce the experimental position of the $f$-states.

Using just SCAN, we have a gap opened between Cd and As states at Fermi, however, once the gap is open this is filled by the $f$-bands and we do not report the results for SCAN without $U$.


\begin{figure}[t!]
\centering
\includegraphics[width=6.3cm,angle=270]{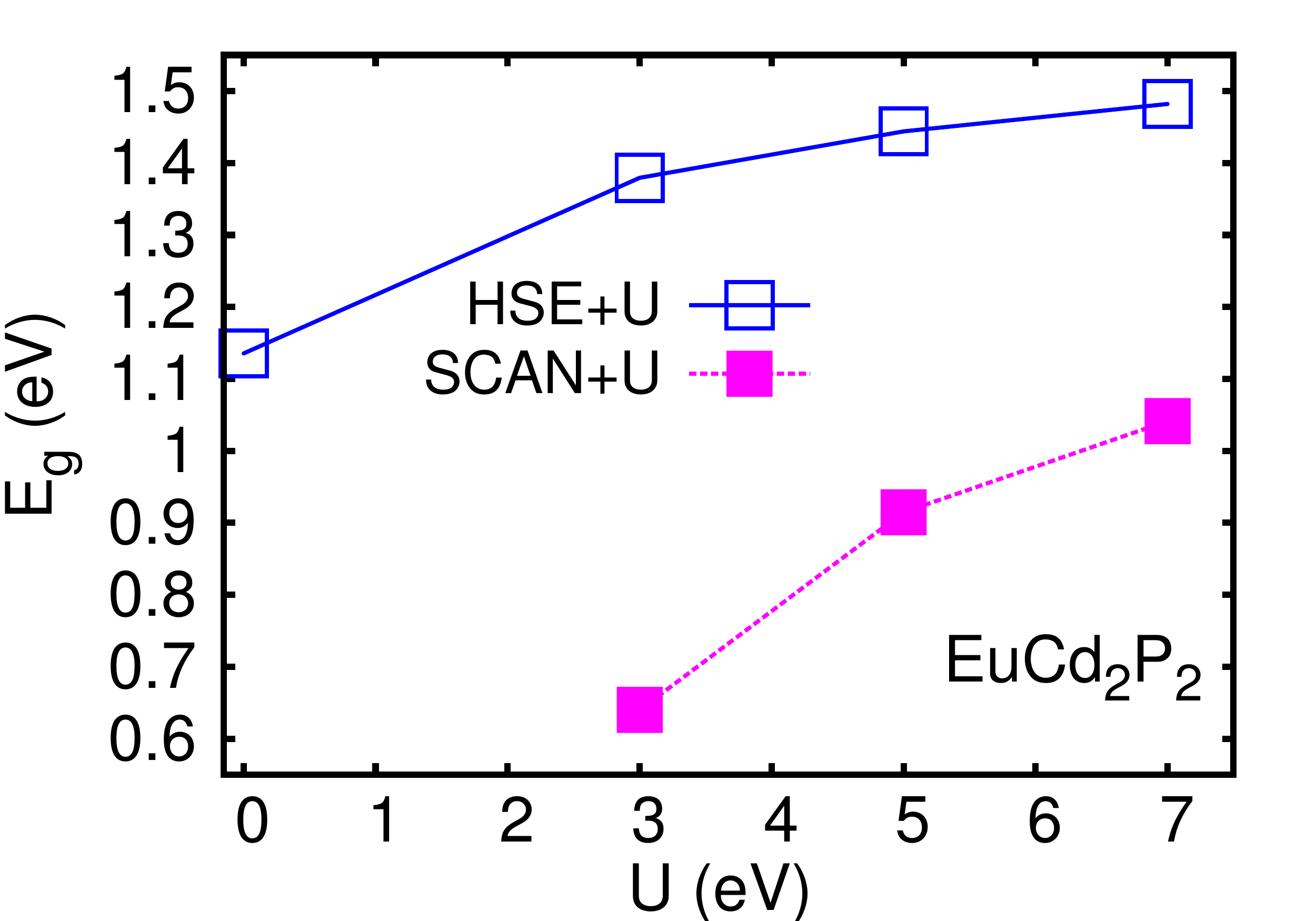}
\caption{$E_g$ for EuCd$_2$P$_2$ with SOC as a function of the Coulomb repulsion in the AFM$a$ configuration by considering HSE+$U$ (blue solid line) and SCAN+$U$ (purple dashed line).}
\label{GAP_SOC_HYBRID_EuCdP}
\end{figure}

\section*{Appendix E: Other members of the E\lowercase{u}C\lowercase{d}$_2$A\lowercase{s}$_2$ family and other families of compounds}

We performed the calculations also for EuCd$_2$P$_2$, EuCd$_2$Sb$_2$ and EuCd$_2$Bi$_2$ that have the same space group of EuCd$_2$As$_2$, namely P$\bar{3}$m1 (No.\,164).
In Fig.~\ref{GAP_SOC_HYBRID_EuCdP}, we show $E_g$ for EuCd$_2$P$_2$ with SOC as a function of the Coulomb repulsion in the A-type AFM ground state configuration with spins in the $a$-$b$ plane by considering HSE+$U$ and SCAN+$U$ functionals. The system is trivial for all the functionals used.
Heavier atoms increase volume, SOC and bandwidth, all properties that increase the band inversion.

\begin{figure}[t!]
\centering
\includegraphics[width=6.3cm,angle=270]{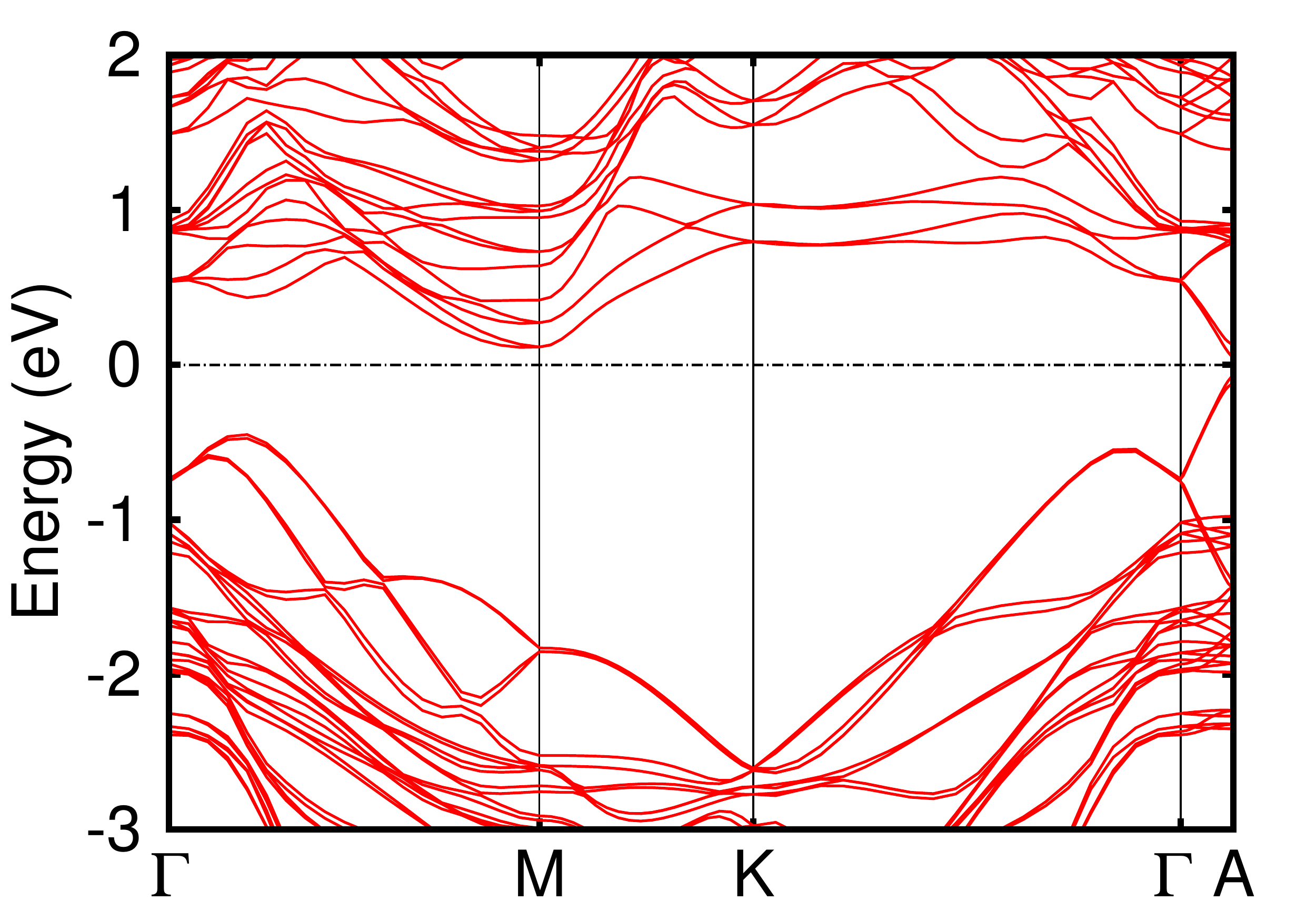}
\caption{Band structure of EuIn$_2$Sb$_2$ in the FM$c$ configuration by considering HSE+$U$ with $U=7$\,eV. The Fermi level is set at zero energy.}
\label{BAND_SOC_HYBRID_EuInSb}
\end{figure}

\begin{figure}[t!]
\centering
\includegraphics[width=6.3cm,angle=270]{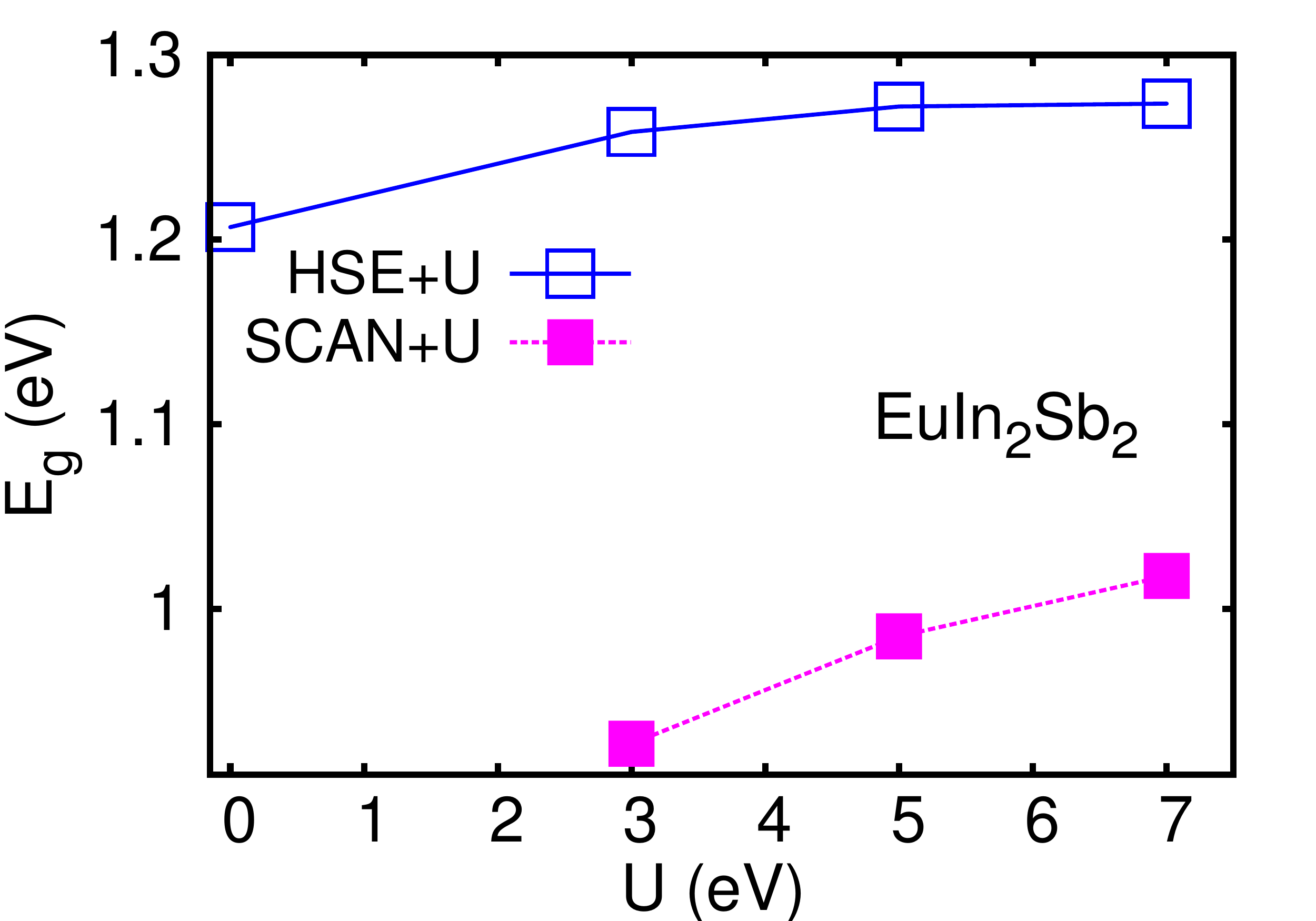}
\caption{Band gap $E_g$ for EuIn$_2$Sb$_2$ as a function of the Coulomb repulsion energy $U$ in the FM$c$ configuration by considering HSE+$U$ (blue solid line) and SCAN+$U$ (purple dashed line).}
\label{GAP_SOC_HYBRID_EuInSb}
\end{figure}

\begin{figure}[t!]
\centering
\includegraphics[width=6.2cm,angle=270]{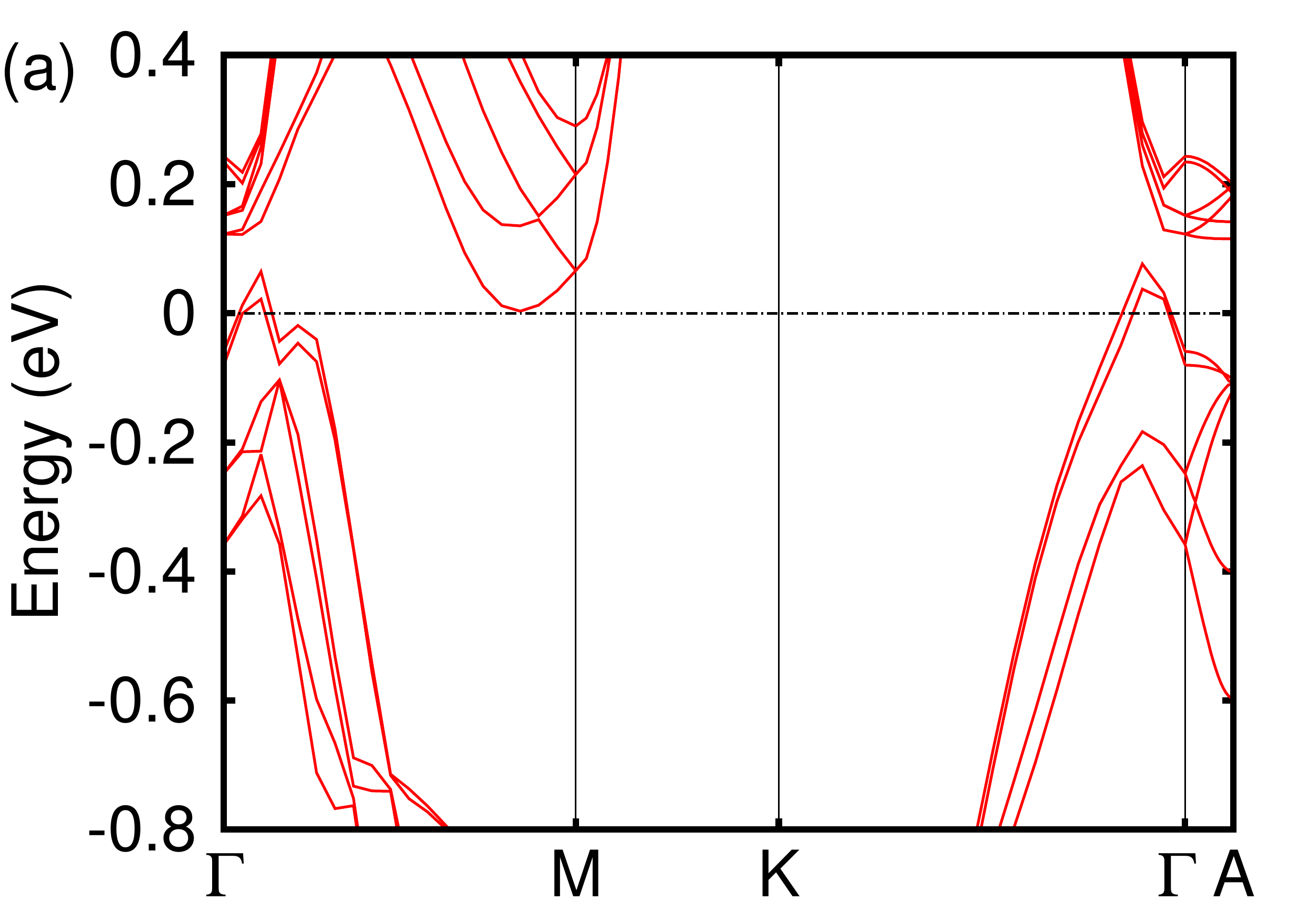}
\includegraphics[width=6.2cm,angle=270]{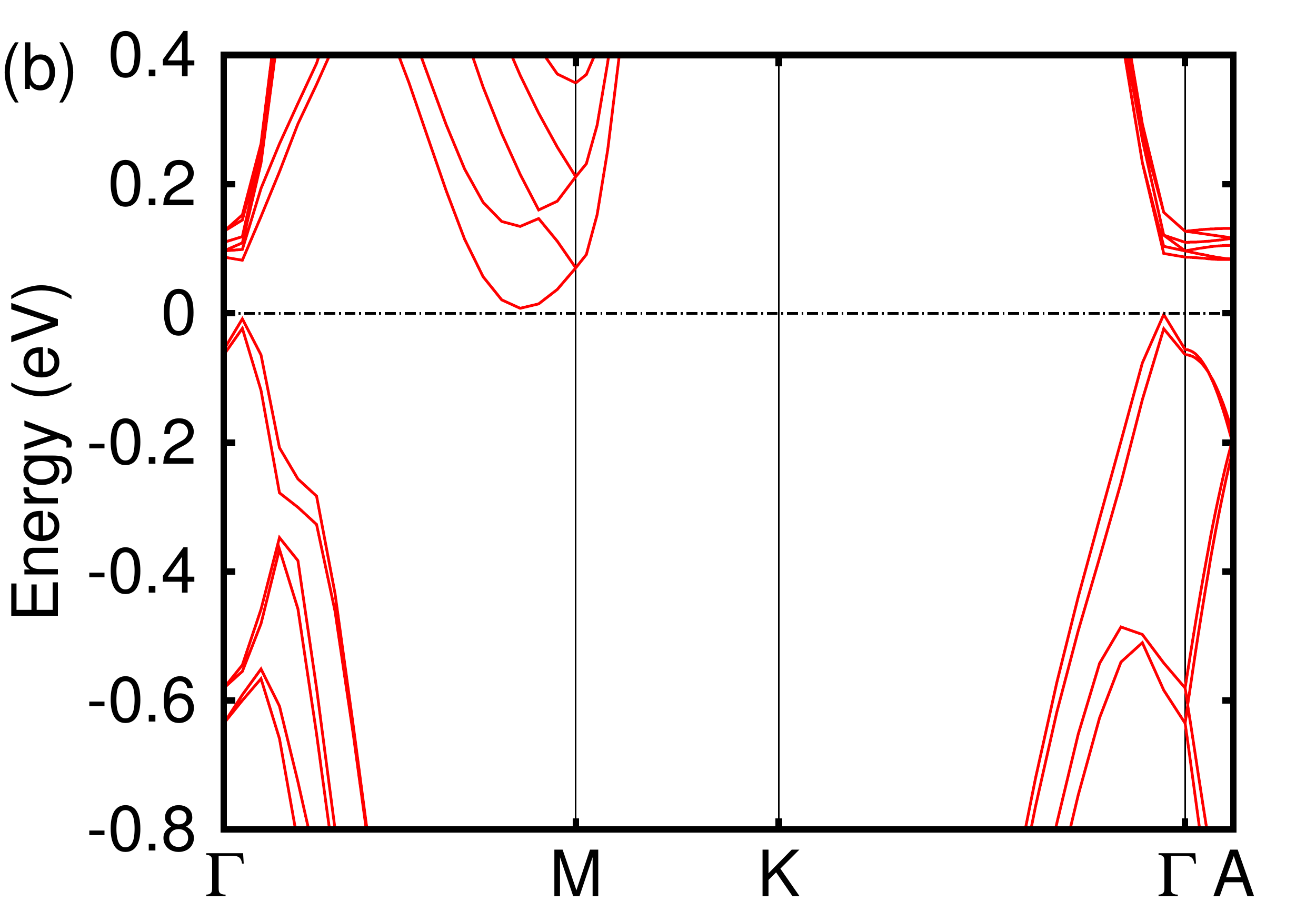}
\includegraphics[width=6.2cm,angle=270]{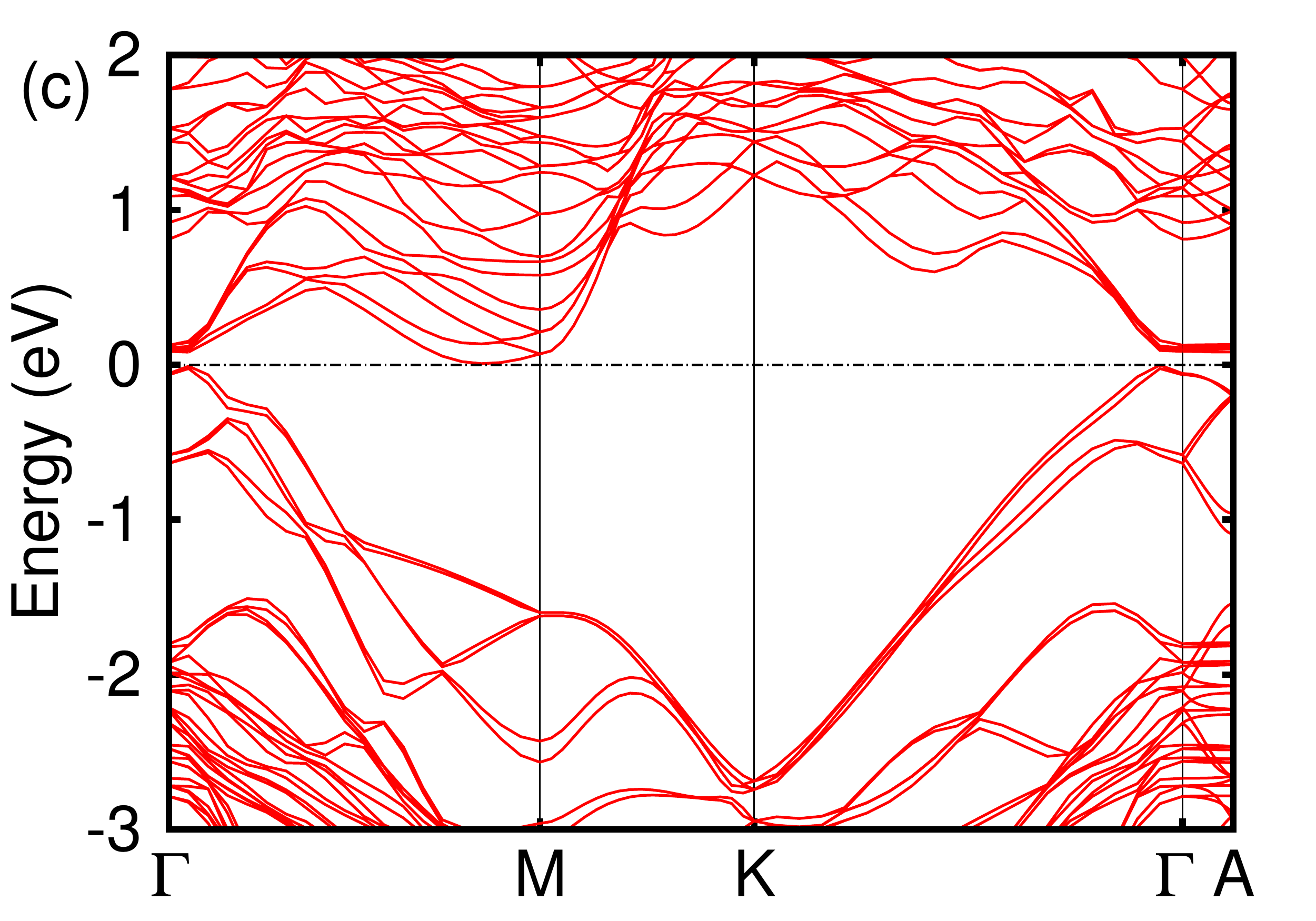}
\caption{Band structure of EuSn$_2$As$_2$ in the FM$c$ configuration and with $U=7$\,eV by considering (a) GGA+$U$ and (b) HSE+$U$  in the Fermi level vicinity. (c)  HSE+$U$  in a broader energy range. The Fermi level is set at zero energy. }
\label{BAND_SOC_HYBRID_EuSnAs}
\end{figure}

\begin{figure}[t!]
\centering
\includegraphics[width=6.3cm,angle=270]{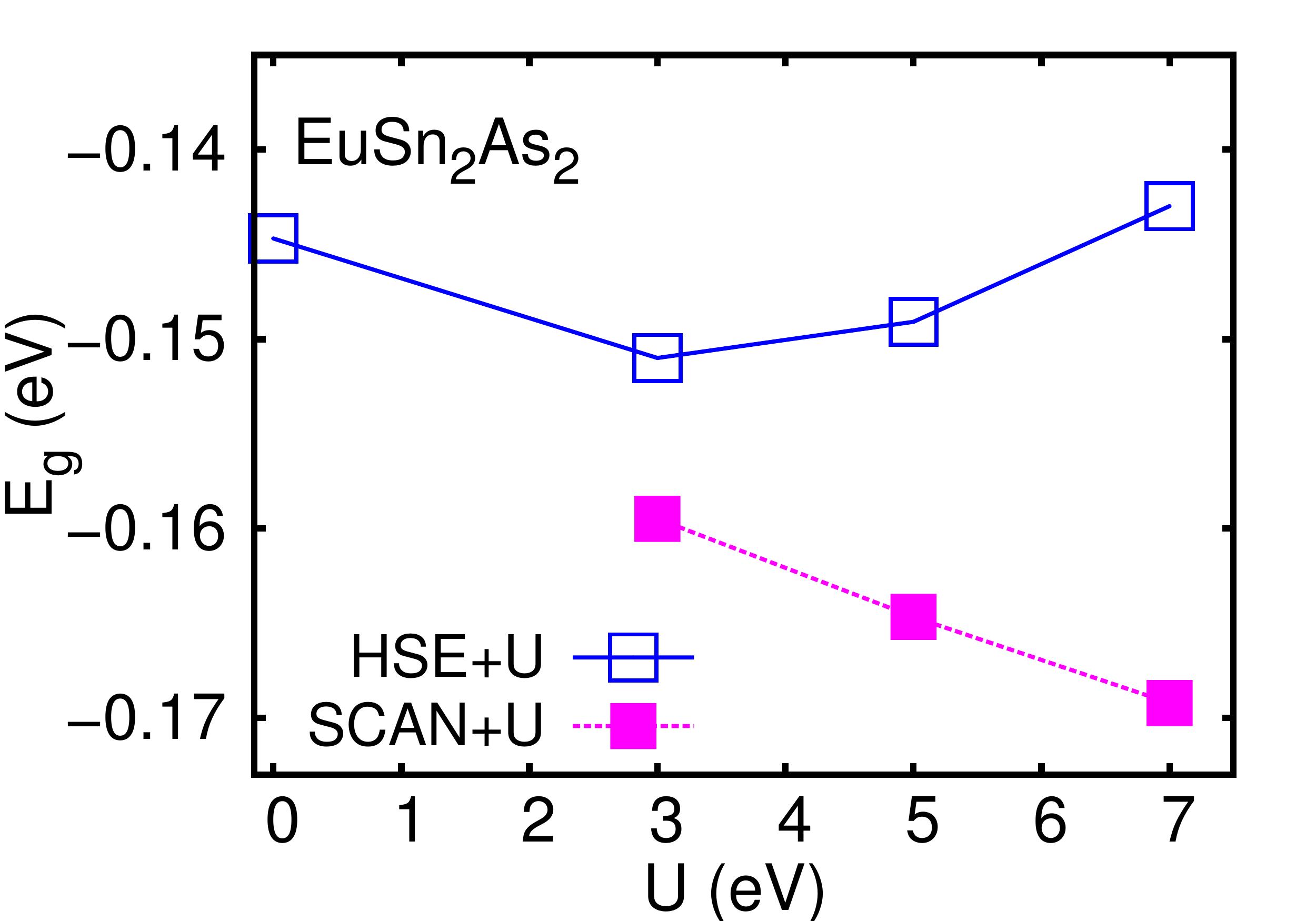}
\caption{Band gap $E_g$ for EuSn$_2$As$_2$ as a function of the Coulomb repulsion energy $U$ in the FM$c$ configuration by considering HSE+$U$ (blue solid line) and SCAN+$U$ (purple dashed line).}
\label{GAP_SOC_HYBRID_EuSnAs}
\end{figure}

Going beyond EuCd$_2$X$_2$ compounds, we investigated also other similar materials. EuIn$_2$As$_2$ is particularly known to be one of the few bulk axion insulators not protected by inversion symmetry proposed until now. Establishing its topological properties would be relevant for future investigations.
EuIn$_2$As$_2$ crystallizes in a hexagonal Bravais lattice with space group P6$_3$/mmc (No. 194) \cite{Regmi20,Goforth08,Rosa12}.
It was proposed that EuIn$_2$As$_2$ with space group P6$_3$/mmc could host higher-order topology \cite{Xu2019higher}, however, within HSE this is a trivial insulator.

Adding heavier atoms in search of topology, going from EuIn$_2$As$_2$ to EuIn$_2$Sb$_2$, it was predicted a change of the space group to R$\bar{3}$m \cite{Kirklin2015,Wang21npj}.
In this structural phase, we are able to study just the FM configuration within HSE due to computational limitations.
We find that EuIn$_2$Sb$_2$ in this structural phase is a robust trivial semimetal as we can see from the plot of the band structure in Fig.~\ref{BAND_SOC_HYBRID_EuInSb} and from the direct gap at the $\Gamma$ point in Fig.~\ref{GAP_SOC_HYBRID_EuInSb}.
An alloy of EuIn$_2$As$_{2-x}$Sb$_x$ that would keep the P6$_3$/mmc  (No.\,194) space group could be topological.

Another compound in the R$\bar{3}$mc phase is EuSn$_2$As$_2$.
We report the band structure in Fig.~\ref{BAND_SOC_HYBRID_EuSnAs}.
We plot the band gap as a function of $U$ and we obtain that we have an inverted band gap even within HSE. While in GGA+$U$, EuSn$_2$As$_2$ is a topological semimetal, in the HSE approach the system becomes an insulator.
This system is the only one without $s$-electrons in the conduction band bottom. As shown in Fig.~\ref{GAP_SOC_HYBRID_EuSnAs}, the band gap of EuSn$_2$As$_2$ shows a non-monotonic dependence on $U$, which can be attributed to the increased complexity due to the presence of several different manifolds of orbitals near the Fermi level.


\section*{Appendix F: E\lowercase{u}C\lowercase{d}$_2$S\lowercase{b}B\lowercase{i} preserving inversion symmetry}

\begin{figure}[t!]
\centering
\includegraphics[width=0.6\linewidth]{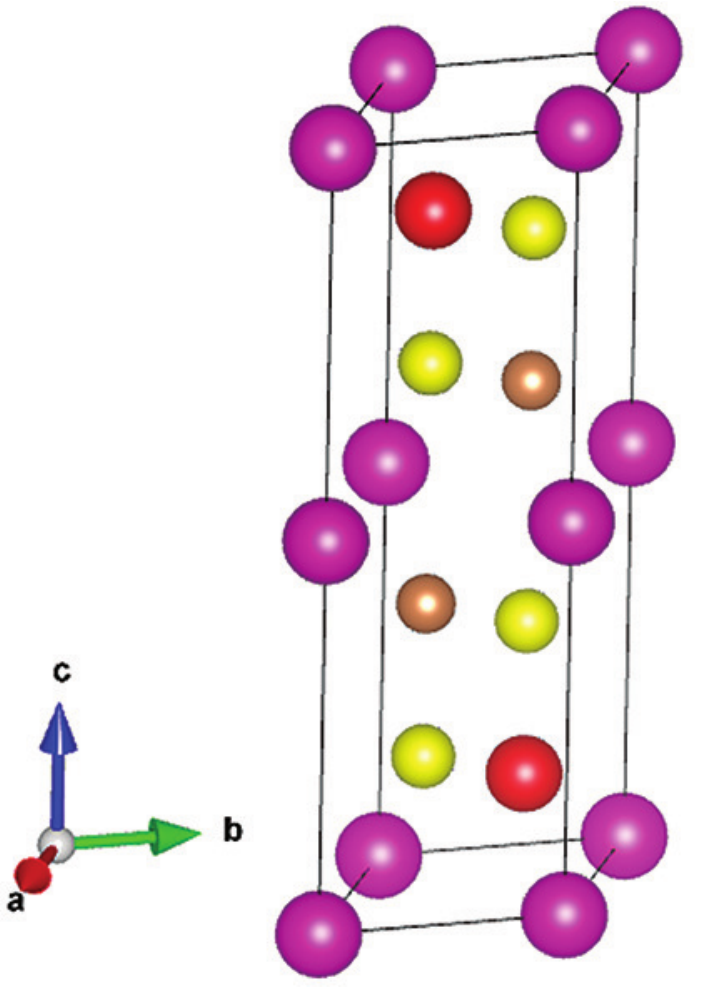}
\caption{Crystal structure of EuCd$_2$SbBi. Purple, yellow, brown and red balls denote Eu, Cd, Sb, and Bi atoms, respectively. In this crystal structure, the inversion symmetry is preserved.}
\label{Crystal_structure_EuCd2SbBi}
\end{figure}

\begin{figure}[t!]
\centering
\includegraphics[width=6.3cm,angle=270]{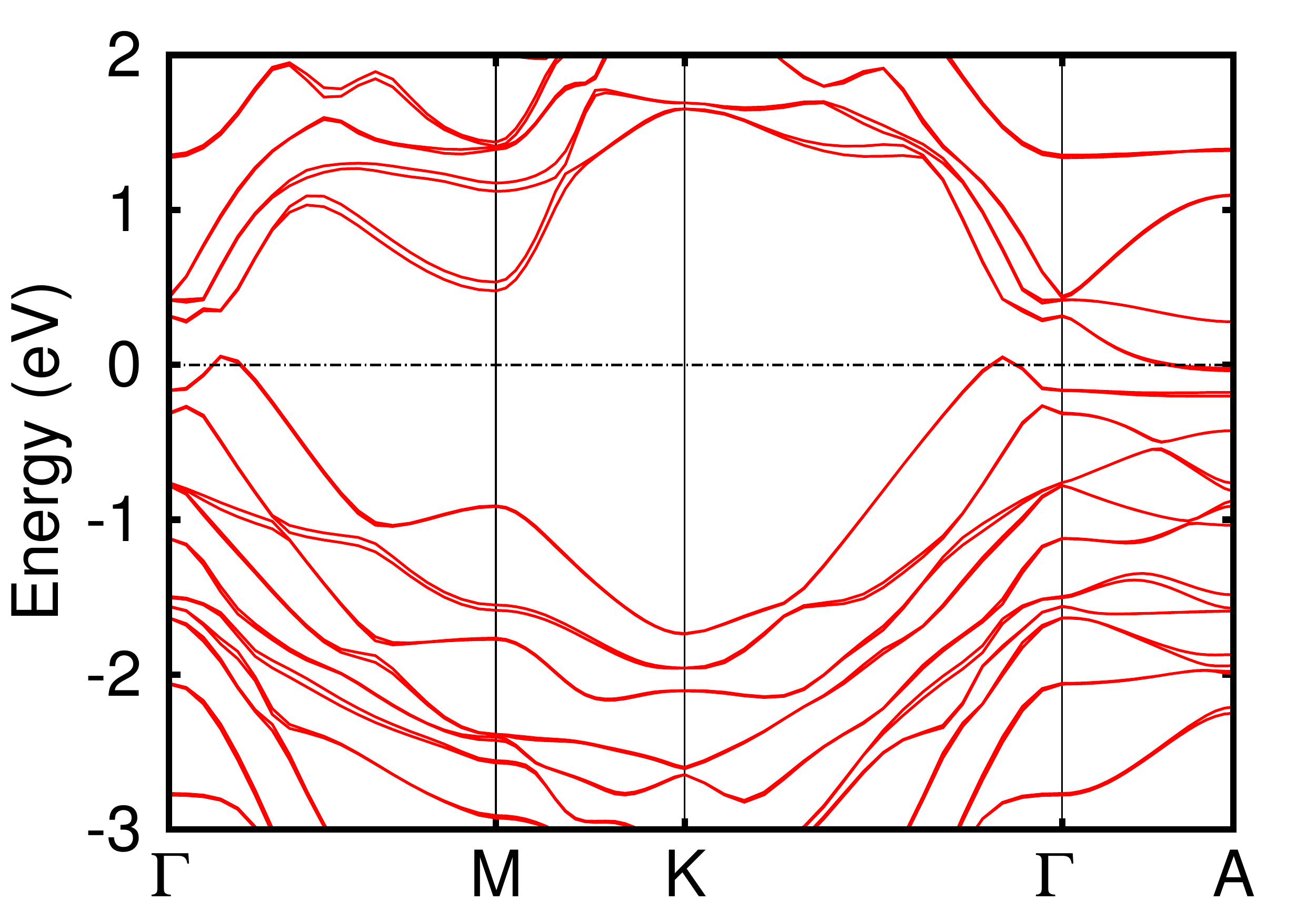}
\caption{Band structure of EuCd$_2$SbBi in the AFM$a$ configuration by considering HSE+$U$ with $U=7$\,eV between -3 and 2\,eV. The Fermi level is set at zero energy. }
\label{BAND_SOC_HYBRID_EuCdBiSb}
\end{figure}

We propose a new compound EuCd$_2$SbBi preserving the inversion symmetry with the inversion center on the Eu atoms. The lattice constants that we get are $a=4.85$\,{\AA} and $c=7.85$\,{\AA} which are close to the average of the lattice parameters of the two pristine compounds. The crystal structure is reported in Fig.~\ref{Crystal_structure_EuCd2SbBi}. This is a new topological semimetal. The band structure shown in Fig.~\ref{BAND_SOC_HYBRID_EuCdBiSb} has the same features of the topological phase of EuCd$_2$Bi$_2$ and very likely the same topology within HSE.

\bibliography{HgTe_td_28July2023}

\begin{thebibliography}{70}%
\makeatletter
\providecommand \@ifxundefined [1]{%
 \@ifx{#1\undefined}
}%
\providecommand \@ifnum [1]{%
 \ifnum #1\expandafter \@firstoftwo
 \else \expandafter \@secondoftwo
 \fi
}%
\providecommand \@ifx [1]{%
 \ifx #1\expandafter \@firstoftwo
 \else \expandafter \@secondoftwo
 \fi
}%
\providecommand \natexlab [1]{#1}%
\providecommand \enquote  [1]{``#1''}%
\providecommand \bibnamefont  [1]{#1}%
\providecommand \bibfnamefont [1]{#1}%
\providecommand \citenamefont [1]{#1}%
\providecommand \href@noop [0]{\@secondoftwo}%
\providecommand \href [0]{\begingroup \@sanitize@url \@href}%
\providecommand \@href[1]{\@@startlink{#1}\@@href}%
\providecommand \@@href[1]{\endgroup#1\@@endlink}%
\providecommand \@sanitize@url [0]{\catcode `\\12\catcode `\$12\catcode
  `\&12\catcode `\#12\catcode `\^12\catcode `\_12\catcode `\%12\relax}%
\providecommand \@@startlink[1]{}%
\providecommand \@@endlink[0]{}%
\providecommand \url  [0]{\begingroup\@sanitize@url \@url }%
\providecommand \@url [1]{\endgroup\@href {#1}{\urlprefix }}%
\providecommand \urlprefix  [0]{URL }%
\providecommand \Eprint [0]{\href }%
\providecommand \doibase [0]{https://doi.org/}%
\providecommand \selectlanguage [0]{\@gobble}%
\providecommand \bibinfo  [0]{\@secondoftwo}%
\providecommand \bibfield  [0]{\@secondoftwo}%
\providecommand \translation [1]{[#1]}%
\providecommand \BibitemOpen [0]{}%
\providecommand \bibitemStop [0]{}%
\providecommand \bibitemNoStop [0]{.\EOS\space}%
\providecommand \EOS [0]{\spacefactor3000\relax}%
\providecommand \BibitemShut  [1]{\csname bibitem#1\endcsname}%
\let\auto@bib@innerbib\@empty
\bibitem [{\citenamefont {Bernevig}\ \emph {et~al.}(2022)\citenamefont
  {Bernevig}, \citenamefont {Felser},\ and\ \citenamefont
  {Beidenkopf}}]{Bernevig:2022_N}%
  \BibitemOpen
  \bibfield  {author} {\bibinfo {author} {\bibfnamefont {B.~A.}\ \bibnamefont
  {Bernevig}}, \bibinfo {author} {\bibfnamefont {C.}~\bibnamefont {Felser}},\
  and\ \bibinfo {author} {\bibfnamefont {H.}~\bibnamefont {Beidenkopf}},\
  }\bibfield  {title} {\bibinfo {title} {Progress and prospects in magnetic
  topological materials},\ }\href {https://doi.org/10.1038/s41586-021-04105-x}
  {\bibfield  {journal} {\bibinfo  {journal} {Nature}\ }\textbf {\bibinfo
  {volume} {603}},\ \bibinfo {pages} {41} (\bibinfo {year} {2022})}\BibitemShut
  {NoStop}%
\bibitem [{\citenamefont {Xu}\ \emph {et~al.}(2020)\citenamefont {Xu},
  \citenamefont {Elcoro}, \citenamefont {Song}, \citenamefont {Wiede},
  \citenamefont {Vergniory}, \citenamefont {Regnault}, \citenamefont {Chen},
  \citenamefont {Felser},\ and\ \citenamefont {Bernevig}}]{Xu:2020_N}%
  \BibitemOpen
  \bibfield  {author} {\bibinfo {author} {\bibfnamefont {Y.}~\bibnamefont
  {Xu}}, \bibinfo {author} {\bibfnamefont {L.}~\bibnamefont {Elcoro}}, \bibinfo
  {author} {\bibfnamefont {Z.-D.}\ \bibnamefont {Song}}, \bibinfo {author}
  {\bibfnamefont {B.~J.}\ \bibnamefont {Wiede}}, \bibinfo {author}
  {\bibfnamefont {M.~G.}\ \bibnamefont {Vergniory}}, \bibinfo {author}
  {\bibfnamefont {N.}~\bibnamefont {Regnault}}, \bibinfo {author}
  {\bibfnamefont {Y.}~\bibnamefont {Chen}}, \bibinfo {author} {\bibfnamefont
  {C.}~\bibnamefont {Felser}},\ and\ \bibinfo {author} {\bibfnamefont {B.~A.}\
  \bibnamefont {Bernevig}},\ }\bibfield  {title} {\bibinfo {title}
  {High-throughput calculations of magnetic topological materials nature
  research},\ }\href {https://doi.org/10.1038/s41586-020-2837-0} {\bibfield
  {journal} {\bibinfo  {journal} {Nature}\ }\textbf {\bibinfo {volume} {586}},\
  \bibinfo {pages} {702} (\bibinfo {year} {2020})}\BibitemShut {NoStop}%
\bibitem [{\citenamefont {Bechstedt}\ \emph {et~al.}(2009)\citenamefont
  {Bechstedt}, \citenamefont {Fuchs},\ and\ \citenamefont
  {Kresse}}]{Bechstedt:2009_pssb}%
  \BibitemOpen
  \bibfield  {author} {\bibinfo {author} {\bibfnamefont {F.}~\bibnamefont
  {Bechstedt}}, \bibinfo {author} {\bibfnamefont {F.}~\bibnamefont {Fuchs}},\
  and\ \bibinfo {author} {\bibfnamefont {G.}~\bibnamefont {Kresse}},\
  }\bibfield  {title} {\bibinfo {title} {Ab-initio theory of semiconductor band
  structures: New developments and progress},\ }\href
  {https://doi.org/10.1002/pssb.200945074} {\bibfield  {journal} {\bibinfo
  {journal} {phys. stat. sol. (b)}\ }\textbf {\bibinfo {volume} {246}},\
  \bibinfo {pages} {1877} (\bibinfo {year} {2009})}\BibitemShut {NoStop}%
\bibitem [{\citenamefont {Schlipf}\ \emph {et~al.}(2013)\citenamefont
  {Schlipf}, \citenamefont {Betzinger}, \citenamefont {Le\v{z}ai\'{c}},
  \citenamefont {Friedrich},\ and\ \citenamefont {Bl\"ugel}}]{Schlipf13}%
  \BibitemOpen
  \bibfield  {author} {\bibinfo {author} {\bibfnamefont {M.}~\bibnamefont
  {Schlipf}}, \bibinfo {author} {\bibfnamefont {M.}~\bibnamefont {Betzinger}},
  \bibinfo {author} {\bibfnamefont {M.}~\bibnamefont {Le\v{z}ai\'{c}}},
  \bibinfo {author} {\bibfnamefont {C.}~\bibnamefont {Friedrich}},\ and\
  \bibinfo {author} {\bibfnamefont {S.}~\bibnamefont {Bl\"ugel}},\ }\bibfield
  {title} {\bibinfo {title} {Structural, electronic, and magnetic properties of
  the europium chalcogenides: A hybrid-functional {DFT} study},\ }\href
  {https://doi.org/10.1103/PhysRevB.88.094433} {\bibfield  {journal} {\bibinfo
  {journal} {Phys. Rev. B}\ }\textbf {\bibinfo {volume} {88}},\ \bibinfo
  {pages} {094433} (\bibinfo {year} {2013})}\BibitemShut {NoStop}%
\bibitem [{\citenamefont {Garza}\ and\ \citenamefont
  {Scuseria}(2016)}]{Garza:2016_JPChL}%
  \BibitemOpen
  \bibfield  {author} {\bibinfo {author} {\bibfnamefont {A.~J.}\ \bibnamefont
  {Garza}}\ and\ \bibinfo {author} {\bibfnamefont {G.~E.}\ \bibnamefont
  {Scuseria}},\ }\bibfield  {title} {\bibinfo {title} {Predicting band gaps
  with hybrid density functionals},\ }\href
  {https://doi.org/10.1021/acs.jpclett.6b01807} {\bibfield  {journal} {\bibinfo
   {journal} {J. Phys. Chem. Lett.}\ }\textbf {\bibinfo {volume} {7}},\
  \bibinfo {pages} {4165} (\bibinfo {year} {2016})}\BibitemShut {NoStop}%
\bibitem [{\citenamefont {Schellenberg}\ \emph {et~al.}(2011)\citenamefont
  {Schellenberg}, \citenamefont {Pfannenschmidt}, \citenamefont {Eul},
  \citenamefont {Schwickert},\ and\ \citenamefont
  {P\"ottgen}}]{Schellenberg:2011_ZAAC}%
  \BibitemOpen
  \bibfield  {author} {\bibinfo {author} {\bibfnamefont {I.}~\bibnamefont
  {Schellenberg}}, \bibinfo {author} {\bibfnamefont {U.}~\bibnamefont
  {Pfannenschmidt}}, \bibinfo {author} {\bibfnamefont {M.}~\bibnamefont {Eul}},
  \bibinfo {author} {\bibfnamefont {C.}~\bibnamefont {Schwickert}},\ and\
  \bibinfo {author} {\bibfnamefont {R.}~\bibnamefont {P\"ottgen}},\ }\bibfield
  {title} {\bibinfo {title} {A {$^{121}$Sb} and {$^{151}$Eu} {M\"ossbauer}
  spectroscopic investigation of {EuCd$_2$X$_2$} {(X = P, As, Sb)} and
  {YbCd$_2$Sb$_2$}},\ }\href {https://doi.org/10.1002/zaac.201100179}
  {\bibfield  {journal} {\bibinfo  {journal} {Z. Anorg. Allg. Chem}\ }\textbf
  {\bibinfo {volume} {637}},\ \bibinfo {pages} {1863} (\bibinfo {year}
  {2011})}\BibitemShut {NoStop}%
\bibitem [{\citenamefont {Rahn}\ \emph {et~al.}(2018)\citenamefont {Rahn},
  \citenamefont {Soh}, \citenamefont {Francoual}, \citenamefont {Veiga},
  \citenamefont {Strempfer}, \citenamefont {Mardegan}, \citenamefont {Yan},
  \citenamefont {Guo}, \citenamefont {Shi},\ and\ \citenamefont
  {Boothroyd}}]{Rahn18}%
  \BibitemOpen
  \bibfield  {author} {\bibinfo {author} {\bibfnamefont {M.~C.}\ \bibnamefont
  {Rahn}}, \bibinfo {author} {\bibfnamefont {J.-R.}\ \bibnamefont {Soh}},
  \bibinfo {author} {\bibfnamefont {S.}~\bibnamefont {Francoual}}, \bibinfo
  {author} {\bibfnamefont {L.~S.~I.}\ \bibnamefont {Veiga}}, \bibinfo {author}
  {\bibfnamefont {J.}~\bibnamefont {Strempfer}}, \bibinfo {author}
  {\bibfnamefont {J.}~\bibnamefont {Mardegan}}, \bibinfo {author}
  {\bibfnamefont {D.~Y.}\ \bibnamefont {Yan}}, \bibinfo {author} {\bibfnamefont
  {Y.~F.}\ \bibnamefont {Guo}}, \bibinfo {author} {\bibfnamefont {Y.~G.}\
  \bibnamefont {Shi}},\ and\ \bibinfo {author} {\bibfnamefont {A.~T.}\
  \bibnamefont {Boothroyd}},\ }\bibfield  {title} {\bibinfo {title} {Coupling
  of magnetic order and charge transport in the candidate {Dirac} semimetal
  {EuCd$_2$As$_2$}},\ }\href {https://doi.org/10.1103/PhysRevB.97.214422}
  {\bibfield  {journal} {\bibinfo  {journal} {Phys. Rev. B}\ }\textbf {\bibinfo
  {volume} {97}},\ \bibinfo {pages} {214422} (\bibinfo {year}
  {2018})}\BibitemShut {NoStop}%
\bibitem [{\citenamefont {Krishna}\ \emph {et~al.}(2018)\citenamefont
  {Krishna}, \citenamefont {Nautiyal},\ and\ \citenamefont
  {Maitra}}]{Krishna:2018_PRB}%
  \BibitemOpen
  \bibfield  {author} {\bibinfo {author} {\bibfnamefont {J.}~\bibnamefont
  {Krishna}}, \bibinfo {author} {\bibfnamefont {T.}~\bibnamefont {Nautiyal}},\
  and\ \bibinfo {author} {\bibfnamefont {T.}~\bibnamefont {Maitra}},\
  }\bibfield  {title} {\bibinfo {title} {First-principles study of electronic
  structure, transport, and optical properties of {EuCd$_2$As$_2$}},\ }\href
  {https://doi.org/10.1103/PhysRevB.98.125110} {\bibfield  {journal} {\bibinfo
  {journal} {Phys. Rev. B}\ }\textbf {\bibinfo {volume} {98}},\ \bibinfo
  {pages} {125110} (\bibinfo {year} {2018})}\BibitemShut {NoStop}%
\bibitem [{\citenamefont {Hua}\ \emph {et~al.}(2018)\citenamefont {Hua},
  \citenamefont {Nie}, \citenamefont {Song}, \citenamefont {Yu}, \citenamefont
  {Xu},\ and\ \citenamefont {Yao}}]{Hua:2018_PRB}%
  \BibitemOpen
  \bibfield  {author} {\bibinfo {author} {\bibfnamefont {G.}~\bibnamefont
  {Hua}}, \bibinfo {author} {\bibfnamefont {S.}~\bibnamefont {Nie}}, \bibinfo
  {author} {\bibfnamefont {Z.}~\bibnamefont {Song}}, \bibinfo {author}
  {\bibfnamefont {R.}~\bibnamefont {Yu}}, \bibinfo {author} {\bibfnamefont
  {G.}~\bibnamefont {Xu}},\ and\ \bibinfo {author} {\bibfnamefont
  {K.}~\bibnamefont {Yao}},\ }\bibfield  {title} {\bibinfo {title} {Dirac
  semimetal in type-{IV} magnetic space groups},\ }\href
  {https://doi.org/10.1103/PhysRevB.98.201116} {\bibfield  {journal} {\bibinfo
  {journal} {Phys. Rev. B}\ }\textbf {\bibinfo {volume} {98}},\ \bibinfo
  {pages} {201116} (\bibinfo {year} {2018})}\BibitemShut {NoStop}%
\bibitem [{\citenamefont {Ma}\ \emph {et~al.}(2020)\citenamefont {Ma},
  \citenamefont {Wang}, \citenamefont {Nie}, \citenamefont {Yi}, \citenamefont
  {Xu}, \citenamefont {Li}, \citenamefont {Jandke}, \citenamefont {Wulfhekel},
  \citenamefont {Huang}, \citenamefont {West}, \citenamefont {Richard},
  \citenamefont {Chikina}, \citenamefont {Strocov}, \citenamefont {Mesot},
  \citenamefont {Weng}, \citenamefont {Zhang}, \citenamefont {Shi},
  \citenamefont {Qian}, \citenamefont {Shi},\ and\ \citenamefont
  {Ding}}]{Ma:2020_AM}%
  \BibitemOpen
  \bibfield  {author} {\bibinfo {author} {\bibfnamefont {J.}~\bibnamefont
  {Ma}}, \bibinfo {author} {\bibfnamefont {H.}~\bibnamefont {Wang}}, \bibinfo
  {author} {\bibfnamefont {S.}~\bibnamefont {Nie}}, \bibinfo {author}
  {\bibfnamefont {C.}~\bibnamefont {Yi}}, \bibinfo {author} {\bibfnamefont
  {Y.}~\bibnamefont {Xu}}, \bibinfo {author} {\bibfnamefont {H.}~\bibnamefont
  {Li}}, \bibinfo {author} {\bibfnamefont {J.}~\bibnamefont {Jandke}}, \bibinfo
  {author} {\bibfnamefont {W.}~\bibnamefont {Wulfhekel}}, \bibinfo {author}
  {\bibfnamefont {Y.}~\bibnamefont {Huang}}, \bibinfo {author} {\bibfnamefont
  {D.}~\bibnamefont {West}}, \bibinfo {author} {\bibfnamefont {P.}~\bibnamefont
  {Richard}}, \bibinfo {author} {\bibfnamefont {A.}~\bibnamefont {Chikina}},
  \bibinfo {author} {\bibfnamefont {V.~N.}\ \bibnamefont {Strocov}}, \bibinfo
  {author} {\bibfnamefont {J.}~\bibnamefont {Mesot}}, \bibinfo {author}
  {\bibfnamefont {H.}~\bibnamefont {Weng}}, \bibinfo {author} {\bibfnamefont
  {S.}~\bibnamefont {Zhang}}, \bibinfo {author} {\bibfnamefont
  {Y.}~\bibnamefont {Shi}}, \bibinfo {author} {\bibfnamefont {T.}~\bibnamefont
  {Qian}}, \bibinfo {author} {\bibfnamefont {M.}~\bibnamefont {Shi}},\ and\
  \bibinfo {author} {\bibfnamefont {H.}~\bibnamefont {Ding}},\ }\bibfield
  {title} {\bibinfo {title} {Emergence of nontrivial low-energy {Dirac}
  {Fermi}ons in antiferromagnetic {EuCd$_2$2As$_2$}},\ }\href
  {https://doi.org/https://doi.org/10.1002/adma.201907565} {\bibfield
  {journal} {\bibinfo  {journal} {Adv. Mater.}\ }\textbf {\bibinfo {volume}
  {32}},\ \bibinfo {pages} {1907565} (\bibinfo {year} {2020})}\BibitemShut
  {NoStop}%
\bibitem [{\citenamefont {Cao}\ \emph {et~al.}(2022)\citenamefont {Cao},
  \citenamefont {Yu}, \citenamefont {Leng}, \citenamefont {Yi}, \citenamefont
  {Chen}, \citenamefont {Yang}, \citenamefont {Liu}, \citenamefont {Kong},
  \citenamefont {Li}, \citenamefont {Dong}, \citenamefont {Shi}, \citenamefont
  {Bibes}, \citenamefont {Peng}, \citenamefont {Zang},\ and\ \citenamefont
  {Xiu}}]{Cao_2022_PRR}%
  \BibitemOpen
  \bibfield  {author} {\bibinfo {author} {\bibfnamefont {X.}~\bibnamefont
  {Cao}}, \bibinfo {author} {\bibfnamefont {J.-X.}\ \bibnamefont {Yu}},
  \bibinfo {author} {\bibfnamefont {P.}~\bibnamefont {Leng}}, \bibinfo {author}
  {\bibfnamefont {C.}~\bibnamefont {Yi}}, \bibinfo {author} {\bibfnamefont
  {X.}~\bibnamefont {Chen}}, \bibinfo {author} {\bibfnamefont {Y.}~\bibnamefont
  {Yang}}, \bibinfo {author} {\bibfnamefont {S.}~\bibnamefont {Liu}}, \bibinfo
  {author} {\bibfnamefont {L.}~\bibnamefont {Kong}}, \bibinfo {author}
  {\bibfnamefont {Z.}~\bibnamefont {Li}}, \bibinfo {author} {\bibfnamefont
  {X.}~\bibnamefont {Dong}}, \bibinfo {author} {\bibfnamefont {Y.}~\bibnamefont
  {Shi}}, \bibinfo {author} {\bibfnamefont {M.}~\bibnamefont {Bibes}}, \bibinfo
  {author} {\bibfnamefont {R.}~\bibnamefont {Peng}}, \bibinfo {author}
  {\bibfnamefont {J.}~\bibnamefont {Zang}},\ and\ \bibinfo {author}
  {\bibfnamefont {F.}~\bibnamefont {Xiu}},\ }\bibfield  {title} {\bibinfo
  {title} {Giant nonlinear anomalous {Hall} effect induced by spin-dependent
  band structure evolution},\ }\href
  {https://doi.org/10.1103/PhysRevResearch.4.023100} {\bibfield  {journal}
  {\bibinfo  {journal} {Phys. Rev. Res.}\ }\textbf {\bibinfo {volume} {4}},\
  \bibinfo {pages} {023100} (\bibinfo {year} {2022})}\BibitemShut {NoStop}%
\bibitem [{\citenamefont {Wang}\ \emph
  {et~al.}(2019{\natexlab{a}})\citenamefont {Wang}, \citenamefont {Jo},
  \citenamefont {Kuthanazhi}, \citenamefont {Wu}, \citenamefont {McQueeney},
  \citenamefont {Kaminski},\ and\ \citenamefont {Canfield}}]{Wang:2019_PRB}%
  \BibitemOpen
  \bibfield  {author} {\bibinfo {author} {\bibfnamefont {L.-L.}\ \bibnamefont
  {Wang}}, \bibinfo {author} {\bibfnamefont {N.~H.}\ \bibnamefont {Jo}},
  \bibinfo {author} {\bibfnamefont {B.}~\bibnamefont {Kuthanazhi}}, \bibinfo
  {author} {\bibfnamefont {Y.}~\bibnamefont {Wu}}, \bibinfo {author}
  {\bibfnamefont {R.~J.}\ \bibnamefont {McQueeney}}, \bibinfo {author}
  {\bibfnamefont {A.}~\bibnamefont {Kaminski}},\ and\ \bibinfo {author}
  {\bibfnamefont {P.~C.}\ \bibnamefont {Canfield}},\ }\bibfield  {title}
  {\bibinfo {title} {Single pair of {Weyl} fermions in the half-metallic
  semimetal {EuCd$_2$As$_2$}},\ }\href
  {https://doi.org/10.1103/PhysRevB.99.245147} {\bibfield  {journal} {\bibinfo
  {journal} {Phys. Rev. B}\ }\textbf {\bibinfo {volume} {99}},\ \bibinfo
  {pages} {245147} (\bibinfo {year} {2019}{\natexlab{a}})}\BibitemShut
  {NoStop}%
\bibitem [{\citenamefont {Ma}\ \emph {et~al.}(2019)\citenamefont {Ma},
  \citenamefont {Nie}, \citenamefont {Yi}, \citenamefont {Jandke},
  \citenamefont {Shang}, \citenamefont {Yao}, \citenamefont {Naamneh},
  \citenamefont {Yan}, \citenamefont {Sun}, \citenamefont {Chikina},
  \citenamefont {Strocov}, \citenamefont {Medarde}, \citenamefont {Song},
  \citenamefont {Xiong}, \citenamefont {Xu}, \citenamefont {Wulfhekel},
  \citenamefont {Mesot}, \citenamefont {Reticcioli}, \citenamefont {Franchini},
  \citenamefont {Mudry}, \citenamefont {M\"uller}, \citenamefont {Shi},
  \citenamefont {Qian}, \citenamefont {Ding},\ and\ \citenamefont
  {Shi}}]{Ma:2019_SA}%
  \BibitemOpen
  \bibfield  {author} {\bibinfo {author} {\bibfnamefont {J.-Z.}\ \bibnamefont
  {Ma}}, \bibinfo {author} {\bibfnamefont {S.~M.}\ \bibnamefont {Nie}},
  \bibinfo {author} {\bibfnamefont {C.~J.}\ \bibnamefont {Yi}}, \bibinfo
  {author} {\bibfnamefont {J.}~\bibnamefont {Jandke}}, \bibinfo {author}
  {\bibfnamefont {T.}~\bibnamefont {Shang}}, \bibinfo {author} {\bibfnamefont
  {M.~Y.}\ \bibnamefont {Yao}}, \bibinfo {author} {\bibfnamefont
  {M.}~\bibnamefont {Naamneh}}, \bibinfo {author} {\bibfnamefont {L.~Q.}\
  \bibnamefont {Yan}}, \bibinfo {author} {\bibfnamefont {Y.}~\bibnamefont
  {Sun}}, \bibinfo {author} {\bibfnamefont {A.}~\bibnamefont {Chikina}},
  \bibinfo {author} {\bibfnamefont {V.~N.}\ \bibnamefont {Strocov}}, \bibinfo
  {author} {\bibfnamefont {M.}~\bibnamefont {Medarde}}, \bibinfo {author}
  {\bibfnamefont {M.}~\bibnamefont {Song}}, \bibinfo {author} {\bibfnamefont
  {Y.-M.}\ \bibnamefont {Xiong}}, \bibinfo {author} {\bibfnamefont
  {G.}~\bibnamefont {Xu}}, \bibinfo {author} {\bibfnamefont {W.}~\bibnamefont
  {Wulfhekel}}, \bibinfo {author} {\bibfnamefont {J.}~\bibnamefont {Mesot}},
  \bibinfo {author} {\bibfnamefont {M.}~\bibnamefont {Reticcioli}}, \bibinfo
  {author} {\bibfnamefont {C.}~\bibnamefont {Franchini}}, \bibinfo {author}
  {\bibfnamefont {C.}~\bibnamefont {Mudry}}, \bibinfo {author} {\bibfnamefont
  {M.}~\bibnamefont {M\"uller}}, \bibinfo {author} {\bibfnamefont {Y.~G.}\
  \bibnamefont {Shi}}, \bibinfo {author} {\bibfnamefont {T.}~\bibnamefont
  {Qian}}, \bibinfo {author} {\bibfnamefont {H.}~\bibnamefont {Ding}},\ and\
  \bibinfo {author} {\bibfnamefont {M.}~\bibnamefont {Shi}},\ }\bibfield
  {title} {\bibinfo {title} {Spin fluctuation induced {Weyl} semimetal state in
  the paramagnetic phase of {EuCd$_2$As$_2$}},\ }\href
  {https://doi.org/10.1126/sciadv.aaw4718} {\bibfield  {journal} {\bibinfo
  {journal} {Sci. Adv.}\ }\textbf {\bibinfo {volume} {5}},\ \bibinfo {pages}
  {eaaw4718} (\bibinfo {year} {2019})}\BibitemShut {NoStop}%
\bibitem [{\citenamefont {Taddei}\ \emph {et~al.}(2022)\citenamefont {Taddei},
  \citenamefont {Yin}, \citenamefont {Sanjeewa}, \citenamefont {Li},
  \citenamefont {Xing}, \citenamefont {dela Cruz}, \citenamefont {Phelan},
  \citenamefont {Sefat},\ and\ \citenamefont {Parker}}]{Taddei22}%
  \BibitemOpen
  \bibfield  {author} {\bibinfo {author} {\bibfnamefont {K.~M.}\ \bibnamefont
  {Taddei}}, \bibinfo {author} {\bibfnamefont {L.}~\bibnamefont {Yin}},
  \bibinfo {author} {\bibfnamefont {L.~D.}\ \bibnamefont {Sanjeewa}}, \bibinfo
  {author} {\bibfnamefont {Y.}~\bibnamefont {Li}}, \bibinfo {author}
  {\bibfnamefont {J.}~\bibnamefont {Xing}}, \bibinfo {author} {\bibfnamefont
  {C.}~\bibnamefont {dela Cruz}}, \bibinfo {author} {\bibfnamefont
  {D.}~\bibnamefont {Phelan}}, \bibinfo {author} {\bibfnamefont {A.~S.}\
  \bibnamefont {Sefat}},\ and\ \bibinfo {author} {\bibfnamefont {D.~S.}\
  \bibnamefont {Parker}},\ }\bibfield  {title} {\bibinfo {title} {Single pair
  of {Weyl} nodes in the spin-canted structure of {EuCd$_2$As$_2$}},\ }\href
  {https://doi.org/10.1103/PhysRevB.105.L140401} {\bibfield  {journal}
  {\bibinfo  {journal} {Phys. Rev. B}\ }\textbf {\bibinfo {volume} {105}},\
  \bibinfo {pages} {L140401} (\bibinfo {year} {2022})}\BibitemShut {NoStop}%
\bibitem [{\citenamefont {Wang}\ \emph {et~al.}(2022)\citenamefont {Wang},
  \citenamefont {Li}, \citenamefont {Miao}, \citenamefont {Zhang},
  \citenamefont {Li}, \citenamefont {Zhou}, \citenamefont {Yang}, \citenamefont
  {Yin}, \citenamefont {Cai}, \citenamefont {Song}, \citenamefont {Luo},
  \citenamefont {Chen}, \citenamefont {Mao}, \citenamefont {Zhao},
  \citenamefont {Deng}, \citenamefont {Sun}, \citenamefont {Zhu}, \citenamefont
  {Zhang}, \citenamefont {Yang}, \citenamefont {Wang}, \citenamefont {Zhang},
  \citenamefont {Peng}, \citenamefont {Pan}, \citenamefont {Shi}, \citenamefont
  {Weng}, \citenamefont {Xiang}, \citenamefont {Xu},\ and\ \citenamefont
  {Zhou}}]{Wang:2022_PRB}%
  \BibitemOpen
  \bibfield  {author} {\bibinfo {author} {\bibfnamefont {Y.}~\bibnamefont
  {Wang}}, \bibinfo {author} {\bibfnamefont {C.}~\bibnamefont {Li}}, \bibinfo
  {author} {\bibfnamefont {T.}~\bibnamefont {Miao}}, \bibinfo {author}
  {\bibfnamefont {S.}~\bibnamefont {Zhang}}, \bibinfo {author} {\bibfnamefont
  {Y.}~\bibnamefont {Li}}, \bibinfo {author} {\bibfnamefont {L.}~\bibnamefont
  {Zhou}}, \bibinfo {author} {\bibfnamefont {M.}~\bibnamefont {Yang}}, \bibinfo
  {author} {\bibfnamefont {C.}~\bibnamefont {Yin}}, \bibinfo {author}
  {\bibfnamefont {Y.}~\bibnamefont {Cai}}, \bibinfo {author} {\bibfnamefont
  {C.}~\bibnamefont {Song}}, \bibinfo {author} {\bibfnamefont {H.}~\bibnamefont
  {Luo}}, \bibinfo {author} {\bibfnamefont {H.}~\bibnamefont {Chen}}, \bibinfo
  {author} {\bibfnamefont {H.}~\bibnamefont {Mao}}, \bibinfo {author}
  {\bibfnamefont {L.}~\bibnamefont {Zhao}}, \bibinfo {author} {\bibfnamefont
  {H.}~\bibnamefont {Deng}}, \bibinfo {author} {\bibfnamefont {Y.}~\bibnamefont
  {Sun}}, \bibinfo {author} {\bibfnamefont {C.}~\bibnamefont {Zhu}}, \bibinfo
  {author} {\bibfnamefont {F.}~\bibnamefont {Zhang}}, \bibinfo {author}
  {\bibfnamefont {F.}~\bibnamefont {Yang}}, \bibinfo {author} {\bibfnamefont
  {Z.}~\bibnamefont {Wang}}, \bibinfo {author} {\bibfnamefont {S.}~\bibnamefont
  {Zhang}}, \bibinfo {author} {\bibfnamefont {Q.}~\bibnamefont {Peng}},
  \bibinfo {author} {\bibfnamefont {S.}~\bibnamefont {Pan}}, \bibinfo {author}
  {\bibfnamefont {Y.}~\bibnamefont {Shi}}, \bibinfo {author} {\bibfnamefont
  {H.}~\bibnamefont {Weng}}, \bibinfo {author} {\bibfnamefont {T.}~\bibnamefont
  {Xiang}}, \bibinfo {author} {\bibfnamefont {Z.}~\bibnamefont {Xu}},\ and\
  \bibinfo {author} {\bibfnamefont {X.~J.}\ \bibnamefont {Zhou}},\ }\bibfield
  {title} {\bibinfo {title} {Giant and reversible electronic structure
  evolution in a magnetic topological material
  {${\mathrm{EuCd}}_{2}{\mathrm{As}}_{2}$}},\ }\href
  {https://doi.org/10.1103/PhysRevB.106.085134} {\bibfield  {journal} {\bibinfo
   {journal} {Phys. Rev. B}\ }\textbf {\bibinfo {volume} {106}},\ \bibinfo
  {pages} {085134} (\bibinfo {year} {2022})}\BibitemShut {NoStop}%
\bibitem [{\citenamefont {Niu}\ \emph {et~al.}(2019)\citenamefont {Niu},
  \citenamefont {Mao}, \citenamefont {Hu}, \citenamefont {Huang},\ and\
  \citenamefont {Dai}}]{Niu19}%
  \BibitemOpen
  \bibfield  {author} {\bibinfo {author} {\bibfnamefont {C.}~\bibnamefont
  {Niu}}, \bibinfo {author} {\bibfnamefont {N.}~\bibnamefont {Mao}}, \bibinfo
  {author} {\bibfnamefont {X.}~\bibnamefont {Hu}}, \bibinfo {author}
  {\bibfnamefont {B.}~\bibnamefont {Huang}},\ and\ \bibinfo {author}
  {\bibfnamefont {Y.}~\bibnamefont {Dai}},\ }\bibfield  {title} {\bibinfo
  {title} {Quantum anomalous {Hall} effect and gate-controllable topological
  phase transition in layered ${\mathrm{eucd}}_{2}{\mathrm{as}}_{2}$},\ }\href
  {https://doi.org/10.1103/PhysRevB.99.235119} {\bibfield  {journal} {\bibinfo
  {journal} {Phys. Rev. B}\ }\textbf {\bibinfo {volume} {99}},\ \bibinfo
  {pages} {235119} (\bibinfo {year} {2019})}\BibitemShut {NoStop}%
\bibitem [{\citenamefont {Santos-Cottin}\ \emph {et~al.}(2023)\citenamefont
  {Santos-Cottin}, \citenamefont {Mohelsk\'y}, \citenamefont {Wyzula},
  \citenamefont {Mardel\'e}, \citenamefont {Kapon}, \citenamefont {Nasrallah},
  \citenamefont {Bari\v{s}i\'c}, \citenamefont {\v{Z}ivkovi\'c}, \citenamefont
  {Soh}, \citenamefont {Guo}, \citenamefont {Puppin}, \citenamefont {Dil},
  \citenamefont {Gudac}, \citenamefont {Rukelj}, \citenamefont {Novak},
  \citenamefont {Kuzmenko}, \citenamefont {Homes}, \citenamefont {Dietl},
  \citenamefont {Orlita},\ and\ \citenamefont
  {Akrap}}]{santoscottin2023eucd2as2}%
  \BibitemOpen
  \bibfield  {author} {\bibinfo {author} {\bibfnamefont {D.}~\bibnamefont
  {Santos-Cottin}}, \bibinfo {author} {\bibfnamefont {I.}~\bibnamefont
  {Mohelsk\'y}}, \bibinfo {author} {\bibfnamefont {J.}~\bibnamefont {Wyzula}},
  \bibinfo {author} {\bibfnamefont {F.~L.}\ \bibnamefont {Mardel\'e}}, \bibinfo
  {author} {\bibfnamefont {I.}~\bibnamefont {Kapon}}, \bibinfo {author}
  {\bibfnamefont {S.}~\bibnamefont {Nasrallah}}, \bibinfo {author}
  {\bibfnamefont {N.}~\bibnamefont {Bari\v{s}i\'c}}, \bibinfo {author}
  {\bibfnamefont {I.}~\bibnamefont {\v{Z}ivkovi\'c}}, \bibinfo {author}
  {\bibfnamefont {J.~R.}\ \bibnamefont {Soh}}, \bibinfo {author} {\bibfnamefont
  {F.}~\bibnamefont {Guo}}, \bibinfo {author} {\bibfnamefont {M.}~\bibnamefont
  {Puppin}}, \bibinfo {author} {\bibfnamefont {J.~H.}\ \bibnamefont {Dil}},
  \bibinfo {author} {\bibfnamefont {B.}~\bibnamefont {Gudac}}, \bibinfo
  {author} {\bibfnamefont {Z.}~\bibnamefont {Rukelj}}, \bibinfo {author}
  {\bibfnamefont {M.}~\bibnamefont {Novak}}, \bibinfo {author} {\bibfnamefont
  {A.~B.}\ \bibnamefont {Kuzmenko}}, \bibinfo {author} {\bibfnamefont {C.~C.}\
  \bibnamefont {Homes}}, \bibinfo {author} {\bibfnamefont {T.}~\bibnamefont
  {Dietl}}, \bibinfo {author} {\bibfnamefont {M.}~\bibnamefont {Orlita}},\ and\
  \bibinfo {author} {\bibfnamefont {A.}~\bibnamefont {Akrap}},\ }\href@noop {}
  {\bibinfo {title} {{EuCd$_2$As$_2$}: a magnetic semiconductor}} (\bibinfo
  {year} {2023}),\ \Eprint {https://arxiv.org/abs/2301.08014} {arXiv:2301.08014
  [cond-mat.mtrl-sci]} \BibitemShut {NoStop}%
\bibitem [{\citenamefont {Paier}\ \emph {et~al.}(2006)\citenamefont {Paier},
  \citenamefont {Marsman}, \citenamefont {Hummer}, \citenamefont {Kresse},
  \citenamefont {Gerber},\ and\ \citenamefont {Angyan}}]{Paier06}%
  \BibitemOpen
  \bibfield  {author} {\bibinfo {author} {\bibfnamefont {J.}~\bibnamefont
  {Paier}}, \bibinfo {author} {\bibfnamefont {M.}~\bibnamefont {Marsman}},
  \bibinfo {author} {\bibfnamefont {K.}~\bibnamefont {Hummer}}, \bibinfo
  {author} {\bibfnamefont {G.}~\bibnamefont {Kresse}}, \bibinfo {author}
  {\bibfnamefont {I.~C.}\ \bibnamefont {Gerber}},\ and\ \bibinfo {author}
  {\bibfnamefont {J.~G.}\ \bibnamefont {Angyan}},\ }\bibfield  {title}
  {\bibinfo {title} {Screened hybrid density functionals applied to solids},\
  }\href {https://doi.org/10.1063/1.2187006} {\bibfield  {journal} {\bibinfo
  {journal} {J. Chem. Phys.}\ }\textbf {\bibinfo {volume} {124}},\ \bibinfo
  {pages} {154709} (\bibinfo {year} {2006})}\BibitemShut {NoStop}%
\bibitem [{\citenamefont {Sun}\ \emph {et~al.}(2015)\citenamefont {Sun},
  \citenamefont {Ruzsinszky},\ and\ \citenamefont
  {Perdew}}]{PhysRevLett.115.036402}%
  \BibitemOpen
  \bibfield  {author} {\bibinfo {author} {\bibfnamefont {J.}~\bibnamefont
  {Sun}}, \bibinfo {author} {\bibfnamefont {A.}~\bibnamefont {Ruzsinszky}},\
  and\ \bibinfo {author} {\bibfnamefont {J.~P.}\ \bibnamefont {Perdew}},\
  }\bibfield  {title} {\bibinfo {title} {Strongly constrained and appropriately
  normed semilocal density functional},\ }\href
  {https://doi.org/10.1103/PhysRevLett.115.036402} {\bibfield  {journal}
  {\bibinfo  {journal} {Phys. Rev. Lett.}\ }\textbf {\bibinfo {volume} {115}},\
  \bibinfo {pages} {036402} (\bibinfo {year} {2015})}\BibitemShut {NoStop}%
\bibitem [{\citenamefont {Hussain}\ \emph {et~al.}(2022)\citenamefont
  {Hussain}, \citenamefont {Cuono}, \citenamefont {Islam}, \citenamefont
  {Trajnerowicz}, \citenamefont {Jurenczyk}, \citenamefont {Autieri},\ and\
  \citenamefont {Dietl}}]{Hussain2022electronic}%
  \BibitemOpen
  \bibfield  {author} {\bibinfo {author} {\bibfnamefont {G.}~\bibnamefont
  {Hussain}}, \bibinfo {author} {\bibfnamefont {G.}~\bibnamefont {Cuono}},
  \bibinfo {author} {\bibfnamefont {R.}~\bibnamefont {Islam}}, \bibinfo
  {author} {\bibfnamefont {A.}~\bibnamefont {Trajnerowicz}}, \bibinfo {author}
  {\bibfnamefont {J.}~\bibnamefont {Jurenczyk}}, \bibinfo {author}
  {\bibfnamefont {C.}~\bibnamefont {Autieri}},\ and\ \bibinfo {author}
  {\bibfnamefont {T.}~\bibnamefont {Dietl}},\ }\bibfield  {title} {\bibinfo
  {title} {Electronic and optical properties of
  {InAs/InAs$_{0.625}$Sb$_{0.375}$} superlattices and their application for
  far-infrared detectors},\ }\href {https://doi.org/10.1088/1361-6463/ac984d}
  {\bibfield  {journal} {\bibinfo  {journal} {J. Phys. D: Appl. Phys.}\
  }\textbf {\bibinfo {volume} {55}},\ \bibinfo {pages} {495301} (\bibinfo
  {year} {2022})}\BibitemShut {NoStop}%
\bibitem [{\citenamefont {Yalameha}\ \emph {et~al.}(2023)\citenamefont
  {Yalameha}, \citenamefont {Nourbakhsh}, \citenamefont {Bahramy},\ and\
  \citenamefont {Vashaee}}]{D3CP00005B}%
  \BibitemOpen
  \bibfield  {author} {\bibinfo {author} {\bibfnamefont {S.}~\bibnamefont
  {Yalameha}}, \bibinfo {author} {\bibfnamefont {Z.}~\bibnamefont
  {Nourbakhsh}}, \bibinfo {author} {\bibfnamefont {M.~S.}\ \bibnamefont
  {Bahramy}},\ and\ \bibinfo {author} {\bibfnamefont {D.}~\bibnamefont
  {Vashaee}},\ }\bibfield  {title} {\bibinfo {title} {New insights into band
  inversion and topological phase of {TiNI} monolayer},\ }\href
  {https://doi.org/10.1039/D3CP00005B} {\bibfield  {journal} {\bibinfo
  {journal} {Phys. Chem. Chem. Phys.}\ ,\ } (\bibinfo {year}
  {2023})}\BibitemShut {NoStop}%
\bibitem [{\citenamefont {Dietl}(2008)}]{Dietl:2008_JPSJ}%
  \BibitemOpen
  \bibfield  {author} {\bibinfo {author} {\bibfnamefont {T.}~\bibnamefont
  {Dietl}},\ }\bibfield  {title} {\bibinfo {title} {Interplay between carrier
  localization and magnetism in diluted magnetic and ferromagnetic
  semiconductors},\ }\href {https://doi.org/10.1143/JPSJ.77.031005} {\bibfield
  {journal} {\bibinfo  {journal} {J. Phys. Soc. Japan}\ }\textbf {\bibinfo
  {volume} {77}},\ \bibinfo {pages} {031005} (\bibinfo {year}
  {2008})}\BibitemShut {NoStop}%
\bibitem [{\citenamefont {Shapira}\ \emph {et~al.}(1972)\citenamefont
  {Shapira}, \citenamefont {Foner}, \citenamefont {Oliveira},\ and\
  \citenamefont {Reed}}]{Shapira:1972_PRB}%
  \BibitemOpen
  \bibfield  {author} {\bibinfo {author} {\bibfnamefont {Y.}~\bibnamefont
  {Shapira}}, \bibinfo {author} {\bibfnamefont {S.}~\bibnamefont {Foner}},
  \bibinfo {author} {\bibfnamefont {N.~F.}\ \bibnamefont {Oliveira}},\ and\
  \bibinfo {author} {\bibfnamefont {T.~B.}\ \bibnamefont {Reed}},\ }\bibfield
  {title} {\bibinfo {title} {{EuTe}. {II}. {Resistivity} and {Hall} effect},\
  }\href {https://doi.org/10.1103/PhysRevB.5.2647} {\bibfield  {journal}
  {\bibinfo  {journal} {Phys. Rev. B}\ }\textbf {\bibinfo {volume} {5}},\
  \bibinfo {pages} {2647} (\bibinfo {year} {1972})}\BibitemShut {NoStop}%
\bibitem [{\citenamefont {Wang}\ \emph
  {et~al.}(2021{\natexlab{a}})\citenamefont {Wang}, \citenamefont {Mao},
  \citenamefont {Hu}, \citenamefont {Dai}, \citenamefont {Huang},\ and\
  \citenamefont {Niu}}]{Wang:2021_MH}%
  \BibitemOpen
  \bibfield  {author} {\bibinfo {author} {\bibfnamefont {H.}~\bibnamefont
  {Wang}}, \bibinfo {author} {\bibfnamefont {N.}~\bibnamefont {Mao}}, \bibinfo
  {author} {\bibfnamefont {X.}~\bibnamefont {Hu}}, \bibinfo {author}
  {\bibfnamefont {Y.}~\bibnamefont {Dai}}, \bibinfo {author} {\bibfnamefont
  {B.}~\bibnamefont {Huang}},\ and\ \bibinfo {author} {\bibfnamefont
  {C.}~\bibnamefont {Niu}},\ }\bibfield  {title} {\bibinfo {title} {A magnetic
  topological insulator in two-dimensional {EuCd$_2$Bi$_2$}: giant gap with
  robust topology against magnetic transitions},\ }\href
  {https://doi.org/10.1039/D0MH01214A} {\bibfield  {journal} {\bibinfo
  {journal} {Mater. Horiz.}\ }\textbf {\bibinfo {volume} {8}},\ \bibinfo
  {pages} {956} (\bibinfo {year} {2021}{\natexlab{a}})}\BibitemShut {NoStop}%
\bibitem [{\citenamefont {Riberolles}\ \emph {et~al.}(2021)\citenamefont
  {Riberolles}, \citenamefont {Trevisan}, \citenamefont {Kuthanazhi},
  \citenamefont {Heitmann}, \citenamefont {Ye}, \citenamefont {Johnston},
  \citenamefont {Bud'ko}, \citenamefont {Ryan}, \citenamefont {Canfield},
  \citenamefont {Kreyssig}, \citenamefont {Vishwanath}, \citenamefont
  {McQueeney}, \citenamefont {Wang}, \citenamefont {Orth},\ and\ \citenamefont
  {Ueland}}]{Riberolles2021}%
  \BibitemOpen
  \bibfield  {author} {\bibinfo {author} {\bibfnamefont {S.~X.~M.}\
  \bibnamefont {Riberolles}}, \bibinfo {author} {\bibfnamefont {T.~V.}\
  \bibnamefont {Trevisan}}, \bibinfo {author} {\bibfnamefont {B.}~\bibnamefont
  {Kuthanazhi}}, \bibinfo {author} {\bibfnamefont {T.~W.}\ \bibnamefont
  {Heitmann}}, \bibinfo {author} {\bibfnamefont {F.}~\bibnamefont {Ye}},
  \bibinfo {author} {\bibfnamefont {D.~C.}\ \bibnamefont {Johnston}}, \bibinfo
  {author} {\bibfnamefont {S.~L.}\ \bibnamefont {Bud'ko}}, \bibinfo {author}
  {\bibfnamefont {D.~H.}\ \bibnamefont {Ryan}}, \bibinfo {author}
  {\bibfnamefont {P.~C.}\ \bibnamefont {Canfield}}, \bibinfo {author}
  {\bibfnamefont {A.}~\bibnamefont {Kreyssig}}, \bibinfo {author}
  {\bibfnamefont {A.}~\bibnamefont {Vishwanath}}, \bibinfo {author}
  {\bibfnamefont {R.~J.}\ \bibnamefont {McQueeney}}, \bibinfo {author}
  {\bibfnamefont {L.-L.}\ \bibnamefont {Wang}}, \bibinfo {author}
  {\bibfnamefont {P.~P.}\ \bibnamefont {Orth}},\ and\ \bibinfo {author}
  {\bibfnamefont {B.~G.}\ \bibnamefont {Ueland}},\ }\bibfield  {title}
  {\bibinfo {title} {Magnetic crystalline-symmetry-protected axion
  electrodynamics and field-tunable unpinned {Dirac} cones in
  {EuIn$_2$As$_2$}},\ }\href {https://doi.org/10.1038/s41467-021-21154-y}
  {\bibfield  {journal} {\bibinfo  {journal} {Nature Communications}\ }\textbf
  {\bibinfo {volume} {12}},\ \bibinfo {pages} {999} (\bibinfo {year}
  {2021})}\BibitemShut {NoStop}%
\bibitem [{\citenamefont {Islam}\ \emph {et~al.}(2023)\citenamefont {Islam},
  \citenamefont {Mardanya}, \citenamefont {Lau}, \citenamefont {Cuono},
  \citenamefont {Chang}, \citenamefont {Singh}, \citenamefont {Canali},
  \citenamefont {Dietl},\ and\ \citenamefont {Autieri}}]{Islam2023}%
  \BibitemOpen
  \bibfield  {author} {\bibinfo {author} {\bibfnamefont {R.}~\bibnamefont
  {Islam}}, \bibinfo {author} {\bibfnamefont {S.}~\bibnamefont {Mardanya}},
  \bibinfo {author} {\bibfnamefont {A.}~\bibnamefont {Lau}}, \bibinfo {author}
  {\bibfnamefont {G.}~\bibnamefont {Cuono}}, \bibinfo {author} {\bibfnamefont
  {T.-R.}\ \bibnamefont {Chang}}, \bibinfo {author} {\bibfnamefont
  {B.}~\bibnamefont {Singh}}, \bibinfo {author} {\bibfnamefont {C.~M.}\
  \bibnamefont {Canali}}, \bibinfo {author} {\bibfnamefont {T.}~\bibnamefont
  {Dietl}},\ and\ \bibinfo {author} {\bibfnamefont {C.}~\bibnamefont
  {Autieri}},\ }\bibfield  {title} {\bibinfo {title} {Engineering axion
  insulator and other topological phases in superlattices without inversion
  symmetry},\ }\href {https://doi.org/10.1103/PhysRevB.107.125102} {\bibfield
  {journal} {\bibinfo  {journal} {Phys. Rev. B}\ }\textbf {\bibinfo {volume}
  {107}},\ \bibinfo {pages} {125102} (\bibinfo {year} {2023})}\BibitemShut
  {NoStop}%
\bibitem [{\citenamefont {Guo}\ \emph {et~al.}(2022)\citenamefont {Guo},
  \citenamefont {Huang},\ and\ \citenamefont {Zhang}}]{Guo2022the}%
  \BibitemOpen
  \bibfield  {author} {\bibinfo {author} {\bibfnamefont {W.-T.}\ \bibnamefont
  {Guo}}, \bibinfo {author} {\bibfnamefont {Z.}~\bibnamefont {Huang}},\ and\
  \bibinfo {author} {\bibfnamefont {J.-M.}\ \bibnamefont {Zhang}},\ }\bibfield
  {title} {\bibinfo {title} {The {Zintl} phase compounds aein2as2 (ae = ca{,}
  sr{,} ba): topological phase transition under pressure},\ }\href
  {https://doi.org/10.1039/D2CP01764D} {\bibfield  {journal} {\bibinfo
  {journal} {Phys. Chem. Chem. Phys.}\ }\textbf {\bibinfo {volume} {24}},\
  \bibinfo {pages} {17337} (\bibinfo {year} {2022})}\BibitemShut {NoStop}%
\bibitem [{\citenamefont {Sarkar}\ \emph {et~al.}(2022)\citenamefont {Sarkar},
  \citenamefont {Mardanya}, \citenamefont {Huang}, \citenamefont {Ghosh},
  \citenamefont {Huang}, \citenamefont {Lin}, \citenamefont {Bansil},
  \citenamefont {Chang}, \citenamefont {Agarwal},\ and\ \citenamefont
  {Singh}}]{PhysRevMaterials.6.044204}%
  \BibitemOpen
  \bibfield  {author} {\bibinfo {author} {\bibfnamefont {A.~B.}\ \bibnamefont
  {Sarkar}}, \bibinfo {author} {\bibfnamefont {S.}~\bibnamefont {Mardanya}},
  \bibinfo {author} {\bibfnamefont {S.-M.}\ \bibnamefont {Huang}}, \bibinfo
  {author} {\bibfnamefont {B.}~\bibnamefont {Ghosh}}, \bibinfo {author}
  {\bibfnamefont {C.-Y.}\ \bibnamefont {Huang}}, \bibinfo {author}
  {\bibfnamefont {H.}~\bibnamefont {Lin}}, \bibinfo {author} {\bibfnamefont
  {A.}~\bibnamefont {Bansil}}, \bibinfo {author} {\bibfnamefont {T.-R.}\
  \bibnamefont {Chang}}, \bibinfo {author} {\bibfnamefont {A.}~\bibnamefont
  {Agarwal}},\ and\ \bibinfo {author} {\bibfnamefont {B.}~\bibnamefont
  {Singh}},\ }\bibfield  {title} {\bibinfo {title} {Magnetically tunable
  {Dirac} and {Weyl} fermions in the {Zintl} materials family},\ }\href
  {https://doi.org/10.1103/PhysRevMaterials.6.044204} {\bibfield  {journal}
  {\bibinfo  {journal} {Phys. Rev. Mater.}\ }\textbf {\bibinfo {volume} {6}},\
  \bibinfo {pages} {044204} (\bibinfo {year} {2022})}\BibitemShut {NoStop}%
\bibitem [{\citenamefont {Xu}\ \emph {et~al.}(2019)\citenamefont {Xu},
  \citenamefont {Song}, \citenamefont {Wang}, \citenamefont {Weng},\ and\
  \citenamefont {Dai}}]{Xu2019higher}%
  \BibitemOpen
  \bibfield  {author} {\bibinfo {author} {\bibfnamefont {Y.}~\bibnamefont
  {Xu}}, \bibinfo {author} {\bibfnamefont {Z.}~\bibnamefont {Song}}, \bibinfo
  {author} {\bibfnamefont {Z.}~\bibnamefont {Wang}}, \bibinfo {author}
  {\bibfnamefont {H.}~\bibnamefont {Weng}},\ and\ \bibinfo {author}
  {\bibfnamefont {X.}~\bibnamefont {Dai}},\ }\bibfield  {title} {\bibinfo
  {title} {Higher-order topology of the axion insulator
  ${\mathrm{euin}}_{2}{\mathrm{as}}_{2}$},\ }\href
  {https://doi.org/10.1103/PhysRevLett.122.256402} {\bibfield  {journal}
  {\bibinfo  {journal} {Phys. Rev. Lett.}\ }\textbf {\bibinfo {volume} {122}},\
  \bibinfo {pages} {256402} (\bibinfo {year} {2019})}\BibitemShut {NoStop}%
\bibitem [{\citenamefont {Li}\ \emph {et~al.}(2019)\citenamefont {Li},
  \citenamefont {Gao}, \citenamefont {Duan}, \citenamefont {Xu}, \citenamefont
  {Zhu}, \citenamefont {Tian}, \citenamefont {Gao}, \citenamefont {Fan},
  \citenamefont {Rao}, \citenamefont {Huang}, \citenamefont {Li}, \citenamefont
  {Yan}, \citenamefont {Liu}, \citenamefont {Liu}, \citenamefont {Huang},
  \citenamefont {Li}, \citenamefont {Liu}, \citenamefont {Zhang}, \citenamefont
  {Zhang}, \citenamefont {Kondo}, \citenamefont {Shin}, \citenamefont {Lei},
  \citenamefont {Shi}, \citenamefont {Zhang}, \citenamefont {Weng},
  \citenamefont {Qian},\ and\ \citenamefont {Ding}}]{PhysRevX.9.041039}%
  \BibitemOpen
  \bibfield  {author} {\bibinfo {author} {\bibfnamefont {H.}~\bibnamefont
  {Li}}, \bibinfo {author} {\bibfnamefont {S.-Y.}\ \bibnamefont {Gao}},
  \bibinfo {author} {\bibfnamefont {S.-F.}\ \bibnamefont {Duan}}, \bibinfo
  {author} {\bibfnamefont {Y.-F.}\ \bibnamefont {Xu}}, \bibinfo {author}
  {\bibfnamefont {K.-J.}\ \bibnamefont {Zhu}}, \bibinfo {author} {\bibfnamefont
  {S.-J.}\ \bibnamefont {Tian}}, \bibinfo {author} {\bibfnamefont {J.-C.}\
  \bibnamefont {Gao}}, \bibinfo {author} {\bibfnamefont {W.-H.}\ \bibnamefont
  {Fan}}, \bibinfo {author} {\bibfnamefont {Z.-C.}\ \bibnamefont {Rao}},
  \bibinfo {author} {\bibfnamefont {J.-R.}\ \bibnamefont {Huang}}, \bibinfo
  {author} {\bibfnamefont {J.-J.}\ \bibnamefont {Li}}, \bibinfo {author}
  {\bibfnamefont {D.-Y.}\ \bibnamefont {Yan}}, \bibinfo {author} {\bibfnamefont
  {Z.-T.}\ \bibnamefont {Liu}}, \bibinfo {author} {\bibfnamefont {W.-L.}\
  \bibnamefont {Liu}}, \bibinfo {author} {\bibfnamefont {Y.-B.}\ \bibnamefont
  {Huang}}, \bibinfo {author} {\bibfnamefont {Y.-L.}\ \bibnamefont {Li}},
  \bibinfo {author} {\bibfnamefont {Y.}~\bibnamefont {Liu}}, \bibinfo {author}
  {\bibfnamefont {G.-B.}\ \bibnamefont {Zhang}}, \bibinfo {author}
  {\bibfnamefont {P.}~\bibnamefont {Zhang}}, \bibinfo {author} {\bibfnamefont
  {T.}~\bibnamefont {Kondo}}, \bibinfo {author} {\bibfnamefont
  {S.}~\bibnamefont {Shin}}, \bibinfo {author} {\bibfnamefont {H.-C.}\
  \bibnamefont {Lei}}, \bibinfo {author} {\bibfnamefont {Y.-G.}\ \bibnamefont
  {Shi}}, \bibinfo {author} {\bibfnamefont {W.-T.}\ \bibnamefont {Zhang}},
  \bibinfo {author} {\bibfnamefont {H.-M.}\ \bibnamefont {Weng}}, \bibinfo
  {author} {\bibfnamefont {T.}~\bibnamefont {Qian}},\ and\ \bibinfo {author}
  {\bibfnamefont {H.}~\bibnamefont {Ding}},\ }\bibfield  {title} {\bibinfo
  {title} {{Dirac Surface States in Intrinsic Magnetic Topological Insulators
  ${\mathrm{EuSn}}_{2}{\mathrm{As}}_{2}$ and
  ${\mathrm{MnBi}}_{2n}{\mathrm{Te}}_{3n+1}$}},\ }\href
  {https://doi.org/10.1103/PhysRevX.9.041039} {\bibfield  {journal} {\bibinfo
  {journal} {Phys. Rev. X}\ }\textbf {\bibinfo {volume} {9}},\ \bibinfo {pages}
  {041039} (\bibinfo {year} {2019})}\BibitemShut {NoStop}%
\bibitem [{\citenamefont {Niu}\ \emph {et~al.}(2020)\citenamefont {Niu},
  \citenamefont {Wang}, \citenamefont {Mao}, \citenamefont {Huang},
  \citenamefont {Mokrousov},\ and\ \citenamefont {Dai}}]{Niu20}%
  \BibitemOpen
  \bibfield  {author} {\bibinfo {author} {\bibfnamefont {C.}~\bibnamefont
  {Niu}}, \bibinfo {author} {\bibfnamefont {H.}~\bibnamefont {Wang}}, \bibinfo
  {author} {\bibfnamefont {N.}~\bibnamefont {Mao}}, \bibinfo {author}
  {\bibfnamefont {B.}~\bibnamefont {Huang}}, \bibinfo {author} {\bibfnamefont
  {Y.}~\bibnamefont {Mokrousov}},\ and\ \bibinfo {author} {\bibfnamefont
  {Y.}~\bibnamefont {Dai}},\ }\bibfield  {title} {\bibinfo {title}
  {{Antiferromagnetic Topological Insulator with Nonsymmorphic Protection in
  Two Dimensions}},\ }\href {https://doi.org/10.1103/PhysRevLett.124.066401}
  {\bibfield  {journal} {\bibinfo  {journal} {Phys. Rev. Lett.}\ }\textbf
  {\bibinfo {volume} {124}},\ \bibinfo {pages} {066401} (\bibinfo {year}
  {2020})}\BibitemShut {NoStop}%
\bibitem [{\citenamefont {Mao}\ \emph {et~al.}(2020)\citenamefont {Mao},
  \citenamefont {Wang}, \citenamefont {Hu}, \citenamefont {Niu}, \citenamefont
  {Huang},\ and\ \citenamefont {Dai}}]{Mao20}%
  \BibitemOpen
  \bibfield  {author} {\bibinfo {author} {\bibfnamefont {N.}~\bibnamefont
  {Mao}}, \bibinfo {author} {\bibfnamefont {H.}~\bibnamefont {Wang}}, \bibinfo
  {author} {\bibfnamefont {X.}~\bibnamefont {Hu}}, \bibinfo {author}
  {\bibfnamefont {C.}~\bibnamefont {Niu}}, \bibinfo {author} {\bibfnamefont
  {B.}~\bibnamefont {Huang}},\ and\ \bibinfo {author} {\bibfnamefont
  {Y.}~\bibnamefont {Dai}},\ }\bibfield  {title} {\bibinfo {title}
  {Antiferromagnetic topological insulator in stable exfoliated two-dimensional
  materials},\ }\href {https://doi.org/10.1103/PhysRevB.102.115412} {\bibfield
  {journal} {\bibinfo  {journal} {Phys. Rev. B}\ }\textbf {\bibinfo {volume}
  {102}},\ \bibinfo {pages} {115412} (\bibinfo {year} {2020})}\BibitemShut
  {NoStop}%
\bibitem [{\citenamefont {Wang}\ \emph
  {et~al.}(2021{\natexlab{b}})\citenamefont {Wang}, \citenamefont {Rogers},
  \citenamefont {Yao}, \citenamefont {Nichols}, \citenamefont {Atay},
  \citenamefont {Xu}, \citenamefont {Franklin}, \citenamefont {Sochnikov},
  \citenamefont {Ryan}, \citenamefont {Haskel},\ and\ \citenamefont
  {Tafti}}]{Wang21Advanced}%
  \BibitemOpen
  \bibfield  {author} {\bibinfo {author} {\bibfnamefont {Z.-C.}\ \bibnamefont
  {Wang}}, \bibinfo {author} {\bibfnamefont {J.~D.}\ \bibnamefont {Rogers}},
  \bibinfo {author} {\bibfnamefont {X.}~\bibnamefont {Yao}}, \bibinfo {author}
  {\bibfnamefont {R.}~\bibnamefont {Nichols}}, \bibinfo {author} {\bibfnamefont
  {K.}~\bibnamefont {Atay}}, \bibinfo {author} {\bibfnamefont {B.}~\bibnamefont
  {Xu}}, \bibinfo {author} {\bibfnamefont {J.}~\bibnamefont {Franklin}},
  \bibinfo {author} {\bibfnamefont {I.}~\bibnamefont {Sochnikov}}, \bibinfo
  {author} {\bibfnamefont {P.~J.}\ \bibnamefont {Ryan}}, \bibinfo {author}
  {\bibfnamefont {D.}~\bibnamefont {Haskel}},\ and\ \bibinfo {author}
  {\bibfnamefont {F.}~\bibnamefont {Tafti}},\ }\bibfield  {title} {\bibinfo
  {title} {Colossal magnetoresistance without mixed valence in a layered
  phosphide crystal},\ }\href
  {https://doi.org/https://doi.org/10.1002/adma.202005755} {\bibfield
  {journal} {\bibinfo  {journal} {Advanced Materials}\ }\textbf {\bibinfo
  {volume} {33}},\ \bibinfo {pages} {2005755} (\bibinfo {year}
  {2021}{\natexlab{b}})}\BibitemShut {NoStop}%
\bibitem [{\citenamefont {Artmann}\ \emph {et~al.}(1996)\citenamefont
  {Artmann}, \citenamefont {Mewis}, \citenamefont {Roepke},\ and\ \citenamefont
  {Michels}}]{Artmann96}%
  \BibitemOpen
  \bibfield  {author} {\bibinfo {author} {\bibfnamefont {A.}~\bibnamefont
  {Artmann}}, \bibinfo {author} {\bibfnamefont {A.}~\bibnamefont {Mewis}},
  \bibinfo {author} {\bibfnamefont {M.}~\bibnamefont {Roepke}},\ and\ \bibinfo
  {author} {\bibfnamefont {G.}~\bibnamefont {Michels}},\ }\bibfield  {title}
  {\bibinfo {title} {{AM2X2-Verbindungen mit CaAl2Si2-Struktur. XI. Struktur
  und Eigenschaften der Verbindungen ACd2X2 (A: Eu, Yb; X: P, As, Sb)}},\
  }\href@noop {} {\bibfield  {journal} {\bibinfo  {journal} {Zeitschrift
  f{\"u}r anorganische und allgemeine Chemie}\ }\textbf {\bibinfo {volume}
  {622}},\ \bibinfo {pages} {679} (\bibinfo {year} {1996})}\BibitemShut
  {NoStop}%
\bibitem [{\citenamefont {Krishna}\ \emph {et~al.}(2019)\citenamefont
  {Krishna}, \citenamefont {Sharma},\ and\ \citenamefont
  {Maitra}}]{krishna2019anisotropic}%
  \BibitemOpen
  \bibfield  {author} {\bibinfo {author} {\bibfnamefont {J.}~\bibnamefont
  {Krishna}}, \bibinfo {author} {\bibfnamefont {M.}~\bibnamefont {Sharma}},\
  and\ \bibinfo {author} {\bibfnamefont {T.}~\bibnamefont {Maitra}},\
  }\bibfield  {title} {\bibinfo {title} {Anisotropic thermoelectric properties
  of {EuCd $ \_ $\{$2$\}$ $ As $ \_ $\{$2$\}$ $}: An ab-initio study},\
  }\href@noop {} {\bibfield  {journal} {\bibinfo  {journal} {arXiv preprint
  arXiv:1907.13446}\ } (\bibinfo {year} {2019})}\BibitemShut {NoStop}%
\bibitem [{\citenamefont {Du}\ \emph {et~al.}(2022{\natexlab{a}})\citenamefont
  {Du}, \citenamefont {Chen}, \citenamefont {Wu}, \citenamefont {Shi},
  \citenamefont {Meng}, \citenamefont {Zhang}, \citenamefont {Gong},
  \citenamefont {Chu},\ and\ \citenamefont {Yuan}}]{du2022comparative}%
  \BibitemOpen
  \bibfield  {author} {\bibinfo {author} {\bibfnamefont {Y.}~\bibnamefont
  {Du}}, \bibinfo {author} {\bibfnamefont {J.}~\bibnamefont {Chen}}, \bibinfo
  {author} {\bibfnamefont {W.}~\bibnamefont {Wu}}, \bibinfo {author}
  {\bibfnamefont {Z.}~\bibnamefont {Shi}}, \bibinfo {author} {\bibfnamefont
  {X.}~\bibnamefont {Meng}}, \bibinfo {author} {\bibfnamefont {C.}~\bibnamefont
  {Zhang}}, \bibinfo {author} {\bibfnamefont {S.}~\bibnamefont {Gong}},
  \bibinfo {author} {\bibfnamefont {J.}~\bibnamefont {Chu}},\ and\ \bibinfo
  {author} {\bibfnamefont {X.}~\bibnamefont {Yuan}},\ }\bibfield  {title}
  {\bibinfo {title} {Comparative raman spectroscopy of magnetic topological
  material {EuCd$_2$X$_2$ (X= P, As)}},\ }\href@noop {} {\bibfield  {journal}
  {\bibinfo  {journal} {J. Phys.: Condensed Matter}\ }\textbf {\bibinfo
  {volume} {34}},\ \bibinfo {pages} {224001} (\bibinfo {year}
  {2022}{\natexlab{a}})}\BibitemShut {NoStop}%
\bibitem [{\citenamefont {{\v{S}}mejkal}\ \emph {et~al.}(2022)\citenamefont
  {{\v{S}}mejkal}, \citenamefont {Sinova},\ and\ \citenamefont
  {Jungwirth}}]{Smejkal:2022_PRX}%
  \BibitemOpen
  \bibfield  {author} {\bibinfo {author} {\bibfnamefont {L.}~\bibnamefont
  {{\v{S}}mejkal}}, \bibinfo {author} {\bibfnamefont {J.}~\bibnamefont
  {Sinova}},\ and\ \bibinfo {author} {\bibfnamefont {T.}~\bibnamefont
  {Jungwirth}},\ }\bibfield  {title} {\bibinfo {title} {{Emerging Research
  Landscape of Altermagnetism}},\ }\href
  {https://doi.org/10.1103/PhysRevX.12.040501} {\bibfield  {journal} {\bibinfo
  {journal} {Phys. Rev. X}\ }\textbf {\bibinfo {volume} {12}},\ \bibinfo
  {pages} {040501} (\bibinfo {year} {2022})}\BibitemShut {NoStop}%
\bibitem [{\citenamefont {\ifmmode~\check{S}\else \v{S}\fi{}mejkal}\ \emph
  {et~al.}(2022)\citenamefont {\ifmmode~\check{S}\else \v{S}\fi{}mejkal},
  \citenamefont {Sinova},\ and\ \citenamefont {Jungwirth}}]{Smejkal22beyond}%
  \BibitemOpen
  \bibfield  {author} {\bibinfo {author} {\bibfnamefont {L.}~\bibnamefont
  {\ifmmode~\check{S}\else \v{S}\fi{}mejkal}}, \bibinfo {author} {\bibfnamefont
  {J.}~\bibnamefont {Sinova}},\ and\ \bibinfo {author} {\bibfnamefont
  {T.}~\bibnamefont {Jungwirth}},\ }\bibfield  {title} {\bibinfo {title}
  {Beyond conventional ferromagnetism and antiferromagnetism: A phase with
  nonrelativistic spin and crystal rotation symmetry},\ }\href
  {https://doi.org/10.1103/PhysRevX.12.031042} {\bibfield  {journal} {\bibinfo
  {journal} {Phys. Rev. X}\ }\textbf {\bibinfo {volume} {12}},\ \bibinfo
  {pages} {031042} (\bibinfo {year} {2022})}\BibitemShut {NoStop}%
\bibitem [{\citenamefont {Gong}\ \emph {et~al.}(2022)\citenamefont {Gong},
  \citenamefont {Sar}, \citenamefont {Friedman}, \citenamefont {Kaczorowski},
  \citenamefont {Abdel~Razek}, \citenamefont {Lee},\ and\ \citenamefont
  {Aynajian}}]{Gong22}%
  \BibitemOpen
  \bibfield  {author} {\bibinfo {author} {\bibfnamefont {M.}~\bibnamefont
  {Gong}}, \bibinfo {author} {\bibfnamefont {D.}~\bibnamefont {Sar}}, \bibinfo
  {author} {\bibfnamefont {J.}~\bibnamefont {Friedman}}, \bibinfo {author}
  {\bibfnamefont {D.}~\bibnamefont {Kaczorowski}}, \bibinfo {author}
  {\bibfnamefont {S.}~\bibnamefont {Abdel~Razek}}, \bibinfo {author}
  {\bibfnamefont {W.-C.}\ \bibnamefont {Lee}},\ and\ \bibinfo {author}
  {\bibfnamefont {P.}~\bibnamefont {Aynajian}},\ }\bibfield  {title} {\bibinfo
  {title} {Surface state evolution induced by magnetic order in axion insulator
  candidate {${\mathrm{EuIn}}_{2}{\mathrm{As}}_{2}$}},\ }\href
  {https://doi.org/10.1103/PhysRevB.106.125156} {\bibfield  {journal} {\bibinfo
   {journal} {Phys. Rev. B}\ }\textbf {\bibinfo {volume} {106}},\ \bibinfo
  {pages} {125156} (\bibinfo {year} {2022})}\BibitemShut {NoStop}%
\bibitem [{\citenamefont {Sato}\ \emph {et~al.}(2020)\citenamefont {Sato},
  \citenamefont {Wang}, \citenamefont {Takane}, \citenamefont {Souma},
  \citenamefont {Cui}, \citenamefont {Li}, \citenamefont {Nakayama},
  \citenamefont {Kawakami}, \citenamefont {Kubota}, \citenamefont {Cacho},
  \citenamefont {Kim}, \citenamefont {Arab}, \citenamefont {Strocov},
  \citenamefont {Yao},\ and\ \citenamefont {Takahashi}}]{Sato20}%
  \BibitemOpen
  \bibfield  {author} {\bibinfo {author} {\bibfnamefont {T.}~\bibnamefont
  {Sato}}, \bibinfo {author} {\bibfnamefont {Z.}~\bibnamefont {Wang}}, \bibinfo
  {author} {\bibfnamefont {D.}~\bibnamefont {Takane}}, \bibinfo {author}
  {\bibfnamefont {S.}~\bibnamefont {Souma}}, \bibinfo {author} {\bibfnamefont
  {C.}~\bibnamefont {Cui}}, \bibinfo {author} {\bibfnamefont {Y.}~\bibnamefont
  {Li}}, \bibinfo {author} {\bibfnamefont {K.}~\bibnamefont {Nakayama}},
  \bibinfo {author} {\bibfnamefont {T.}~\bibnamefont {Kawakami}}, \bibinfo
  {author} {\bibfnamefont {Y.}~\bibnamefont {Kubota}}, \bibinfo {author}
  {\bibfnamefont {C.}~\bibnamefont {Cacho}}, \bibinfo {author} {\bibfnamefont
  {T.~K.}\ \bibnamefont {Kim}}, \bibinfo {author} {\bibfnamefont
  {A.}~\bibnamefont {Arab}}, \bibinfo {author} {\bibfnamefont {V.~N.}\
  \bibnamefont {Strocov}}, \bibinfo {author} {\bibfnamefont {Y.}~\bibnamefont
  {Yao}},\ and\ \bibinfo {author} {\bibfnamefont {T.}~\bibnamefont
  {Takahashi}},\ }\bibfield  {title} {\bibinfo {title} {Signature of band
  inversion in the antiferromagnetic phase of axion insulator candidate
  ${\mathrm{euin}}_{2}{\mathrm{as}}_{2}$},\ }\href
  {https://doi.org/10.1103/PhysRevResearch.2.033342} {\bibfield  {journal}
  {\bibinfo  {journal} {Phys. Rev. Research}\ }\textbf {\bibinfo {volume}
  {2}},\ \bibinfo {pages} {033342} (\bibinfo {year} {2020})}\BibitemShut
  {NoStop}%
\bibitem [{\citenamefont {Regmi}\ \emph {et~al.}(2020)\citenamefont {Regmi},
  \citenamefont {Hosen}, \citenamefont {Ghosh}, \citenamefont {Singh},
  \citenamefont {Dhakal}, \citenamefont {Sims}, \citenamefont {Wang},
  \citenamefont {Kabir}, \citenamefont {Dimitri}, \citenamefont {Liu},
  \citenamefont {Agarwal}, \citenamefont {Lin}, \citenamefont {Kaczorowski},
  \citenamefont {Bansil},\ and\ \citenamefont {Neupane}}]{Regmi20}%
  \BibitemOpen
  \bibfield  {author} {\bibinfo {author} {\bibfnamefont {S.}~\bibnamefont
  {Regmi}}, \bibinfo {author} {\bibfnamefont {M.~M.}\ \bibnamefont {Hosen}},
  \bibinfo {author} {\bibfnamefont {B.}~\bibnamefont {Ghosh}}, \bibinfo
  {author} {\bibfnamefont {B.}~\bibnamefont {Singh}}, \bibinfo {author}
  {\bibfnamefont {G.}~\bibnamefont {Dhakal}}, \bibinfo {author} {\bibfnamefont
  {C.}~\bibnamefont {Sims}}, \bibinfo {author} {\bibfnamefont {B.}~\bibnamefont
  {Wang}}, \bibinfo {author} {\bibfnamefont {F.}~\bibnamefont {Kabir}},
  \bibinfo {author} {\bibfnamefont {K.}~\bibnamefont {Dimitri}}, \bibinfo
  {author} {\bibfnamefont {Y.}~\bibnamefont {Liu}}, \bibinfo {author}
  {\bibfnamefont {A.}~\bibnamefont {Agarwal}}, \bibinfo {author} {\bibfnamefont
  {H.}~\bibnamefont {Lin}}, \bibinfo {author} {\bibfnamefont {D.}~\bibnamefont
  {Kaczorowski}}, \bibinfo {author} {\bibfnamefont {A.}~\bibnamefont
  {Bansil}},\ and\ \bibinfo {author} {\bibfnamefont {M.}~\bibnamefont
  {Neupane}},\ }\bibfield  {title} {\bibinfo {title} {Temperature-dependent
  electronic structure in a higher-order topological insulator candidate
  {$\mathrm{Eu}{\mathrm{In}}_{2}{\mathrm{As}}_{2}$}},\ }\href
  {https://doi.org/10.1103/PhysRevB.102.165153} {\bibfield  {journal} {\bibinfo
   {journal} {Phys. Rev. B}\ }\textbf {\bibinfo {volume} {102}},\ \bibinfo
  {pages} {165153} (\bibinfo {year} {2020})}\BibitemShut {NoStop}%
\bibitem [{\citenamefont {Goforth}\ \emph {et~al.}(2008)\citenamefont
  {Goforth}, \citenamefont {Klavins}, \citenamefont {Fettinger},\ and\
  \citenamefont {Kauzlarich}}]{Goforth08}%
  \BibitemOpen
  \bibfield  {author} {\bibinfo {author} {\bibfnamefont {A.~M.}\ \bibnamefont
  {Goforth}}, \bibinfo {author} {\bibfnamefont {P.}~\bibnamefont {Klavins}},
  \bibinfo {author} {\bibfnamefont {J.~C.}\ \bibnamefont {Fettinger}},\ and\
  \bibinfo {author} {\bibfnamefont {S.~M.}\ \bibnamefont {Kauzlarich}},\
  }\bibfield  {title} {\bibinfo {title} {Magnetic properties and negative
  colossal magnetoresistance of the rare earth {Zintl} phase
  {$\mathrm{Eu}{\mathrm{In}}_{2}{\mathrm{As}}_{2}$}},\ }\bibfield  {journal}
  {\bibinfo  {journal} {Inorganic Chemistry}\ }\textbf {\bibinfo {volume}
  {47}},\ \href {https://doi.org/10.1021/ic801290u} {10.1021/ic801290u}
  (\bibinfo {year} {2008})\BibitemShut {NoStop}%
\bibitem [{\citenamefont {Rosa}\ \emph {et~al.}(2012)\citenamefont {Rosa},
  \citenamefont {Adriano}, \citenamefont {Garitezi}, \citenamefont {Ribeiro},
  \citenamefont {Fisk},\ and\ \citenamefont {Pagliuso}}]{Rosa12}%
  \BibitemOpen
  \bibfield  {author} {\bibinfo {author} {\bibfnamefont {P.~F.~S.}\
  \bibnamefont {Rosa}}, \bibinfo {author} {\bibfnamefont {C.}~\bibnamefont
  {Adriano}}, \bibinfo {author} {\bibfnamefont {T.~M.}\ \bibnamefont
  {Garitezi}}, \bibinfo {author} {\bibfnamefont {R.~A.}\ \bibnamefont
  {Ribeiro}}, \bibinfo {author} {\bibfnamefont {Z.}~\bibnamefont {Fisk}},\ and\
  \bibinfo {author} {\bibfnamefont {P.~G.}\ \bibnamefont {Pagliuso}},\
  }\bibfield  {title} {\bibinfo {title} {Electron spin resonance of the
  intermetallic antiferromagnet {EuIn${}_{2}$As${}_{2}$}},\ }\href
  {https://doi.org/10.1103/PhysRevB.86.094408} {\bibfield  {journal} {\bibinfo
  {journal} {Phys. Rev. B}\ }\textbf {\bibinfo {volume} {86}},\ \bibinfo
  {pages} {094408} (\bibinfo {year} {2012})}\BibitemShut {NoStop}%
\bibitem [{\citenamefont {Wang}\ \emph
  {et~al.}(2021{\natexlab{c}})\citenamefont {Wang}, \citenamefont {Botti},\
  and\ \citenamefont {Marques}}]{Wang21npj}%
  \BibitemOpen
  \bibfield  {author} {\bibinfo {author} {\bibfnamefont {H.}~\bibnamefont
  {Wang}}, \bibinfo {author} {\bibfnamefont {S.}~\bibnamefont {Botti}},\ and\
  \bibinfo {author} {\bibfnamefont {M.}~\bibnamefont {Marques}},\ }\bibfield
  {title} {\bibinfo {title} {Predicting stable crystalline compounds using
  chemical similarity},\ }\bibfield  {journal} {\bibinfo  {journal} {npj
  Comput. Mater.}\ }\textbf {\bibinfo {volume} {7}},\ \href
  {https://doi.org/10.1038/s41524-020-00481-6} {10.1038/s41524-020-00481-6}
  (\bibinfo {year} {2021}{\natexlab{c}})\BibitemShut {NoStop}%
\bibitem [{\citenamefont {Arguilla}\ \emph {et~al.}(2017)\citenamefont
  {Arguilla}, \citenamefont {Cultrara}, \citenamefont {Baum}, \citenamefont
  {Jiang}, \citenamefont {Ross},\ and\ \citenamefont
  {Goldberger}}]{Arguilla17}%
  \BibitemOpen
  \bibfield  {author} {\bibinfo {author} {\bibfnamefont {M.~Q.}\ \bibnamefont
  {Arguilla}}, \bibinfo {author} {\bibfnamefont {N.~D.}\ \bibnamefont
  {Cultrara}}, \bibinfo {author} {\bibfnamefont {Z.~J.}\ \bibnamefont {Baum}},
  \bibinfo {author} {\bibfnamefont {S.}~\bibnamefont {Jiang}}, \bibinfo
  {author} {\bibfnamefont {R.~D.}\ \bibnamefont {Ross}},\ and\ \bibinfo
  {author} {\bibfnamefont {J.~E.}\ \bibnamefont {Goldberger}},\ }\bibfield
  {title} {\bibinfo {title} {{EuSn$_2$As$_2$}: an exfoliatable magnetic layered
  {Zintl-Klemm} phase},\ }\href {https://doi.org/10.1039/C6QI00476H} {\bibfield
   {journal} {\bibinfo  {journal} {Inorg. Chem. Front.}\ }\textbf {\bibinfo
  {volume} {4}},\ \bibinfo {pages} {378} (\bibinfo {year} {2017})}\BibitemShut
  {NoStop}%
\bibitem [{\citenamefont {Kresse}\ and\ \citenamefont
  {Hafner}(1993)}]{Kresse93}%
  \BibitemOpen
  \bibfield  {author} {\bibinfo {author} {\bibfnamefont {G.}~\bibnamefont
  {Kresse}}\ and\ \bibinfo {author} {\bibfnamefont {J.}~\bibnamefont
  {Hafner}},\ }\bibfield  {title} {\bibinfo {title} {Ab initio molecular
  dynamics for liquid metals},\ }\href
  {https://doi.org/10.1103/PhysRevB.47.558} {\bibfield  {journal} {\bibinfo
  {journal} {Phys. Rev. B}\ }\textbf {\bibinfo {volume} {47}},\ \bibinfo
  {pages} {558} (\bibinfo {year} {1993})}\BibitemShut {NoStop}%
\bibitem [{\citenamefont {Kresse}\ and\ \citenamefont
  {Furthmuller}(1996)}]{Kresse96}%
  \BibitemOpen
  \bibfield  {author} {\bibinfo {author} {\bibfnamefont {G.}~\bibnamefont
  {Kresse}}\ and\ \bibinfo {author} {\bibfnamefont {J.}~\bibnamefont
  {Furthmuller}},\ }\bibfield  {title} {\bibinfo {title} {Efficiency of
  ab-initio total energy calculations for metals and semiconductors using a
  plane-wave basis set},\ }\href
  {https://doi.org/https://doi.org/10.1016/0927-0256(96)00008-0} {\bibfield
  {journal} {\bibinfo  {journal} {Computational Materials Science}\ }\textbf
  {\bibinfo {volume} {6}},\ \bibinfo {pages} {15} (\bibinfo {year}
  {1996})}\BibitemShut {NoStop}%
\bibitem [{\citenamefont {Kresse}\ and\ \citenamefont
  {Furthm\"uller}(1996)}]{Kresse96b}%
  \BibitemOpen
  \bibfield  {author} {\bibinfo {author} {\bibfnamefont {G.}~\bibnamefont
  {Kresse}}\ and\ \bibinfo {author} {\bibfnamefont {J.}~\bibnamefont
  {Furthm\"uller}},\ }\bibfield  {title} {\bibinfo {title} {Efficient iterative
  schemes for ab initio total-energy calculations using a plane-wave basis
  set},\ }\href {https://doi.org/10.1103/PhysRevB.54.11169} {\bibfield
  {journal} {\bibinfo  {journal} {Phys. Rev. B}\ }\textbf {\bibinfo {volume}
  {54}},\ \bibinfo {pages} {11169} (\bibinfo {year} {1996})}\BibitemShut
  {NoStop}%
\bibitem [{\citenamefont {Kresse}\ and\ \citenamefont
  {Joubert}(1999)}]{Kresse99}%
  \BibitemOpen
  \bibfield  {author} {\bibinfo {author} {\bibfnamefont {G.}~\bibnamefont
  {Kresse}}\ and\ \bibinfo {author} {\bibfnamefont {D.}~\bibnamefont
  {Joubert}},\ }\bibfield  {title} {\bibinfo {title} {From ultrasoft
  pseudopotentials to the projector augmented-wave method},\ }\href
  {https://doi.org/10.1103/PhysRevB.59.1758} {\bibfield  {journal} {\bibinfo
  {journal} {Phys. Rev. B}\ }\textbf {\bibinfo {volume} {59}},\ \bibinfo
  {pages} {1758} (\bibinfo {year} {1999})}\BibitemShut {NoStop}%
\bibitem [{\citenamefont {Wadge}\ \emph {et~al.}(2022)\citenamefont {Wadge},
  \citenamefont {Grabecki}, \citenamefont {Autieri}, \citenamefont {Kowalski},
  \citenamefont {Iwanowski}, \citenamefont {Cuono}, \citenamefont {Islam},
  \citenamefont {Canali}, \citenamefont {Dybko}, \citenamefont {Hruban},
  \citenamefont {{\L}usakowski}, \citenamefont {Wojciechowski}, \citenamefont
  {Diduszko}, \citenamefont {Lynnyk}, \citenamefont {Olszowska}, \citenamefont
  {Rosmus}, \citenamefont {Ko{\l}odziej},\ and\ \citenamefont
  {Wi{\'{s}}niewski}}]{wadge2021electronic}%
  \BibitemOpen
  \bibfield  {author} {\bibinfo {author} {\bibfnamefont {A.~S.}\ \bibnamefont
  {Wadge}}, \bibinfo {author} {\bibfnamefont {G.}~\bibnamefont {Grabecki}},
  \bibinfo {author} {\bibfnamefont {C.}~\bibnamefont {Autieri}}, \bibinfo
  {author} {\bibfnamefont {B.~J.}\ \bibnamefont {Kowalski}}, \bibinfo {author}
  {\bibfnamefont {P.}~\bibnamefont {Iwanowski}}, \bibinfo {author}
  {\bibfnamefont {G.}~\bibnamefont {Cuono}}, \bibinfo {author} {\bibfnamefont
  {M.~F.}\ \bibnamefont {Islam}}, \bibinfo {author} {\bibfnamefont {C.~M.}\
  \bibnamefont {Canali}}, \bibinfo {author} {\bibfnamefont {K.}~\bibnamefont
  {Dybko}}, \bibinfo {author} {\bibfnamefont {A.}~\bibnamefont {Hruban}},
  \bibinfo {author} {\bibfnamefont {A.}~\bibnamefont {{\L}usakowski}}, \bibinfo
  {author} {\bibfnamefont {T.}~\bibnamefont {Wojciechowski}}, \bibinfo {author}
  {\bibfnamefont {R.}~\bibnamefont {Diduszko}}, \bibinfo {author}
  {\bibfnamefont {A.}~\bibnamefont {Lynnyk}}, \bibinfo {author} {\bibfnamefont
  {N.}~\bibnamefont {Olszowska}}, \bibinfo {author} {\bibfnamefont
  {M.}~\bibnamefont {Rosmus}}, \bibinfo {author} {\bibfnamefont
  {J.}~\bibnamefont {Ko{\l}odziej}},\ and\ \bibinfo {author} {\bibfnamefont
  {A.}~\bibnamefont {Wi{\'{s}}niewski}},\ }\bibfield  {title} {\bibinfo {title}
  {Electronic properties of {TaAs}2 topological semimetal investigated by
  transport and {ARPES}},\ }\href {https://doi.org/10.1088/1361-648x/ac43fe}
  {\bibfield  {journal} {\bibinfo  {journal} {Journal of Physics: Condensed
  Matter}\ }\textbf {\bibinfo {volume} {34}},\ \bibinfo {pages} {125601}
  (\bibinfo {year} {2022})}\BibitemShut {NoStop}%
\bibitem [{\citenamefont {Togo}\ and\ \citenamefont {Tanaka}(2015)}]{phonopy}%
  \BibitemOpen
  \bibfield  {author} {\bibinfo {author} {\bibfnamefont {A.}~\bibnamefont
  {Togo}}\ and\ \bibinfo {author} {\bibfnamefont {I.}~\bibnamefont {Tanaka}},\
  }\bibfield  {title} {\bibinfo {title} {First principles phonon calculations
  in materials science},\ }\href@noop {} {\bibfield  {journal} {\bibinfo
  {journal} {Scr. Mater.}\ }\textbf {\bibinfo {volume} {108}},\ \bibinfo
  {pages} {1} (\bibinfo {year} {2015})}\BibitemShut {NoStop}%
\bibitem [{\citenamefont {Togo}(2023)}]{phonopy-phono3py-JPSJ}%
  \BibitemOpen
  \bibfield  {author} {\bibinfo {author} {\bibfnamefont {A.}~\bibnamefont
  {Togo}},\ }\bibfield  {title} {\bibinfo {title} {First-principles phonon
  calculations with phonopy and phono3py},\ }\href
  {https://doi.org/10.7566/JPSJ.92.012001} {\bibfield  {journal} {\bibinfo
  {journal} {J. Phys. Soc. Jpn.}\ }\textbf {\bibinfo {volume} {92}},\ \bibinfo
  {pages} {012001} (\bibinfo {year} {2023})}\BibitemShut {NoStop}%
\bibitem [{\citenamefont {Perdew}\ \emph {et~al.}(1996)\citenamefont {Perdew},
  \citenamefont {Burke},\ and\ \citenamefont {Ernzerhof}}]{Perdew96}%
  \BibitemOpen
  \bibfield  {author} {\bibinfo {author} {\bibfnamefont {J.~P.}\ \bibnamefont
  {Perdew}}, \bibinfo {author} {\bibfnamefont {K.}~\bibnamefont {Burke}},\ and\
  \bibinfo {author} {\bibfnamefont {M.}~\bibnamefont {Ernzerhof}},\ }\bibfield
  {title} {\bibinfo {title} {{Generalized Gradient Approximation Made
  Simple}},\ }\href {https://doi.org/10.1103/PhysRevLett.77.3865} {\bibfield
  {journal} {\bibinfo  {journal} {Phys. Rev. Lett.}\ }\textbf {\bibinfo
  {volume} {77}},\ \bibinfo {pages} {3865} (\bibinfo {year}
  {1996})}\BibitemShut {NoStop}%
\bibitem [{\citenamefont {Soh}\ \emph {et~al.}(2019)\citenamefont {Soh},
  \citenamefont {de~Juan}, \citenamefont {Vergniory}, \citenamefont
  {Schr\"oter}, \citenamefont {Rahn}, \citenamefont {Yan}, \citenamefont
  {Jiang}, \citenamefont {Bristow}, \citenamefont {Reiss}, \citenamefont
  {Blandy}, \citenamefont {Guo}, \citenamefont {Shi}, \citenamefont {Kim},
  \citenamefont {McCollam}, \citenamefont {Simon}, \citenamefont {Chen},
  \citenamefont {Coldea},\ and\ \citenamefont {Boothroyd}}]{Soh19}%
  \BibitemOpen
  \bibfield  {author} {\bibinfo {author} {\bibfnamefont {J.-R.}\ \bibnamefont
  {Soh}}, \bibinfo {author} {\bibfnamefont {F.}~\bibnamefont {de~Juan}},
  \bibinfo {author} {\bibfnamefont {M.~G.}\ \bibnamefont {Vergniory}}, \bibinfo
  {author} {\bibfnamefont {N.~B.~M.}\ \bibnamefont {Schr\"oter}}, \bibinfo
  {author} {\bibfnamefont {M.~C.}\ \bibnamefont {Rahn}}, \bibinfo {author}
  {\bibfnamefont {D.~Y.}\ \bibnamefont {Yan}}, \bibinfo {author} {\bibfnamefont
  {J.}~\bibnamefont {Jiang}}, \bibinfo {author} {\bibfnamefont
  {M.}~\bibnamefont {Bristow}}, \bibinfo {author} {\bibfnamefont
  {P.}~\bibnamefont {Reiss}}, \bibinfo {author} {\bibfnamefont {J.~N.}\
  \bibnamefont {Blandy}}, \bibinfo {author} {\bibfnamefont {Y.~F.}\
  \bibnamefont {Guo}}, \bibinfo {author} {\bibfnamefont {Y.~G.}\ \bibnamefont
  {Shi}}, \bibinfo {author} {\bibfnamefont {T.~K.}\ \bibnamefont {Kim}},
  \bibinfo {author} {\bibfnamefont {A.}~\bibnamefont {McCollam}}, \bibinfo
  {author} {\bibfnamefont {S.~H.}\ \bibnamefont {Simon}}, \bibinfo {author}
  {\bibfnamefont {Y.}~\bibnamefont {Chen}}, \bibinfo {author} {\bibfnamefont
  {A.~I.}\ \bibnamefont {Coldea}},\ and\ \bibinfo {author} {\bibfnamefont
  {A.~T.}\ \bibnamefont {Boothroyd}},\ }\bibfield  {title} {\bibinfo {title}
  {Ideal {Weyl} semimetal induced by magnetic exchange},\ }\href
  {https://doi.org/10.1103/PhysRevB.100.201102} {\bibfield  {journal} {\bibinfo
   {journal} {Phys. Rev. B}\ }\textbf {\bibinfo {volume} {100}},\ \bibinfo
  {pages} {201102} (\bibinfo {year} {2019})}\BibitemShut {NoStop}%
\bibitem [{\citenamefont {Wang}\ \emph
  {et~al.}(2019{\natexlab{b}})\citenamefont {Wang}, \citenamefont {Jo},
  \citenamefont {Kuthanazhi}, \citenamefont {Wu}, \citenamefont {McQueeney},
  \citenamefont {Kaminski},\ and\ \citenamefont {Canfield}}]{Wang19}%
  \BibitemOpen
  \bibfield  {author} {\bibinfo {author} {\bibfnamefont {L.-L.}\ \bibnamefont
  {Wang}}, \bibinfo {author} {\bibfnamefont {N.~H.}\ \bibnamefont {Jo}},
  \bibinfo {author} {\bibfnamefont {B.}~\bibnamefont {Kuthanazhi}}, \bibinfo
  {author} {\bibfnamefont {Y.}~\bibnamefont {Wu}}, \bibinfo {author}
  {\bibfnamefont {R.~J.}\ \bibnamefont {McQueeney}}, \bibinfo {author}
  {\bibfnamefont {A.}~\bibnamefont {Kaminski}},\ and\ \bibinfo {author}
  {\bibfnamefont {P.~C.}\ \bibnamefont {Canfield}},\ }\bibfield  {title}
  {\bibinfo {title} {Single pair of {Weyl} fermions in the half-metallic
  semimetal {EuCd$_2$As$_2$}},\ }\href
  {https://doi.org/10.1103/PhysRevB.99.245147} {\bibfield  {journal} {\bibinfo
  {journal} {Phys. Rev. B}\ }\textbf {\bibinfo {volume} {99}},\ \bibinfo
  {pages} {245147} (\bibinfo {year} {2019}{\natexlab{b}})}\BibitemShut
  {NoStop}%
\bibitem [{\citenamefont {Gati}\ \emph {et~al.}(2021)\citenamefont {Gati},
  \citenamefont {Bud'ko}, \citenamefont {Wang}, \citenamefont {Valadkhani},
  \citenamefont {Gupta}, \citenamefont {Kuthanazhi}, \citenamefont {Xiang},
  \citenamefont {Wilde}, \citenamefont {Sapkota}, \citenamefont {Guguchia},
  \citenamefont {Khasanov}, \citenamefont {Valent\'{\i}},\ and\ \citenamefont
  {Canfield}}]{Gati21}%
  \BibitemOpen
  \bibfield  {author} {\bibinfo {author} {\bibfnamefont {E.}~\bibnamefont
  {Gati}}, \bibinfo {author} {\bibfnamefont {S.~L.}\ \bibnamefont {Bud'ko}},
  \bibinfo {author} {\bibfnamefont {L.-L.}\ \bibnamefont {Wang}}, \bibinfo
  {author} {\bibfnamefont {A.}~\bibnamefont {Valadkhani}}, \bibinfo {author}
  {\bibfnamefont {R.}~\bibnamefont {Gupta}}, \bibinfo {author} {\bibfnamefont
  {B.}~\bibnamefont {Kuthanazhi}}, \bibinfo {author} {\bibfnamefont
  {L.}~\bibnamefont {Xiang}}, \bibinfo {author} {\bibfnamefont {J.~M.}\
  \bibnamefont {Wilde}}, \bibinfo {author} {\bibfnamefont {A.}~\bibnamefont
  {Sapkota}}, \bibinfo {author} {\bibfnamefont {Z.}~\bibnamefont {Guguchia}},
  \bibinfo {author} {\bibfnamefont {R.}~\bibnamefont {Khasanov}}, \bibinfo
  {author} {\bibfnamefont {R.}~\bibnamefont {Valent\'{\i}}},\ and\ \bibinfo
  {author} {\bibfnamefont {P.~C.}\ \bibnamefont {Canfield}},\ }\bibfield
  {title} {\bibinfo {title} {Pressure-induced ferromagnetism in the topological
  semimetal {EuCd$_2$As$_2$}},\ }\href
  {https://doi.org/10.1103/PhysRevB.104.155124} {\bibfield  {journal} {\bibinfo
   {journal} {Phys. Rev. B}\ }\textbf {\bibinfo {volume} {104}},\ \bibinfo
  {pages} {155124} (\bibinfo {year} {2021})}\BibitemShut {NoStop}%
\bibitem [{\citenamefont {Du}\ \emph {et~al.}(2022{\natexlab{b}})\citenamefont
  {Du}, \citenamefont {Yang}, \citenamefont {Nie}, \citenamefont {Wu},
  \citenamefont {Li}, \citenamefont {Luo}, \citenamefont {Chen}, \citenamefont
  {Su}, \citenamefont {Smidman}, \citenamefont {Shi}, \citenamefont {Cao},
  \citenamefont {Steglich}, \citenamefont {Song},\ and\ \citenamefont
  {Yuan}}]{Du22}%
  \BibitemOpen
  \bibfield  {author} {\bibinfo {author} {\bibfnamefont {F.}~\bibnamefont
  {Du}}, \bibinfo {author} {\bibfnamefont {L.}~\bibnamefont {Yang}}, \bibinfo
  {author} {\bibfnamefont {Z.}~\bibnamefont {Nie}}, \bibinfo {author}
  {\bibfnamefont {N.}~\bibnamefont {Wu}}, \bibinfo {author} {\bibfnamefont
  {Y.}~\bibnamefont {Li}}, \bibinfo {author} {\bibfnamefont {S.}~\bibnamefont
  {Luo}}, \bibinfo {author} {\bibfnamefont {Y.}~\bibnamefont {Chen}}, \bibinfo
  {author} {\bibfnamefont {D.}~\bibnamefont {Su}}, \bibinfo {author}
  {\bibfnamefont {M.}~\bibnamefont {Smidman}}, \bibinfo {author} {\bibfnamefont
  {Y.}~\bibnamefont {Shi}}, \bibinfo {author} {\bibfnamefont {C.}~\bibnamefont
  {Cao}}, \bibinfo {author} {\bibfnamefont {F.}~\bibnamefont {Steglich}},
  \bibinfo {author} {\bibfnamefont {Y.}~\bibnamefont {Song}},\ and\ \bibinfo
  {author} {\bibfnamefont {H.}~\bibnamefont {Yuan}},\ }\bibfield  {title}
  {\bibinfo {title} {Consecutive topological phase transitions and colossal
  magnetoresistance in a magnetic topological semimetal},\ }\bibfield
  {journal} {\bibinfo  {journal} {npj Quantum Mater.}\ }\textbf {\bibinfo
  {volume} {7}},\ \href {https://doi.org/10.1038/s41535-022-00468-0}
  {10.1038/s41535-022-00468-0} (\bibinfo {year}
  {2022}{\natexlab{b}})\BibitemShut {NoStop}%
\bibitem [{\citenamefont {Yu}\ \emph {et~al.}(2020)\citenamefont {Yu},
  \citenamefont {Yang}, \citenamefont {Wu},\ and\ \citenamefont
  {Marom}}]{Yu2020}%
  \BibitemOpen
  \bibfield  {author} {\bibinfo {author} {\bibfnamefont {M.}~\bibnamefont
  {Yu}}, \bibinfo {author} {\bibfnamefont {S.}~\bibnamefont {Yang}}, \bibinfo
  {author} {\bibfnamefont {C.}~\bibnamefont {Wu}},\ and\ \bibinfo {author}
  {\bibfnamefont {N.}~\bibnamefont {Marom}},\ }\bibfield  {title} {\bibinfo
  {title} {Machine learning the {Hubbard} {U} parameter in {DFT}+{U} using
  {Bayesian} optimization},\ }\href
  {https://doi.org/10.1038/s41524-020-00446-9} {\bibfield  {journal} {\bibinfo
  {journal} {npj Computational Materials}\ }\textbf {\bibinfo {volume} {6}},\
  \bibinfo {pages} {180} (\bibinfo {year} {2020})}\BibitemShut {NoStop}%
\bibitem [{\citenamefont {Larson}\ and\ \citenamefont
  {Lambrecht}(2006)}]{Larson_2006}%
  \BibitemOpen
  \bibfield  {author} {\bibinfo {author} {\bibfnamefont {P.}~\bibnamefont
  {Larson}}\ and\ \bibinfo {author} {\bibfnamefont {W.~R.~L.}\ \bibnamefont
  {Lambrecht}},\ }\bibfield  {title} {\bibinfo {title} {Electronic structure
  and magnetism of europium chalcogenides in comparison with gadolinium
  nitride},\ }\href {https://doi.org/10.1088/0953-8984/18/49/024} {\bibfield
  {journal} {\bibinfo  {journal} {Journal of Physics: Condensed Matter}\
  }\textbf {\bibinfo {volume} {18}},\ \bibinfo {pages} {11333} (\bibinfo {year}
  {2006})}\BibitemShut {NoStop}%
\bibitem [{\citenamefont {Liu}\ \emph {et~al.}(2022)\citenamefont {Liu},
  \citenamefont {Zhang}, \citenamefont {Nie}, \citenamefont {Liu},
  \citenamefont {Sun}, \citenamefont {Wang}, \citenamefont {Ding},
  \citenamefont {Jiang}, \citenamefont {Sun}, \citenamefont {Xue},
  \citenamefont {Huang}, \citenamefont {Su}, \citenamefont {Yang},
  \citenamefont {Jiang}, \citenamefont {Lu}, \citenamefont {Yuan},
  \citenamefont {Cho}, \citenamefont {Liu}, \citenamefont {Liu}, \citenamefont
  {Ye}, \citenamefont {Zhang}, \citenamefont {Weng}, \citenamefont {Liu},
  \citenamefont {Guo}, \citenamefont {Wang},\ and\ \citenamefont
  {Shen}}]{PhysRevLett.129.166402}%
  \BibitemOpen
  \bibfield  {author} {\bibinfo {author} {\bibfnamefont {W.~L.}\ \bibnamefont
  {Liu}}, \bibinfo {author} {\bibfnamefont {X.}~\bibnamefont {Zhang}}, \bibinfo
  {author} {\bibfnamefont {S.~M.}\ \bibnamefont {Nie}}, \bibinfo {author}
  {\bibfnamefont {Z.~T.}\ \bibnamefont {Liu}}, \bibinfo {author} {\bibfnamefont
  {X.~Y.}\ \bibnamefont {Sun}}, \bibinfo {author} {\bibfnamefont {H.~Y.}\
  \bibnamefont {Wang}}, \bibinfo {author} {\bibfnamefont {J.~Y.}\ \bibnamefont
  {Ding}}, \bibinfo {author} {\bibfnamefont {Q.}~\bibnamefont {Jiang}},
  \bibinfo {author} {\bibfnamefont {L.}~\bibnamefont {Sun}}, \bibinfo {author}
  {\bibfnamefont {F.~H.}\ \bibnamefont {Xue}}, \bibinfo {author} {\bibfnamefont
  {Z.}~\bibnamefont {Huang}}, \bibinfo {author} {\bibfnamefont
  {H.}~\bibnamefont {Su}}, \bibinfo {author} {\bibfnamefont {Y.~C.}\
  \bibnamefont {Yang}}, \bibinfo {author} {\bibfnamefont {Z.~C.}\ \bibnamefont
  {Jiang}}, \bibinfo {author} {\bibfnamefont {X.~L.}\ \bibnamefont {Lu}},
  \bibinfo {author} {\bibfnamefont {J.}~\bibnamefont {Yuan}}, \bibinfo {author}
  {\bibfnamefont {S.}~\bibnamefont {Cho}}, \bibinfo {author} {\bibfnamefont
  {J.~S.}\ \bibnamefont {Liu}}, \bibinfo {author} {\bibfnamefont {Z.~H.}\
  \bibnamefont {Liu}}, \bibinfo {author} {\bibfnamefont {M.}~\bibnamefont
  {Ye}}, \bibinfo {author} {\bibfnamefont {S.~L.}\ \bibnamefont {Zhang}},
  \bibinfo {author} {\bibfnamefont {H.~M.}\ \bibnamefont {Weng}}, \bibinfo
  {author} {\bibfnamefont {Z.}~\bibnamefont {Liu}}, \bibinfo {author}
  {\bibfnamefont {Y.~F.}\ \bibnamefont {Guo}}, \bibinfo {author} {\bibfnamefont
  {Z.~J.}\ \bibnamefont {Wang}},\ and\ \bibinfo {author} {\bibfnamefont
  {D.~W.}\ \bibnamefont {Shen}},\ }\bibfield  {title} {\bibinfo {title}
  {Spontaneous ferromagnetism induced topological transition in
  ${\mathrm{eub}}_{6}$},\ }\href
  {https://doi.org/10.1103/PhysRevLett.129.166402} {\bibfield  {journal}
  {\bibinfo  {journal} {Phys. Rev. Lett.}\ }\textbf {\bibinfo {volume} {129}},\
  \bibinfo {pages} {166402} (\bibinfo {year} {2022})}\BibitemShut {NoStop}%
\bibitem [{\citenamefont {Su}\ \emph {et~al.}(2020)\citenamefont {Su},
  \citenamefont {Gong}, \citenamefont {Shi}, \citenamefont {Yang},
  \citenamefont {Wang}, \citenamefont {Xia}, \citenamefont {Yu}, \citenamefont
  {Guo}, \citenamefont {Wang}, \citenamefont {Ding}, \citenamefont {Xu},
  \citenamefont {Li}, \citenamefont {Wang}, \citenamefont {Zou}, \citenamefont
  {Yu}, \citenamefont {Zhu}, \citenamefont {Chen}, \citenamefont {Liu},
  \citenamefont {Liu}, \citenamefont {Li},\ and\ \citenamefont
  {Guo}}]{doi:10.1063/1.5129467}%
  \BibitemOpen
  \bibfield  {author} {\bibinfo {author} {\bibfnamefont {H.}~\bibnamefont
  {Su}}, \bibinfo {author} {\bibfnamefont {B.}~\bibnamefont {Gong}}, \bibinfo
  {author} {\bibfnamefont {W.}~\bibnamefont {Shi}}, \bibinfo {author}
  {\bibfnamefont {H.}~\bibnamefont {Yang}}, \bibinfo {author} {\bibfnamefont
  {H.}~\bibnamefont {Wang}}, \bibinfo {author} {\bibfnamefont {W.}~\bibnamefont
  {Xia}}, \bibinfo {author} {\bibfnamefont {Z.}~\bibnamefont {Yu}}, \bibinfo
  {author} {\bibfnamefont {P.-J.}\ \bibnamefont {Guo}}, \bibinfo {author}
  {\bibfnamefont {J.}~\bibnamefont {Wang}}, \bibinfo {author} {\bibfnamefont
  {L.}~\bibnamefont {Ding}}, \bibinfo {author} {\bibfnamefont {L.}~\bibnamefont
  {Xu}}, \bibinfo {author} {\bibfnamefont {X.}~\bibnamefont {Li}}, \bibinfo
  {author} {\bibfnamefont {X.}~\bibnamefont {Wang}}, \bibinfo {author}
  {\bibfnamefont {Z.}~\bibnamefont {Zou}}, \bibinfo {author} {\bibfnamefont
  {N.}~\bibnamefont {Yu}}, \bibinfo {author} {\bibfnamefont {Z.}~\bibnamefont
  {Zhu}}, \bibinfo {author} {\bibfnamefont {Y.}~\bibnamefont {Chen}}, \bibinfo
  {author} {\bibfnamefont {Z.}~\bibnamefont {Liu}}, \bibinfo {author}
  {\bibfnamefont {K.}~\bibnamefont {Liu}}, \bibinfo {author} {\bibfnamefont
  {G.}~\bibnamefont {Li}},\ and\ \bibinfo {author} {\bibfnamefont
  {Y.}~\bibnamefont {Guo}},\ }\bibfield  {title} {\bibinfo {title} {Magnetic
  exchange induced {Weyl }state in a semimetal {EuCd$_2$Sb$_2$}},\ }\href
  {https://doi.org/10.1063/1.5129467} {\bibfield  {journal} {\bibinfo
  {journal} {APL Materials}\ }\textbf {\bibinfo {volume} {8}},\ \bibinfo
  {pages} {011109} (\bibinfo {year} {2020})}\BibitemShut {NoStop}%
\bibitem [{\citenamefont {Cococcioni}\ and\ \citenamefont
  {de~Gironcoli}(2005)}]{Cococcioni2005}%
  \BibitemOpen
  \bibfield  {author} {\bibinfo {author} {\bibfnamefont {M.}~\bibnamefont
  {Cococcioni}}\ and\ \bibinfo {author} {\bibfnamefont {S.}~\bibnamefont
  {de~Gironcoli}},\ }\bibfield  {title} {\bibinfo {title} {Linear response
  approach to the calculation of the effective interaction parameters in the
  $\mathrm{LDA}+\mathrm{U}$ method},\ }\href
  {https://doi.org/10.1103/PhysRevB.71.035105} {\bibfield  {journal} {\bibinfo
  {journal} {Phys. Rev. B}\ }\textbf {\bibinfo {volume} {71}},\ \bibinfo
  {pages} {035105} (\bibinfo {year} {2005})}\BibitemShut {NoStop}%
\bibitem [{\citenamefont {Agapito}\ \emph {et~al.}(2015)\citenamefont
  {Agapito}, \citenamefont {Curtarolo},\ and\ \citenamefont
  {Buongiorno~Nardelli}}]{Agapito2015}%
  \BibitemOpen
  \bibfield  {author} {\bibinfo {author} {\bibfnamefont {L.~A.}\ \bibnamefont
  {Agapito}}, \bibinfo {author} {\bibfnamefont {S.}~\bibnamefont {Curtarolo}},\
  and\ \bibinfo {author} {\bibfnamefont {M.}~\bibnamefont
  {Buongiorno~Nardelli}},\ }\bibfield  {title} {\bibinfo {title}
  {{Reformulation of $\mathrm{DFT}$+{U} as a Pseudohybrid {Hubbard} Density
  Functional for Accelerated Materials Discovery}},\ }\href
  {https://doi.org/10.1103/PhysRevX.5.011006} {\bibfield  {journal} {\bibinfo
  {journal} {Phys. Rev. X}\ }\textbf {\bibinfo {volume} {5}},\ \bibinfo {pages}
  {011006} (\bibinfo {year} {2015})}\BibitemShut {NoStop}%
\bibitem [{\citenamefont {Aras}\ and\ \citenamefont {Kilic}(2014)}]{Aras14}%
  \BibitemOpen
  \bibfield  {author} {\bibinfo {author} {\bibfnamefont {M.}~\bibnamefont
  {Aras}}\ and\ \bibinfo {author} {\bibfnamefont {C.}~\bibnamefont {Kilic}},\
  }\bibfield  {title} {\bibinfo {title} {Combined hybrid functional and {DFT+U}
  calculations for metal chalcogenides},\ }\href
  {https://doi.org/10.1063/1.4890458} {\bibfield  {journal} {\bibinfo
  {journal} {J. Chem. Phys.}\ }\textbf {\bibinfo {volume} {141}},\ \bibinfo
  {pages} {044106} (\bibinfo {year} {2014})}\BibitemShut {NoStop}%
\bibitem [{\citenamefont {Sai~Gautam}\ and\ \citenamefont
  {Carter}(2018)}]{Sai18}%
  \BibitemOpen
  \bibfield  {author} {\bibinfo {author} {\bibfnamefont {G.}~\bibnamefont
  {Sai~Gautam}}\ and\ \bibinfo {author} {\bibfnamefont {E.~A.}\ \bibnamefont
  {Carter}},\ }\bibfield  {title} {\bibinfo {title} {Evaluating transition
  metal oxides within {DFT-SCAN} and {$\text{SCAN}+U$} frameworks for solar
  thermochemical applications},\ }\href
  {https://doi.org/10.1103/PhysRevMaterials.2.095401} {\bibfield  {journal}
  {\bibinfo  {journal} {Phys. Rev. Materials}\ }\textbf {\bibinfo {volume}
  {2}},\ \bibinfo {pages} {095401} (\bibinfo {year} {2018})}\BibitemShut
  {NoStop}%
\bibitem [{\citenamefont {Long}\ \emph {et~al.}(2020)\citenamefont {Long},
  \citenamefont {Sai~Gautam},\ and\ \citenamefont {Carter}}]{Long20}%
  \BibitemOpen
  \bibfield  {author} {\bibinfo {author} {\bibfnamefont {O.~Y.}\ \bibnamefont
  {Long}}, \bibinfo {author} {\bibfnamefont {G.}~\bibnamefont {Sai~Gautam}},\
  and\ \bibinfo {author} {\bibfnamefont {E.~A.}\ \bibnamefont {Carter}},\
  }\bibfield  {title} {\bibinfo {title} {Evaluating optimal {$U$} for $3d$
  transition-metal oxides within the {SCAN+$U$} framework},\ }\href
  {https://doi.org/10.1103/PhysRevMaterials.4.045401} {\bibfield  {journal}
  {\bibinfo  {journal} {Phys. Rev. Materials}\ }\textbf {\bibinfo {volume}
  {4}},\ \bibinfo {pages} {045401} (\bibinfo {year} {2020})}\BibitemShut
  {NoStop}%
\bibitem [{\citenamefont {Autieri}\ \emph {et~al.}(2021)\citenamefont
  {Autieri}, \citenamefont {\'Sliwa}, \citenamefont {Islam}, \citenamefont
  {Cuono},\ and\ \citenamefont {Dietl}}]{Autieri:2021_PRB}%
  \BibitemOpen
  \bibfield  {author} {\bibinfo {author} {\bibfnamefont {C.}~\bibnamefont
  {Autieri}}, \bibinfo {author} {\bibfnamefont {C.}~\bibnamefont {\'Sliwa}},
  \bibinfo {author} {\bibfnamefont {R.}~\bibnamefont {Islam}}, \bibinfo
  {author} {\bibfnamefont {G.}~\bibnamefont {Cuono}},\ and\ \bibinfo {author}
  {\bibfnamefont {T.}~\bibnamefont {Dietl}},\ }\bibfield  {title} {\bibinfo
  {title} {Momentum-resolved spin splitting in {Mn}-doped trivial {CdTe} and
  topological {HgTe} semiconductors},\ }\href
  {https://doi.org/10.1103/PhysRevB.103.115209} {\bibfield  {journal} {\bibinfo
   {journal} {Phys. Rev. B}\ }\textbf {\bibinfo {volume} {103}},\ \bibinfo
  {pages} {115209} (\bibinfo {year} {2021})}\BibitemShut {NoStop}%
\bibitem [{\citenamefont {Dietl}\ \emph {et~al.}(1994)\citenamefont {Dietl},
  \citenamefont {\ifmmode~\acute{S}\else \'{S}\fi{}liwa}, \citenamefont
  {Bauer},\ and\ \citenamefont {Pascher}}]{Dietl:1994_PRB}%
  \BibitemOpen
  \bibfield  {author} {\bibinfo {author} {\bibfnamefont {T.}~\bibnamefont
  {Dietl}}, \bibinfo {author} {\bibfnamefont {C.}~\bibnamefont
  {\ifmmode~\acute{S}\else \'{S}\fi{}liwa}}, \bibinfo {author} {\bibfnamefont
  {G.}~\bibnamefont {Bauer}},\ and\ \bibinfo {author} {\bibfnamefont
  {H.}~\bibnamefont {Pascher}},\ }\bibfield  {title} {\bibinfo {title}
  {Mechanisms of exchange interactions between carriers and mn or eu spins in
  lead chalcogenides},\ }\href {https://doi.org/10.1103/PhysRevB.49.2230}
  {\bibfield  {journal} {\bibinfo  {journal} {Phys. Rev. B}\ }\textbf {\bibinfo
  {volume} {49}},\ \bibinfo {pages} {2230} (\bibinfo {year}
  {1994})}\BibitemShut {NoStop}%
\bibitem [{\citenamefont {Atta-Fynn}\ and\ \citenamefont {Ray}(2009)}]{Atta09}%
  \BibitemOpen
  \bibfield  {author} {\bibinfo {author} {\bibfnamefont {R.}~\bibnamefont
  {Atta-Fynn}}\ and\ \bibinfo {author} {\bibfnamefont {A.~K.}\ \bibnamefont
  {Ray}},\ }\bibfield  {title} {\bibinfo {title} {Does hybrid density
  functional theory predict a non-magnetic ground state for $\delta$-{Pu}?},\
  }\href {https://doi.org/10.1209/0295-5075/85/27008} {\bibfield  {journal}
  {\bibinfo  {journal} {Europhys. Lett.}\ }\textbf {\bibinfo {volume} {85}},\
  \bibinfo {pages} {27008} (\bibinfo {year} {2009})}\BibitemShut {NoStop}%
\bibitem [{\citenamefont {Kirklin}\ \emph {et~al.}(2015)\citenamefont
  {Kirklin}, \citenamefont {Saal}, \citenamefont {Meredig}, \citenamefont
  {Thompson}, \citenamefont {Doak}, \citenamefont {Aykol}, \citenamefont
  {R{\"u}hl},\ and\ \citenamefont {Wolverton}}]{Kirklin2015}%
  \BibitemOpen
  \bibfield  {author} {\bibinfo {author} {\bibfnamefont {S.}~\bibnamefont
  {Kirklin}}, \bibinfo {author} {\bibfnamefont {J.~E.}\ \bibnamefont {Saal}},
  \bibinfo {author} {\bibfnamefont {B.}~\bibnamefont {Meredig}}, \bibinfo
  {author} {\bibfnamefont {A.}~\bibnamefont {Thompson}}, \bibinfo {author}
  {\bibfnamefont {J.~W.}\ \bibnamefont {Doak}}, \bibinfo {author}
  {\bibfnamefont {M.}~\bibnamefont {Aykol}}, \bibinfo {author} {\bibfnamefont
  {S.}~\bibnamefont {R{\"u}hl}},\ and\ \bibinfo {author} {\bibfnamefont
  {C.}~\bibnamefont {Wolverton}},\ }\bibfield  {title} {\bibinfo {title} {The
  open quantum materials database ({OQMD}): assessing the accuracy of {DFT}
  formation energies},\ }\href {https://doi.org/10.1038/npjcompumats.2015.10}
  {\bibfield  {journal} {\bibinfo  {journal} {npj Comp. Mater.}\ }\textbf
  {\bibinfo {volume} {1}},\ \bibinfo {pages} {15010} (\bibinfo {year}
  {2015})}\BibitemShut {NoStop}%
\end{thebibliography}%
\end{document}